\documentclass[aps,prb,reprint,amssymb,10pt]{revtex4-2}

\usepackage{amsmath,amssymb,color,braket,enumitem}
\usepackage{bm}
\usepackage{hyperref}
\usepackage{graphicx}
\usepackage[dvipsnames]{xcolor}

\DeclareMathOperator{\Tr}{Tr}

\begin{document}

\title{Functional renormalization group for extremely correlated electrons}

\author{Jonas Arnold}
\affiliation{Institut f\"{u}r Theoretische Physik, Universit\"{a}t Frankfurt,  Max-von-Laue Stra{\ss}e 1, 60438 Frankfurt, Germany}

\author{Peter Kopietz}
\affiliation{Institut f\"{u}r Theoretische Physik, Universit\"{a}t Frankfurt,  Max-von-Laue Stra{\ss}e 1, 60438 Frankfurt, Germany}

\author{Andreas R\"{u}ckriegel}
\affiliation{Institut f\"{u}r Theoretische Physik, Universit\"{a}t Frankfurt,  Max-von-Laue Stra{\ss}e 1, 60438 Frankfurt, Germany}

\date{July 23, 2026}

\begin{abstract}
At strong on-site repulsion $ U $,
the fermionic Hubbard model realizes an extremely correlated electron system.
In this regime, it is natural to derive the 
low-energy physics  with the help of  non-canonical operators acting on a projected Hilbert space without double occupancies.
Using a strong-coupling functional renormalization group technique,
we study the physics of such extreme correlations in the strict $ U = \infty $ limit,
where only kinematic interactions due to the Hilbert space projection remain.
For nearest-neighbor hopping on a square lattice,
we find that the electronic spectrum is significantly renormalized,
with bandwidth and quasi-particle residue strongly decreasing 
with increasing electron density.
On the other hand,
damping and particle-hole asymmetry increase,
while a polaronic continuum forms in the hole sector, 
below the single-particle band.
Fermi liquid phenomenology applies only at low densities,
where the system remains paramagnetic.
At higher densities,
we find a bad metal with strong magnetic correlations,
indicating that the ground state is the Nagaoka ferromagnet at high densities 
and a stripe antiferromagnet at intermediate densities.
Both in the paramagnetic and the ferromagnetic regimes,
we observe a violation of Luttinger's theorem.
\end{abstract}

\maketitle


\section{Introduction}






%
Strongly correlated electrons are of central interest in
modern condensed matter physics. In spite of many decades of intense research,
fundamental questions in this field remain open due a lack of controlled methods for dealing with strong interactions in dimensions larger than one.
The predictive power of numerical approaches is  limited by the
exponentially growing dimension the Hilbert space with the system size, while
analytical methods often suffer from uncontrolled or biased approximations.
A notable exception is the fermionic functional 
renormalization group (FRG)~\cite{Salmhofer2001, Honerkamp2001, Kopietz2001, Kopietz2010, Metzner2012, Dupuis2021}
which in the past 25 years has been  established as an unbiased theoretical tool for detecting  instabilities of interacting electrons on a lattice for weak to intermediate strength of the interaction~\cite{Halboth2000, Halboth2000b, Salmhofer2004, Ossadnik2008, Husemann2009, Husemann2012, Taranto2014, Lichtenstein2017, Vilardi2017, Honerkamp2018, Vilardi2019, Ehrlich2020, Hille2020, Honerkamp2022}.
%
%
Unfortunately, in the strong-coupling regime the inevitable truncations of the  formally exact FRG flow equations break down so that up until now
it has not been possible to apply 
FRG methods to models for correlated electrons in the regime where
the interaction is much larger than any other energy scale. 
%
In this work, we use our recently developed \cite{Rueckriegel2023} formulation of the FRG in terms of Hubbard X-operators \cite{Fulde1995,Izyumov1988,Ovchinnikov2004}
dubbed X-FRG to transcend  this limitation.

While the  X-FRG can be applied to any lattice model for strongly correlated electrons, it is
especially useful for models defined on projected Hilbert spaces where standard weak coupling methods fail. To  motivate these models, let us start from the 
Hubbard model, which ever since its inception~\cite{Hubbard1963}, 
has been  established as minimal model for 
electronic correlations in solids
encompassing such diverse phenomena as metal-insulator transitions,
ferro- and antiferromagnetism,
superconductivity,
and the Luttinger quantum liquid \cite{Auerbach1994, Fulde1995, Fazekas1999, Ovchinnikov2004}.
Its Hamiltonian reads
\begin{equation}\label{eq:normalHubbard}
	\mathcal{H}_H = \sum_{ij}\sum_\sigma t_{ij}c_{i\sigma }^\dagger  c_{j\sigma} + U \sum_i n_{i\uparrow} n_{i \downarrow}-\mu \sum_i \sum_\sigma n_{i\sigma} , 
\end{equation}
where $ n_{i\sigma} = c_{i\sigma}^\dagger c_{i\sigma} $ and
$c_{ i \sigma }$ are canonical fermionic operators 
that annihilate a spin-$ \sigma $ electron 
($ \sigma = \uparrow , \downarrow = + , - $)
at lattice site $ \bm{R}_i $.
These electrons are allowed to hop between lattice sites with amplitude $ t_{ i j } $,
while the  on-site repulsion $ U > 0 $ penalizes double occupancy of lattice sites.
The Hubbard model represents a crude approximation of the screened Coulomb interaction in real materials.
The total electron density is adjusted via the chemical potential $ \mu $.
At weak to intermediate values of the interaction $U$,
the Hubbard model \eqref{eq:normalHubbard} is by now well understood.
In this regime,
Landau's Fermi liquid theory holds in dimensions greater than one as long as there is no Fermi surface nesting~\cite{Virosztek1990, Schaefer2015, Rohringer2016, Simkovic2020, Kim2020},
so that the physics is perturbatively accessible from the $ U = 0 $ limit of non-interacting electrons~\cite{Shankar1994}. 
To transcend perturbation theory in the  bare interaction 
elaborate resummation schemes have been developed,
such as diagrammatic 
Monte Carlo \cite{Svistunov2010} or the 
FRG \cite{Wetterich1993, Berges2002, Pawlowski2007, Kopietz2010, Metzner2012, Dupuis2021}. 
These methods offer an unbiased means  of accessing non-perturbative physics in the regime where the interaction is not too large.

The situation is quite different in the strong-coupling regime, where $U$ is by far the largest energy scale in the system. To emphasize this, the term 
``extremely correlated'' for this regime has been coined by Shastry \cite{Shastry2010}.
Then doubly occupied lattice sites are effectively eliminated so that
the low-energy physics of the Hubbard model \eqref{eq:normalHubbard} at large enough $ U $ 
takes place in a projected Hilbert space without doubly occupied sites.
The fact the low-energy states of the Hubbard model \eqref{eq:normalHubbard} at small and large $ U $ live in Hilbert spaces with different dimensions implies that
one generally cannot obtain the strong-coupling physics  of the Hubbard model by means of a continuous extrapolation of the 
weak-coupling physics~\cite{Chiappe1993}.
Thus, 
any technique that is rooted in a perturbation expansion around the $ U = 0 $ limit is bound to fail for sufficiently large $ U $.
A more natural strong-coupling starting point is obtained by splitting the canonical fermionic operators into two parts,
$ c_{ i \sigma } 
= h_{ i \sigma }
+ d_{ i \sigma } $,
where
\begin{equation} \label{eq:h_via_c}
	h_{ i \sigma } =  
	\left(
	1 - n_{ i \bar{\sigma} }
	\right)
	c_{ i \sigma }
\end{equation}
creates a hole with spin $ \bar{\sigma} = - \sigma $
and
$ d_{ i \sigma } = n_{ i \bar{\sigma} } c_{ i \sigma } $
destroys a double occupancy~\cite{footnote_holon_and_Xoperators}.
We refer to the associated quasi-particles as holons and doublons.
At strong coupling,
the latter are high-energy excitations,
and only the former remain in the low-energy,
projected Hilbert space.
The effective Hamiltonian in this space is given by the so-called $ t $-$ J $ model \cite{Rice1988, Fulde1995, Fazekas1999, Ovchinnikov2004}
which has attracted a lot of attention in recent decades as a minimal model for the physics of high-temperature superconducting cuprates \cite{Dessau1993, Norman1998, Shen2005, Kanigel2007, Anderson2008, Anderson2009, Casey2011, Yang2011, Keimer2015, Sachdev2025}.
In particular,
it is believed that the Hilbert space projection is important for such unconventional features as the pseudogap and a disjointed Fermi surface made up of Fermi arcs  \cite{Norman1998,Sherman2003,Simkovic2024,Shen2005,Kanigel2007,Yang2011,Keimer2015,Sachdev2025}.
Therefore,
it is important to develop a deeper understanding of the physics of projected electrons.
To that end,
one should consider  the strict $ U = \infty $ limit first \cite{Anderson2008, Anderson2009, Casey2011}.
Then the exchange interaction $ J $ in the $t $-$J$ model vanishes
and the Hubbard model \eqref{eq:normalHubbard} reduces to the $ t $ model with Hamiltonian
\begin{equation}\label{eq:tModel}
	\mathcal{H} 
	=
	\sum_{ i j } \sum_\sigma
	t_{ i j }
	h_{ i \sigma }^\dagger 
	h_{ j \sigma } 
	- \mu \sum_i  \sum_\sigma
	\tilde{n}_{ i \sigma } ,
\end{equation}
where
$ \tilde{n}_{ i \sigma }
= h_{ i \sigma }^\dagger h_{ i \sigma }
= n_{ i \sigma }
\left(
1 - n_{ i \bar{\sigma} }
\right) $
is the occupation number in the projected Hilbert space.
The $ t $ model \eqref{eq:tModel} contains only the kinetic energy due to the hopping of electrons.
However,
it is nonetheless highly non-trivial because the hopping is correlated;
it implicitly contains a kinematic interaction due to the Hilbert space projection of the holon operators \eqref{eq:h_via_c}.
This interaction,
expressed in the non-canonical anti-commutation relation
\begin{equation}
	h_{ i \sigma } 
	h_{ j \sigma' }^\dagger  
	+
	h_{ j \sigma' }^\dagger
	h_{ i \sigma } 
	=
	\delta_{ i j } \left[
	\delta_{ \sigma \sigma' }
	\left(
	1 - \tilde{n}_{ i \bar{ \sigma } }
	\right)
	+ 
	\delta_{ \sigma \bar{ \sigma }' }
	h_{ i \bar{ \sigma } }^\dagger 
	h_{ i \sigma } 
	\right] ,
	\label{eq:holon_algebra}
\end{equation}
ensures that
$ \tilde{n}_{ i \uparrow }
+ \tilde{n}_{ i \downarrow }
= 0 , 1 $
and therefore cannot be treated in perturbation theory.
The non-trivial physics engendered by the projected Hilbert space of the $t$ model~\eqref{eq:tModel}
is exemplified strikingly by the Nagaoka theorem \cite{Nagaoka1965, Nagaoka1966, Kollar1996, Tasaki1998}, which guarantees that 
in dimensions greater than one and
on a bipartite lattice with periodic boundary conditions,
the ground state of the Hubbard model at infinite repulsion is ferromagnetic when the system  
is doped with a single hole away from half-filling.
For many decades
the question of whether Nagaoka ferromagnetism survives for finite hole doping 
in the thermodynamic limit has been the subject of intense research
\cite{Roth1967, Brinkman1970,  Aruiac1990, Elser1990, Kotrla1990, Izyumov1990, Shastry1990, vonderLinden1991, Putikka1992, Hanisch1993, Chiappe1993, Zotos1993, Wurth1995, Kollar1996, Kuzmin1997, Obermeier1997, Tasaki1998, Becca2001, Coleman2002, Zitzler2002, Park2008, Kumar2008, Baroni2011, Liu2012, Maska2012, Kochetov2017, Blesio2019, Morera2023, Samajdar2024a, Newby2025, Sharma2025}.
In view of the fact that the ground state of the Hubbard model on a bipartite lattice at half filling is known to be antiferromagnetic for arbitrary $ U $ \cite{Lieb1989}, 
the fate of Nagaoka ferromagnetism for finite hole doping is by no means obvious.
The renewed interest in Nagaoka ferromagnetism is also motivated by
recent experimental progress realizing the $t$ model on optical lattices~\cite{Cheuk2016, Dehollain2019, Bohrdt2021, Spar2021, Lebrat2024, Prichard2024, Kendrick2025}.
%
%
Unfortunately, numerical studies of Nagaoka ferromagnetism  suffer from severe finite-size issues because its emergence is accompanied by a complete reorganization of the ground state at arbitrary distances. 
It would therefore be desirable to have a controlled and unbiased method to address strong-coupling,
extremely correlated physics directly in the thermodynamic limit.
For that purpose,
strong-coupling and high-temperature expansions appear to offer a convenient starting point.
They set out from the exactly solvable atomic limit and incorporate the non-local hopping $ t_{ i j } $ as perturbation.
However,
these expansions also suffer from finite-size effects:
For nearest-neighbor hopping an $ n $-th order term in such a series only contains correlations between at most $ n + 1 $ lattice sites.
To access the long-range,
non-local physics implied by the Nagaoka theorem,
we therefore expect that a resummation of the series to infinite order is necessary.
Such a resummation on the other hand requires a convenient formalism for constructing and classifying the individual terms of the expansion,
like diagrammatic perturbation theory does in the weak-coupling limit.
Over the decades,
it has been realized several times that similar diagrammatic languages in fact exist also for strong-coupling and high-temperature expansions \cite{Zaitsev1976, Izyumov1990, Metzner1991, Izyumov1992,Pairault1998, Pairault2000, Ovchinnikov2004, Izyumov2005, Perepelitsky2015, Carlstroem2021, Rueckriegel2023}.
Unfortunately,
the proposed diagrammatic rules tend to be rather complicated,
because at each vertex there is a Hubbard atom with non-trivial quantum dynamics.
This may be the reason why
the diagrammatic strong-coupling expansions have never gained widespread popularity,
despite some promising results \cite{Zaitsev1976, Izyumov1990, Izyumov1992, Izyumov1994, Khatami2013, Khatami2014, Mai2018, Rueckriegel2023, Carlstroem2025}.
An alternative approach to the strong-coupling problem based on diagrammatic extension of dynamical mean field theory has recently also gained attention~\cite{Rohringer2018}.

\begin{figure}%
\includegraphics[width=\linewidth]{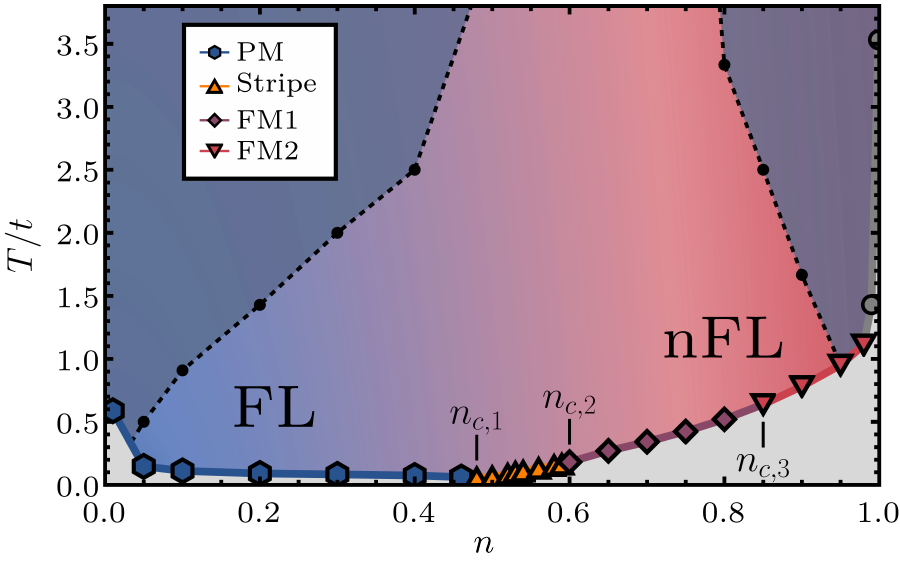}
\caption{Temperature-density phase diagram of the $t$ model obtained from the numerical solution of the X-FRG flow equations. 
Colored symbols show the temperature where the flow breaks down for a given density $ n $.
Black circles indicate the temperature below which the system becomes conducting.
Connecting lines are guides to the eye.
At densities 
$ n < n_{ c , 1 } = 0.48 $,
the $ t $ model remains a paramagnetic (PM) Fermi liquid (FL).
Flow breakdown in this regime,
as well as for $ n \gtrsim 0.95 $ (gray circle), 
is due to the finite number of Matsubara frequencies in our numerical implementation.
For $ n \ge n_{ c , 1 } $
the $ t $ model exhibits non-Fermi liquid behavior (nFL).
The flow breaks down in this density range because of physical instabilities indicative of magnetically ordered ground states---a stripe antiferromagnet for 
$ n_{ c, 1 } \le n < n_{ c , 2 } = 0.6 $,
and two disinct ferromagnets 
for $ n_{ c , 2 } \le n < n_{ c , 3 } = 0.85 $ (FM1)
and $ n_{ c , 3 } \le n < 1 $ (FM2).
The latter are separated by a Lifshitz transition of the electronic Fermi surface.}
\label{fig:PhaseDiag}
\end{figure}%
In this work,
we show that in the framework of our recently developed X-FRG \cite{Rueckriegel2023},
it is possible to combine the advantages of the unbiased strong-coupling expansion with 
the full power of the established FRG formalism for canonical fermions.
This provides the sought-for unbiased and non-perturbative approach to the physics of 
extremely correlated electrons
in the thermodynamic limit.
To illustrate this,
we apply the X-FRG to the $ t $ model \eqref{eq:tModel} on the square lattice,
with nearest-neighbor hopping of strength $ t > 0 $.
Our key physical results,
already presented briefly in Ref.~\cite{Arnold2025},
are summarized in the phase diagram shown in Fig.~\ref{fig:PhaseDiag}.
In particular,
we find that the Nagaoka ferromagnet exists also in the thermodynamic limit at finite hole doping,
in good agreement with most other studies \cite{Shastry1990, Izyumov1990, vonderLinden1991, Hanisch1993, Wurth1995, Obermeier1997, Becca2001, Park2008, Kumar2008, Baroni2011, Liu2012, Blesio2019, Newby2025}.
Moreover,
both from the divergence of the ferromagnetic correlation length and from the Fermi surface topology,
we have evidence for the existence of a second,
distinct ferromagnetic phase at lower densities \cite{vonderLinden1991, Chiappe1993, Hanisch1993, Wurth1995, Zitzler2002, Coleman2002, Becca2001}.
At still lower densities,
around quarter filling
we find antiferromagnetic stripe order,
before the system enters a paramagnetic regime at low densities.
Beyond the magnetic order,
we also compute the momentum-resolved electronic spectral function.
It reveals that with the onset of magnetic correlations at elevated densities,
the $ t $ model transitions from a Fermi liquid
to an incoherent bad metal \cite{Haule2003, Park2008, Wang2018}.
The latter is characterized by a marked particle-hole asymmetry and large band tails in the hole sector,
which we associate with long-range spin polaron states \cite{Brinkman1970, White2001, Maska2012, Lebrat2024, Prichard2024, Samajdar2024a}.
Another hallmark of strong correlations that we observe 
is the apparent breakdown of Luttinger's theorem \cite{Luttinger1960, Stephan1991, Singh1992, Putikka1998, Oshikawa2000, Kokalj2007, Shastry2010, Kozik2015, Seki2017, Quinn2018, Shastry2019, Osborne2021, Rueckriegel2023},
which states that the volume enclosed by the Fermi surface is equal to the electronic density.

The remainder of this paper is organized as follows:
In Sec.~\ref{sec:X-FRG},
we present the basic building blocks of the X-FRG approach to the $ t $ model \eqref{eq:tModel},
as well as our truncation strategy for the flow equations.
The central Sec.~\ref{sec:results} discusses the results obtained from the numerical solution of the truncated X-FRG flow in depth.
In Sec.~\ref{sec:conclusions} we summarize our main results and identify a number of 
open problems which can be solved  by means of extensions of the methods developed in this work.
Additional technical details and consistency checks are relegated to six appendices.

\section{Fermionic X-FRG for the $ t $ model}

\label{sec:X-FRG}

\subsection{The X-FRG framework}

The basic idea of the X-FRG \cite{Rueckriegel2023} is to continuously deform the hopping amplitudes,
$ t_{ i j } \to 
t_{ \Lambda , i j } $.
Here,
$ \Lambda \in [ 0 , 1 ] $ parametrizes the deformation,
which is chosen such that
$ t_{ \Lambda = 0 , i j } = 0 $ and
$ t_{ \Lambda = 1 , i j } = t_{ i j } $.
Thus,
our X-FRG flow starts from the exactly solvable limit of isolated Hubbard atoms,
thereby taking the holon algebra \eqref{eq:holon_algebra} into account \textit{exactly}.
In practice,
we choose a simple multiplicative deformation
$ t_{ \Lambda , i j } 
= \Lambda t_{ i j } $.
This has the advantage that the X-FRG flow can,
upon suitable rescaling, 
be interpreted as a flow of the temperature 
$ T = 1 / \beta $;
see Appendix~\ref{app:temperature}.
For later convenience,
we also deform the chemical potential,
$ \mu \to 
\mu_\Lambda = \mu_0 + \delta \mu_\Lambda  $,
where
\begin{equation}
\label{eq:mu0}
\mu_0 = T \ln \left( \frac{ n }{ 2 - 2 n } \right) 
\end{equation}
is the atomic limit for a given holon density $ n $.
The X-FRG of the $ t $ model then sets out from the following deformed generating functional of imaginary-time ordered connected holon correlation functions:
\begin{widetext}
\begin{align}
\mathcal{G}_\Lambda [ \bar{j} , j ]
- \frac{ 2 }{ 3 } \beta N \delta \mu_\Lambda
= {} &
\ln \Tr 
\left[
e^{ 
\beta \mu_0 
\sum_{ i \sigma } \tilde{ n }_{ i \sigma } 
}
\mathcal{T}
\exp \left\{
- 
\int_0^\beta \textrm{d} \tau
\sum_{ i j \sigma }
\left( 
t_{ \Lambda , i j } -
\frac{ 2 }{ 3 } \delta \mu_\Lambda 
\delta_{ i j }
\right)
\frac{ 1 }{ 2 }
\sum_{ p = \pm }
h_{ i \sigma }^\dagger ( \tau + p 0^+ )
h_{ j \sigma } ( \tau )
\right\}
\right.
\nonumber\\[.2cm]
& 
\hspace{2.7cm}
\times \left.
\exp \left\{
\int_0^\beta \textrm{d} \tau
\sum_{ i \vphantom{j} \sigma } 
\left[
\bar{j}_{ i \sigma } ( \tau )
h_{ i \sigma } ( \tau )
+
h_{ i \sigma }^\dagger ( \tau )
j_{ i \sigma } ( \tau )
\right]
\right\}
\right] \; .
\label{eq:generating_functional_G}
\end{align}
Here,
$ h_{ i \sigma } ( \tau ) =
e^{ \mu_0 \tau } h_{ i \sigma } $
are the holon operators in the interaction representation with respect to the atomic limit,
$ \mathcal{T} $
denotes imaginary-time ordering,
$ j_{ i \sigma } ( \tau ) $ and
$ \bar{j}_{ i \sigma } ( \tau ) $
are Grassmann source fields that anti-commute with the holon operators,
and we choose a symmetric regularization $ \pm 0^+ $ to resolve equal-time ambiguities~\cite{footnote_reg}. 
This is necessary in order to evaluate non-convergent Matsubara sums numerically later on,
which for a finite number of Matsubara frequencies automatically result in the symmetric regularization.
Next,
we introduce the generating functional of one-line irreducible holon vertices as shifted Legendre transform of Eq.~\eqref{eq:generating_functional_G},
\begin{subequations}
\label{eq:generating_functional_vertices}
\begin{align}
\Gamma_\Lambda [ \bar{ \psi } , \psi ]
= {} &
\int_0^\beta \textrm{d} \tau
\sum_{ i \sigma } 
\left[
\bar{j}_{ i \sigma } ( \tau )
\psi_{ i \sigma } ( \tau )
+
\bar{\psi}_{ i \sigma } ( \tau )
j_{ i \sigma } ( \tau )
\right]
- \mathcal{G}_\Lambda [ \bar{j} , j ]
+ \frac{ 2 }{ 3 } \beta N \delta \mu_\Lambda
\nonumber\\[.1cm]
&
- \int_0^\beta \textrm{d} \tau
\sum_{ i j \sigma }
\left( 
t_{ \Lambda , i j } -
\frac{ 2 }{ 3 } \delta \mu_\Lambda 
\delta_{ i j }
\right)
\frac{ 1 }{ 2 }
\sum_{ p = \pm }
\bar{ \psi }_{ i \sigma } ( \tau + p 0^+ )
\psi_{ j \sigma } ( \tau )
\\[.2cm]
= {} &
\beta N \tilde{f}_\Lambda
+ \int_{ K } \sum_{ \sigma }
\left[
- \frac{ i \omega + \mu_0 }{ Z }
+ \Sigma_\Lambda ( K )
\right]
\bar{ \psi }_{ K \sigma }
\psi_{ K \sigma }
\nonumber\\[.1cm]
&
+ \frac{ 1 }{ 2 }
\int_{ K_1' K_2' K_2 K_1 }
\sum_{ \sigma_1 \sigma_2 }
\delta_{ K_1' + K_2' , K_2 + K_1 }
U_\Lambda ( K_1' , K_2' ; K_2 , K_1 )
\bar{ \psi }_{ K_1' \sigma_1 }
\bar{ \psi }_{ K_2' \sigma_2 }
\psi_{ K_2 \sigma_2 }
\psi_{ K_1 \sigma_1 }
+ \ldots \; ,
\end{align}
\end{subequations}
\end{widetext}
where
$ K = ( \bm{k} , \omega ) $
collects crystal momentum $ \bm{k} $
and fermionic Matsubara frequency $ \omega $,
and we have introduced the integral symbol
$ \int_K 
= ( \beta N )^{ - 1 } \sum_{ \bm{k} \omega } $ 
and the delta symbol
$ \delta_{ K , K' }
= \beta N 
\delta_{ \bm{k} , \bm{k}' }
\delta_{ \omega , \omega' } $.
It is also understood that the sources are obtained as functionals of the average holon fields 
$ \psi_{ i \sigma } ( \tau ) $ and
$ \bar{ \psi }_{ i \sigma } ( \tau ) $
via inversion of
\begin{subequations}
\begin{align}
\psi_{ i \sigma } ( \tau )
& =
\int_K 
e^{ i \bm{k} \cdot \bm{R}_i - i \omega \tau }
\psi_{ K \sigma }
= 
\frac{ 
\delta \mathcal{G}_\Lambda [ \bar{j} , j ] 
}{
\delta \bar{j}_{ i \sigma } ( \tau )
}
\; ,
\\[.1cm]
\bar{\psi}_{ i \sigma } ( \tau )
& =
\int_K 
e^{ i \bm{k} \cdot \bm{R}_i - i \omega \tau }
\bar{\psi}_{ K \sigma }
= -
\frac{ 
\delta \mathcal{G}_\Lambda [ \bar{j} , j ] 
}{
\delta j_{ i \sigma } ( \tau )
} 
\; .
\end{align}
\end{subequations}
The holon propagator then reads
\begin{subequations}
\label{eq:holon_propagator}
\begin{align}
G_\Lambda ( K ) 
& = 
-
\int_0^\beta \textrm{d} \tau \sum_i 
e^{ - i \bm{k} \cdot \bm{R}_i + i \omega \tau }
\Braket{ 
\mathcal{T}
h_{ i \sigma } ( \tau ) 
h_{ j \sigma }^\dagger ( 0 )
} 
\\
& =
\frac{ Z }{ 
i \omega + \mu_0 
- Z \left( 
t_{ \Lambda , \bm{k} }
- 2 \delta \mu_\Lambda / 3 
\right)
- Z \Sigma_\Lambda ( K )
} \; ,
\end{align}
\end{subequations}
with the quasi-particle residue 
$ Z = 1 - n / 2  $
of the Hubbard atom,
and the holon self-energy
$ \Sigma_\Lambda ( K ) $.
The latter contains the hopping induced correlation effects that are generated at $ \Lambda > 0 $.
Since these become irrelevant at high energies or short times 
where the frequency is much larger than the hopping amplitude 
($ | \omega | \gg t $),
the high-frequency asymptotic of the holon propagator \eqref{eq:holon_propagator} is
$ G_\Lambda ( K ) \sim Z / i \omega $,
which is density dependent.
To fix the high-energy tail of the propagator during the flow, 
and thereby reduce the regularization dependence of Matsubara sums appearing later on in Eqs.~\eqref{eq:Pi_dot} and \eqref{eq:P_dot},
we now choose the flowing correction 
$ \delta \mu_\Lambda $
to the chemical potential such that
the density $ n $ remains constant during the flow.
This means that we keep  the integral
\begin{subequations}
\begin{align}
\int_K \cos ( \omega 0^+ ) G_\Lambda ( K )
= {} &
\frac{ 1 }{ 2 }
\Braket{
h_{ i \sigma }^\dagger h_{ i \sigma } -
h_{ i \sigma } h_{ i \sigma }^\dagger
}
\\
= {} & 
\frac{ 1 }{ 2 } \left( 
\frac{ 3 }{ 2 } n - 1
\right)
\end{align}
\end{subequations}
constant,
yielding the following flow equation for the chemical potential shift 
$ \delta \mu_\Lambda $:
\begin{equation}\label{eq:counter-term}
\partial_\Lambda \delta \mu_\Lambda =
\frac{ 3
\int_K \cos ( \omega 0^+ )  
G_\Lambda^2 ( K )
\partial_\Lambda \left[ 
t_{ \Lambda , \bm{k} } +
\Sigma_\Lambda ( K )
\right]
}{ 2
\int_K \cos ( \omega 0^+ ) 
G_\Lambda^2 ( K ) 
} \; .
\end{equation}
The holon self-energy $\Sigma_{\Lambda} ( K )$ 
in Eq.~(\ref{eq:holon_propagator}) satisfies the exact flow equation~\cite{Kopietz2010,Rueckriegel2023}
 \begin{align}
 \partial_{\Lambda} \Sigma_{\Lambda} ( K )
  = \int_{ K'} \dot{G}_{\Lambda} ( K' ) 
  \bigl[ & 2 U_{\Lambda} ( K , K'; K', K ) 
 \nonumber
  \\  
    & \hspace{-2mm} - U_{\Lambda} ( K, K' ; K, K' ) \bigr],
    \label{eq:flowself}
  \end{align}
which is shown graphically in Fig.~\ref{fig:su2flow} (a). 
Here, 
$\dot{G}_{\Lambda} ( K )$ is the so-called single-scale propagator, 
which in our deformation scheme is given by
\begin{equation} 
\dot{G}_{\Lambda} ( K ) = 
\cos ( \omega 0^+) 
G_{\Lambda}^2 ( K ) 
\partial_{\Lambda}
( t_{\Lambda, \bm{k}} - 2 \delta \mu_{\Lambda} /3 ) .
\label{eq:singlescale}
\end{equation}

\subsection{Climbing ladders}
\begin{figure*}%
\includegraphics[width=\linewidth]{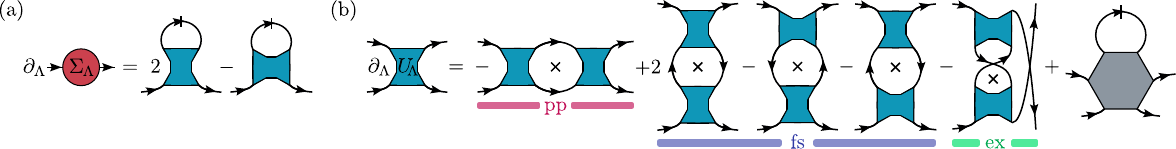}
\caption{Diagrammatic flow equations of 
(a) holon self-energy 
$ \Sigma_\Lambda ( K ) $ and 
(b) two-body interaction vertex
$ U_\Lambda ( K_1' , K_2' ; K_2 , K_1 ) $.
The holon propagator 
$ G_\Lambda ( K ) $
is represented by a line with an arrow.
If the line is slashed,
it has to be replaced by the corresponding single-scale propagator
$ \dot{G}_\Lambda ( K )
= \cos ( \omega 0^+) G_\Lambda^2 ( K ) \partial_\Lambda
( t_{ \Lambda , \bm{k} } 
- 2 \delta \mu_\Lambda / 3 ) $.
Closed loops with a cross inside mean a sum of diagrams where each 
$ G_\Lambda ( K ) $ is in turn replaced by $ \dot{G}_\Lambda ( K ) $.}
\label{fig:su2flow}
\end{figure*}%
The flow of the self-energy 
$ \Sigma_\Lambda ( K ) $
is governed by the 
2-body interaction vertex
$ U_\Lambda ( K_1' , K_2' ; K_2 , K_1 ) $,
which is itself determined by three types of  ladder diagrams and the 3-body interaction vertex as shown graphically in  Fig.~\ref{fig:su2flow}~(b).
Neglecting for simplicity the 3-body vertex, the flow of  the effective interaction is determined by the
following closed flow equation:
\begin{align}
\partial_{\Lambda}  
U_{\Lambda} ( K_1^{\prime} , K_2^{\prime} ; K_2 , K_1 ) 
& =  
{\cal{T}}_{\Lambda}^{\rm pp} (  K_1^{\prime} , K_2^{\prime} ; K_2 , K_1 ) 
 \nonumber
 \\ 
 & +
 {\cal{T}}_{\Lambda}^{\rm fs} (  K_1^{\prime} , K_2^{\prime} ; K_2 , K_1 ) 
 \nonumber
 \\ 
 & +
 {\cal{T}}_{\Lambda}^{\rm ex} (  K_1^{\prime} , K_2^{\prime} ; K_2 , K_1 ) .
\label{eq:Uflow}
\end{align}
The three terms on the right-hand side represent distinct  renormalizations of the
effective interaction in
the
particle-particle channel (pp),
in the forward-scattering channel (fs), and in the 
exchange-scattering channel (ex).
Each of this channels depends on one of the following characteristic bosonic 
momentum-frequency transfers,
\begin{subequations}
\begin{align}
Q_\textrm{pp} & = K_1 + K_2
= K_1' + K_2' \; ,
\\
Q_\textrm{fs} & = K_1' - K_1
= K_2 - K_2' \; ,
\\
Q_\textrm{ex} & = K_1' - K_2
= K_1 - K_2' \; .
\end{align}
\end{subequations}
Note that here and in the following,
we denote bosonic momentum-frequency transfers as
$ Q = ( \bm{q} , \Omega ) $.
The contribution from the particle-particle channel reads
 \begin{align}\label{eq: Tpp}
&  {\cal{T}}_\Lambda^{\rm pp} (  K_1^{\prime} , K_2^{\prime} ; K_2 , K_1 ) 
  = - \int_K  \left[ G_{\Lambda} ( K ) G_{\Lambda} (Q_{\rm pp} - K ) \right]^{\bullet}
 \nonumber
 \\
 & \times U_{\Lambda} ( K_1^{\prime} , K_2^{\prime} ; Q_{\rm pp} - K , K )
    U_{\Lambda} ( K, Q_{\rm pp} - K; K_2 , K_1),
 \end{align} 
where we have introduced the notation
 \begin{align}
 & \left[ G_{\Lambda} ( K ) G_{\Lambda} ( K^{\prime}) \right]^{\bullet}
=  \dot{G}_{\Lambda} ( K ) G_{\Lambda} ( K^{\prime}) + 
  {G}_{\Lambda} ( K ) \dot{G}_{\Lambda} ( K^{\prime}).
  \label{eq:productrule}
  \end{align}
The forward scattering channel gives rise to the following contribution to the
flow of the effective interaction:
\begin{align}\label{eq: Tfs}
&  {\cal{T}}_{\Lambda}^{\rm fs} (  K_1^{\prime} , K_2^{\prime} ; K_2 , K_1 ) 
  =  \int_K  \left[ G_{\Lambda} ( K ) G_{\Lambda} ( K - Q_{\rm fs} ) \right]^{\bullet}
 \nonumber
 \\
 & \times \bigl[ 2 U_{\Lambda} ( K_1^{\prime} , K - Q_{\rm fs};  K , K_1 )
    U_{\Lambda} ( K_2^{\prime}, K; K- Q_{\rm fs} ,  K_2)
 \nonumber
 \\
    & \; \; -  U_{\Lambda} ( K_1^{\prime} , K - Q_{\rm fs};  K_1 , K )
    U_{\Lambda} ( K_2^{\prime}, K; K- Q_{\rm fs} ,  K_2)
     \nonumber
 \\
    & \; \; -  U_{\Lambda} ( K_1^{\prime} , K - Q_{\rm fs};  K , K_1 )
    U_{\Lambda} ( K_2^{\prime}, K; K_2 , K- Q_{\rm fs} )
  \bigr].    
 \end{align}
Finally, the contribution from exchange scattering is
\begin{align}\label{eq: Tex}
&  {\cal{T}}_{\Lambda}^{\rm ex} (  K_1^{\prime} , K_2^{\prime} ; K_2 , K_1 ) 
  = - \int_K  \left[ G_{\Lambda} ( K ) G_{\Lambda} ( K - Q_{\rm ex}) \right]^{\bullet}
 \nonumber
 \\
 & \times U_{\Lambda} ( K_1^{\prime} , K -  Q_{\rm ex};  K_2 , K )
    U_{\Lambda} ( K, K_2^\prime ; K-  Q_{\rm ex} , K_1).
 \end{align}

In the X-FRG,
the initial condition of the flow is determined by the correlation functions of the Hubbard atom,
and consequently contains the full information about the local holon algebra \eqref{eq:holon_algebra}.
This is encoded in the non-trivial frequency dependence of the initial vertex functions.
As the hopping becomes negligible for large enough frequencies,
the non-trivial dynamics of the Hubbard atom vertices necessarily also gives the correct high-frequency asymptotics of the vertices.
In order to obtain sensible results,
it is therefore vital to treat this frequency dependence as accurately as possible,
and only dress it with hopping-dependent corrections. 
The initial 2-body interaction vertex is explicitly given by \cite{Rueckriegel2023}
\begin{align}
&
Z^2 U_0 ( \omega_1', \omega_2' ; \omega_2 , \omega_1 ) =
- G_0^{ - 1 } ( \omega_2 )
- G_0^{ - 1 } ( \omega_1 )
\nonumber\\
&
+ 
\left(
\frac{ n^2 }{ 4 } 
\beta \delta_{ \Omega_\textrm{fs} , 0 }
+
\frac{ n }{ 2 } 
\beta \delta_{ \Omega_\textrm{ex} , 0 }
\right)
G_0^{ - 1 } ( \omega_2 )
G_0^{ - 1 } ( \omega_1 ) ,
\label{eq:U_0_scalar}
\end{align}
with the inverse Hubbard atom propagator 
 \begin{equation}
 G_0^{-1} ( \omega ) = \frac{ i \omega + \mu }{ Z } .
\end{equation}
This suggests an ansatz of the form
\begin{align}
&
U_\Lambda ( K_1' , K_2' ; K_2 , K_1 ) =
\nonumber\\
&
\textbf{v}^\top ( \omega_2 )
\textbf{U}_\Lambda ( K_1' , K_2' ; K_2 , K_1 )
\textbf{v} ( \omega_1 )
,
\label{eq:U_ansatz}
\end{align}
where
\begin{equation}
\label{eq:v}
\textbf{v} ( \omega ) =
\frac{ 1 }{ Z }
\begin{pmatrix}
1 \\
G_0^{ - 1 } ( \omega )
\end{pmatrix}
,
\end{equation}
and  $\textbf{U}_\Lambda ( K_1' , K_2' ; K_2 , K_1 )$
is a $2 \times 2$ matrix.
The initial condition \eqref{eq:U_0_scalar} can then be written as
\begin{equation}
\label{eq:U_0}
\textbf{U}_0 
( \omega_1' , \omega_2' ; \omega_2 , \omega_1 ) =
%
%
\begin{pmatrix}
0 & - 1 \\
- 1 & 
\frac{ n^2 }{ 4 } 
\beta \delta_{ \Omega_\textrm{fs} , 0 }
+
\frac{ n }{ 2 } 
\beta \delta_{ \Omega_\textrm{ex} , 0 }
\end{pmatrix} 
,
\end{equation}
which crucially depends only on bosonic transfer frequencies. 
The flow equations for the components of the $ 2 \times 2 $ matrix 
$ \textbf{U}_\Lambda $
can then be obtained by comparing the coefficients of the $ G_0^{ - 1 } $-terms on both sides of the flow equation \eqref{eq:Uflow} depicted in Fig.~\ref{fig:su2flow} (b).
A similar parametrization of a 2-body interaction vertex has been used previously to describe collective spin dynamics in quantum Heisenberg ferromagnets with the spin FRG \cite{Goll2020}.
In fact,
for system satisfying a generalized Wick theorem like the free spin and the Hubbard atom \cite{Rueckriegel2025},
frequency dependencies of this form are generic
because all correlation functions decompose into products of bare propagators and conserved susceptibilities 
$ \propto \beta \delta_{ \Omega , 0 } $.
The initial condition \eqref{eq:U_0} also highlights the need for a resummation at low temperatures $ T \ll | t | $:
In the flow of the self-energy,
one of the frequency-deltas of \eqref{eq:U_0} is not integrated over,
leading to a term that diverges as $ \beta $ for low temperatures \cite{footnote_beta_previous}.
Thus,
some form of resummation of the 2-body interaction vertex is required in order to obtain meaningful results away from the infinite-$T$ Hubbard atom limit. 
Such a resummation will be developed in the following.

\begin{figure}%
\includegraphics[width=\linewidth]{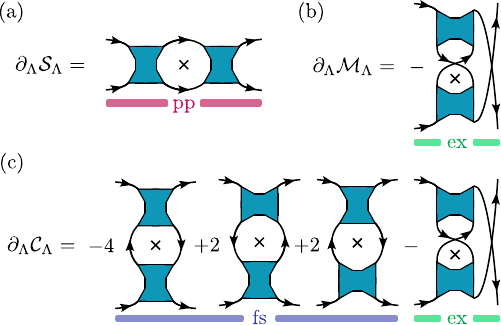}
\caption{Diagrammatic flow equations of 
(a) the superconducting channel
$\mathcal{S}_\Lambda(Q_\textrm{pp}; K_2',K_1)$, 
(b) the magnetic channel
$\mathcal{M}_\Lambda(Q_\textrm{ex}; K_2,K_1)$,
and
(c) the charge channel $\mathcal{C}_\Lambda(Q_\textrm{fs}; K_2,K_1) $.}
\label{fig:channelflows}
\end{figure}

Given the ansatz \eqref{eq:U_ansatz} that factors out the non-transfer, fermionic frequency dependence that is due to the holon algebra \eqref{eq:holon_algebra},
we can apply a channel-decomposition \cite{Husemann2009, Metzner2012, Vilardi2017, Vilardi2019} to the remaining $ 2 \times 2 $ interaction matrix: 
\begin{align}
\textbf{U}_\Lambda ( K_1' , K_2' ; K_2 , K_1 )
= {} &
\textbf{V}
- \textbf{S}_\Lambda ( Q_\textrm{pp} )
+ \textbf{M}_\Lambda ( Q_\textrm{ex}  )
\nonumber\\
& 
+ \frac{ 1 }{ 2 } \left[
\textbf{M}_\Lambda^\top ( Q_\textrm{fs} ) -
\textbf{C}_\Lambda ( Q_\textrm{fs} )
\right] .
\label{eq:U_channels}
\end{align}
Here 
$ \textbf{S}_\Lambda $,
$ \textbf{M}_\Lambda $, and
$ \textbf{C}_\Lambda $
are the 
superconducting, 
magnetic, and 
charge matrices,
respectively.
They relate to the physical
interaction channels via
\begin{subequations}\label{eq: formfactors}
	\begin{align}
		\mathcal{S}_\Lambda(Q_\textrm{pp}; K_2',K_1) ={}& \textbf{v}^\top (\omega_2 ) \textbf{S}_\Lambda(Q_\textrm{pp}) \textbf{v}(\omega_1),\\
		\mathcal{M}_\Lambda(Q_\textrm{ex}; K_2,K_1) ={}& \textbf{v}^\top (\omega_2 ) \textbf{M}_\Lambda(Q_\textrm{ex}) \textbf{v}(\omega_1),\\
		\mathcal{C}_\Lambda(Q_\textrm{fs}; K_2,K_1) ={}& \textbf{v}^\top (\omega_2 ) \textbf{C}_\Lambda(Q_\textrm{fs}) \textbf{v}(\omega_1).
	\end{align}
\end{subequations}
The transfer-frequency independent part of the
initial condition \eqref{eq:U_0} is contained in
\begin{equation}
\textbf{V} =
- 
\begin{pmatrix}
0 & 1 \\
1 & 0
\end{pmatrix}.
\end{equation}
This constant part could in principle be absorbed into any of the channels;
however,
to keep the decomposition \eqref{eq:U_channels} as unbiased as possible,
it should be pulled out. 
The initial conditions of the channels are then given by
\begin{equation}
\label{eq:SMC_0}
\begin{Bmatrix}
\textbf{S}_0 ( \Omega ) \\
\textbf{M}_0 ( \Omega ) \\
\textbf{C}_0 ( \Omega ) 
\end{Bmatrix} 
= 
\frac{ n }{ 2 } \beta \delta_{ \Omega , 0 }
\times
\begin{Bmatrix}
0 \\ 
1 \\
1 - n 
\end{Bmatrix} 
\times 
\begin{pmatrix}
0 & 0 \\
0 & 1 
\end{pmatrix}
.
\end{equation}
Note especially that since both magnetic and charge channels have finite initial conditions,
a single-channel truncation is a priori ruled out for the $ t $ model;
all three channels must be retained.
In order to obtain flow equations for the three channels,
we assign each to the ladder diagrams with the corresponding momentum-frequency transfer;
see Fig.~\ref{fig:channelflows}.
To reduce the numerical complexity of these three flow equations,
we now apply our two main approximations,
which are standard in fermionic FRG:
First,
we only keep the dependence of the channel matrices on the relevant momentum-frequency transfer.
This could be improved by an expansion in form factors 
in Eq.~\eqref{eq: formfactors}
\cite{footnote_d-wave, Husemann2009, Vilardi2017, Vilardi2019, Metzner2012,Profe2022}.
Second,
we retain only the leading instabilities in each channel,
which is necessary to obtain a consistent truncation that retains only the bosonic momentum-frequency transfers.
This amounts to dropping terms of the form
$ \textbf{S}_\Lambda ( Q_\textrm{pp} )
\textbf{S}_\Lambda ( Q_\textrm{pp} + K' ) $,
where $ K' $ is integrated over,
while retaining terms like
$ \textbf{S}_\Lambda^2 ( Q_\textrm{pp} ) $.
With these two simplifications,
our flow is similar to a multi-channel random-phase approximation for the interaction matrix $ \textbf{U}_\Lambda $,
with the crucial difference that the flow self-consistently includes self-energy corrections to all internal propagators.
In contrast to weak-coupling FRG where these self-energy corrections are often neglected \cite{Halboth2000, Halboth2000b, Honerkamp2001, Ossadnik2008, Husemann2009, Kopietz2010, Metzner2012, Husemann2012, Lichtenstein2017, Honerkamp2018, Ehrlich2020, Honerkamp2022},
they turn out to be essential in the limit of extreme correlations considered here 
because the large renormalization of single-particle physics must be taken into account.
They also provide an indirect feedback mechanism between the three channels;
direct feedback via channel-mixing terms in the flow equations is however neglected.
We expect this to be valid for the investigation of the leading instabilities that are encountered upon lowering the temperatures,
provided that there are no competing instabilities in different channels.
Going beyond the leading instabilities towards very low temperatures,
for example to observe Kohn-Luttinger pairing instabilities \cite{Kohn1965}, on the other hand needs a more sophisticated truncation that includes channel-mixing contributions as well \cite{Shankar1994, Husemann2009}.
This is, however, 
beyond the scope of this work.

Taking into account the initial condition \eqref{eq:SMC_0},
the flow of the superconducting channel is within our approximations then solved by 
\begin{equation}
\textbf{S}_\Lambda ( Q )
=
- S_\Lambda ( Q ) \textbf{V} ,
\end{equation}
where the scalar prefactor satisfies
\begin{equation}
\label{eq:S_dot}
\partial_\Lambda S_\Lambda ( Q )
=
\dot{ \Pi }_\Lambda ( Q ) 
\left[
1 + S_\Lambda ( Q )
\right]^2 
 ,
\end{equation}
with the single-scale particle-particle polarization bubble
\begin{align}
\dot{ \Pi }_\Lambda ( Q )
=  {} &
\frac{ 1 }{ Z^2 }
\int_{ K }
\left[
G_{ \Lambda } ( K )
G_{ \Lambda } ( Q - K )
\right]^\bullet
\nonumber\\
& 
\hspace{.6cm}
\times
\left[
G_0^{ - 1 } ( \omega ) +
G_0^{ - 1 } ( \Omega - \omega ) 
\right] .
\label{eq:Pi_dot}
\end{align}
Here, 
$ [ \ldots ]^\bullet $ is defined in Eq.~\eqref{eq:productrule},
i.e., each propagator in the bracket is successively replaced by its single-scale counterpart
$ \dot{G}_\Lambda ( K )$ defined in Eq.~\eqref{eq:singlescale}.
%
%
The solution of the flow equation \eqref{eq:S_dot} for the scalar superconducting channel is 
\begin{equation}
\label{eq:S_sol}
S_\Lambda ( Q ) =
\frac{ 
\Pi_\Lambda ( Q ) 
}{
1 -
\Pi_\Lambda ( Q )
} ,
\end{equation}
where
 \begin{equation}
\Pi_\Lambda ( Q )
= \int_0^\Lambda \textrm{d} \Lambda' 
\dot{ \Pi }_{ \Lambda' } ( Q ).
\end{equation}

Next,
we turn to the magnetic and charge channels.
Their (matrix-valued) flows are within our approximations governed by
\begin{subequations}
\label{eq:MC_dot}
\begin{align}
\partial_\Lambda \textbf{M}_\Lambda ( Q )
& =
- \left[ 
\textbf{V} +
\textbf{M}_\Lambda ( Q )
\right] 
{\bf \dot{ P } }_\Lambda ( Q ) 
\left[ 
\textbf{V} +
\textbf{M}_\Lambda ( Q )
\right] 
 , 
\\
\partial_\Lambda \textbf{C}_\Lambda ( Q )
& =
- \left[ 
- \textbf{V} +
\textbf{C}_\Lambda ( Q )
\right] 
{\bf \dot{ P } }_\Lambda^\top ( Q ) 
\left[ 
- \textbf{V} +
\textbf{C}_\Lambda ( Q )
\right] 
 , 
\end{align}
\end{subequations}
with the matrix of single-scale particle-hole bubbles
\begin{equation}
\label{eq:P_dot}
{\bf \dot{ P } }_\Lambda ( Q )
= 
\int_{ K }
\left[
G_{ \Lambda } ( K )
G_{ \Lambda } ( K - Q )
\right]^\bullet
\textbf{v} ( \omega )
\textbf{v}^\top ( \omega - \Omega )
.
\end{equation}
The formal solutions of the matrix Ricatti equations \eqref{eq:MC_dot} are
\begin{subequations}
\label{eq:MC_sol}
\begin{align}
\textbf{M}_\Lambda ( Q )
= {} &
\left\{
\textbf{1} +
\left[
\textbf{M}_0 ( \Omega ) +
\textbf{V}
\right]
\textbf{P}_\Lambda ( Q )
\right\}^{ - 1 }
\left[
\textbf{M}_0 ( \Omega ) +
\textbf{V}
\right]
\nonumber\\
&
- \textbf{V} ,
\\
\textbf{C}_\Lambda ( Q )
= {} &
\left\{
\textbf{1} +
\left[
\textbf{C}_0 ( \Omega ) -
\textbf{V}
\right]
\textbf{P}_\Lambda^\top ( Q )
\right\}^{ - 1 }
\left[
\textbf{C}_0 ( \Omega ) -
\textbf{V}
\right]
\nonumber\\
&
+ \textbf{V} ,
\end{align}
\end{subequations}
where
 \begin{equation}
 \textbf{P}_\Lambda ( Q )
= \int_0^\Lambda \textrm{d} \Lambda' 
{\bf \dot{ P } }_{ \Lambda' } ( Q ).
 \end{equation}
From these solutions,
it is obvious that we achieved our goal of resumming the $ T \to 0 $ divergence of the initial conditions \eqref{eq:SMC_0}:
In this limit,
the singularities in the numerators and denominators of $ \textbf{M}_\Lambda $ 
and $ \textbf{C}_\Lambda $
cancel,
yielding a finite result.

In truncating the flow equations,
we have so far completely neglected the effect of the 3-body interaction vertex,
represented by the last diagram in Fig.~\ref{fig:su2flow}.
A way to incorporate at least a part of this correction is the Katanin substitution \cite{Katanin04}. 
It amounts to replacing the single-scale derivatives in the polarization bubbles \eqref{eq:Pi_dot} and \eqref{eq:P_dot} by total $ \Lambda $-derivatives,
effectively adding self-energy corrections to each internal line.
The polarization bubbles can then be trivially integrated up,
yielding
\begin{align}
Z^2 \Pi_\Lambda ( Q )
= {} & 
\frac{ i \Omega + 2 \mu_0 }{ Z }
\int_{ K }
G_{ \Lambda } ( K )
G_{ \Lambda } ( Q - K )
- \frac{ 3 }{ 2 } n 
+ 1 ,
\label{eq:Pi_Katanin}
\end{align}
and
\begin{align}
\textbf{P}_\Lambda ( Q )
=  {} &
\int_{ K }
\left[
G_{ \Lambda } ( K )
G_{ \Lambda } ( K - Q )
\right.
\nonumber\\
& \hspace{.7cm} \left.
-
G_{ 0 } ( \omega )
G_{ 0 } ( \omega - \Omega )
\right]
\textbf{v} ( \omega )
\textbf{v}^\top ( \omega - \Omega )
.
\label{eq:P_Katanin}
\end{align}
Note that in some of the above bubbles,
only the Matsubara sums of the differences of flowing and initial propagators converge;
separately they do not because of the additional powers of frequency in $ \textbf{v} ( \omega ) $ 
[Eq.~\eqref{eq:v}].
The convergence furthermore hinges on the fixed $ Z / i \omega $ high-frequency decay of the holon propagator \eqref{eq:holon_propagator}.
In order to avoid regularization-dependent artifacts in the solution of the flow equation,
one should therefore always hold the density $ n $ and hence (ideally) $ Z $ constant during the flow.
With the Katanin substitution,
we heuristically incorporate some 2-loop corrections to $ U_\Lambda $ in an uncontrolled manner,
while others are dropped.
To obtain a consistent truncation,
we next drop all terms that are at least second order in the loop expansion from the numerators of the particle-hole bubbles \eqref{eq:P_Katanin}.
This amounts to approximating
 \begin{align}
G_\Lambda ( K ) G_0^{ - 1 } ( \omega )
& = 1 +  
\left[ 
t_{ \Lambda, \bm{k} } -
2 \delta \mu_\Lambda / 3 + 
\Sigma_\Lambda ( K ) \right] 
G_\Lambda ( K )
 \nonumber
 \\
 &
\approx
1 + \left(
t_{ \Lambda, \bm{k} } -
2 \delta \mu_\Lambda / 3
\right)
G_\Lambda ( K ).
 \end{align}
At constant density $ n $,
we then finally obtain
%
%
	%
	%
		%
		%
		%
		%
		%
		%
		%
		%
		%
		%
		%
		%
	%
	%
%
%
\begin{subequations}
\label{eq:P_1-loop}
\begin{align}
Z^2 P_\Lambda^{ 1 1 } ( Q )
={} &
\beta \delta_{ \Omega , 0 } 
\frac{ n }{ 2 } \left( 1 - n \right)
+
\int_K G_\Lambda ( K ) G_\Lambda ( K - Q ) ,
\\
Z^2 P_\Lambda^{ 1 2 } ( Q )
\approx {} &
\int_K G_\Lambda ( K ) G_\Lambda ( K - Q )
\left(
t_{ \Lambda , \bm{k} - \bm{q} }  - \frac{2}{3} \delta \mu_\Lambda
\right) ,
\\
Z^2 P_\Lambda^{ 2 2 } ( Q )
\approx {} &
2 \int_K t_{ \Lambda , \bm{k} } G_\Lambda ( K ) 
- 
\frac{2}{3} \delta \mu_\Lambda 
\left( \frac{ 3 }{ 2 } n - 1 \right)  
\nonumber\\
&
+
\int_K G_\Lambda ( K ) G_\Lambda ( K - Q )
\left( 
t_{ \Lambda , \bm{k} }  - 
\frac{2}{3} \delta \mu_\Lambda \right)
\nonumber\\
& 
\hphantom{+\int_K}
\times
\left(
t_{ \Lambda , \bm{k} - \bm{q} }  - \frac{2}{3} \delta \mu_\Lambda
\right) ,
\end{align}
\end{subequations}
for the three independent components of the matrix $ \textbf{P}_\Lambda ( Q ) $.
Note that in this approximation,
all Matsubara sums are convergent \cite{footnote_tG_sum}.
We show in Appendix~\ref{app:symmetries} that time-reversal symmetry fixes the remaining component of the matrix $ \textbf{P}_\Lambda ( Q ) $ of particle-hole bubbles to 
\begin{equation}
P_\Lambda^{ 2 1 } ( Q ) =
P_\Lambda^{ 1 2 } ( Q )
+ \frac{ i \Omega }{ Z } 
P_\Lambda^{ 1 1 } ( Q ) .
\label{eq:P21}
\end{equation}
The second term on the right-hand side of Eq.~\eqref{eq:P21} above corresponds to a vertex correction to the loop expansion in Eq.~\eqref{eq:P_1-loop}.
Including this vertex correction,
and thus exact time-reversal symmetry of the 2-body interaction vertex
$ U_\Lambda ( K_1' , K_2' ; K_2 , K_1 ) $,
has proven crucial for the numerical solution of the X-FRG flow of the $ t $ model. 

This completes our truncation of the X-FRG flow.
Given the holon propagator \eqref{eq:holon_propagator},
the channel decomposition \eqref{eq:U_channels},
together with the solutions 
\eqref{eq:S_sol} and 
\eqref{eq:MC_sol}
for the channels and
the polarization bubbles
\eqref{eq:Pi_Katanin}, 
\eqref{eq:P_1-loop},
and \eqref{eq:P21}
yields the 2-body interaction vertex \eqref{eq:U_ansatz} at each $ \Lambda $,
which in turn governs the flow of the holon self-energy via the flow equation \eqref{eq:flowself} shown in Fig.~\ref{fig:su2flow} (a).
In the remainder of this work,
we explore the numerical solution of this flow.

\section{Results}
\label{sec:results}
\subsection{Numerical Implementation}
Within our truncation scheme presented in the preceding section,
only two flow equations must be solved numerically: 
The equation \eqref{eq:flowself} for the momentum- and frequency-dependent self-energy $ \Sigma_\Lambda ( K ) $,
shown diagrammatically in Fig.~\ref{fig:su2flow} (a),
and the flow equation~\eqref{eq:counter-term} for the counter term $ \delta \mu_\Lambda $.
To that end,
we discretize all momentum dependencies on a 
$ 19 \times 19 $ grid within 
the first Brillouin-zone quarter of the two-dimensional square lattice.
Momentum integrations are performed using closed Newton-Cotes quadratures.
To evaluate the Matsubara frequency summations, 
we use 
$ m_{ \textrm{max} } = 100 $ positive fermionic frequencies.
This introduces additional limitations on the lowest temperatures that we can reliably access:
To ensure that the correct high-frequency asymptotics is captured, 
both the chemical potential $ \mu $ and the bandwidth $ W $ must be small compared to the maximum Matsubara frequency
$ \omega_{ \textrm{max} } = 
\pi T ( 2 m_{ \textrm{max} } - 1 ) $.
In practice,
we 
cautiously restrict ourselves to temperatures obeying
$ | \mu | $, $ W \lesssim 
 \omega_{ \textrm{max} } (T)/10$.
Convergence with respect to the the maximum number $ m_{ \textrm{max} } $ of Matsubara frequencies has been checked for all results presented in this paper.
We discuss this source of errors further in Appendix~\ref{app:Merror}.
To obtain the real-frequency retarded self-energy $ \Sigma^R_\Lambda ( \bm{k} , \omega ) 
= \Sigma_\Lambda ( \bm{k} , \omega ) 
|_{ i \omega \rightarrow \omega + i \eta } $, we use the Pad\'{e} analytic continuation procedure discussed in Ref.~\cite{Beach2000}. 
It is based on a high-precision matrix inversion, which we implement using 250 specified digits.
To obtain physical spectral functions, 
we furthermore use varying ranges of positive Matsubara frequencies for the Pad\'{e} approximant;
if needed to resolve sharp features close to zero frequency we also include the first few negative Matsubara frequencies \cite{Schoett2016}.
This yields robust and convergent results that are independent of the small artificial broadening $\eta $, 
which we set to 
$ \eta = 0.001t $
unless mentioned otherwise.

\subsection{Phase diagram}
\label{sec:phase}
\begin{figure*}%
\includegraphics[width=\linewidth]{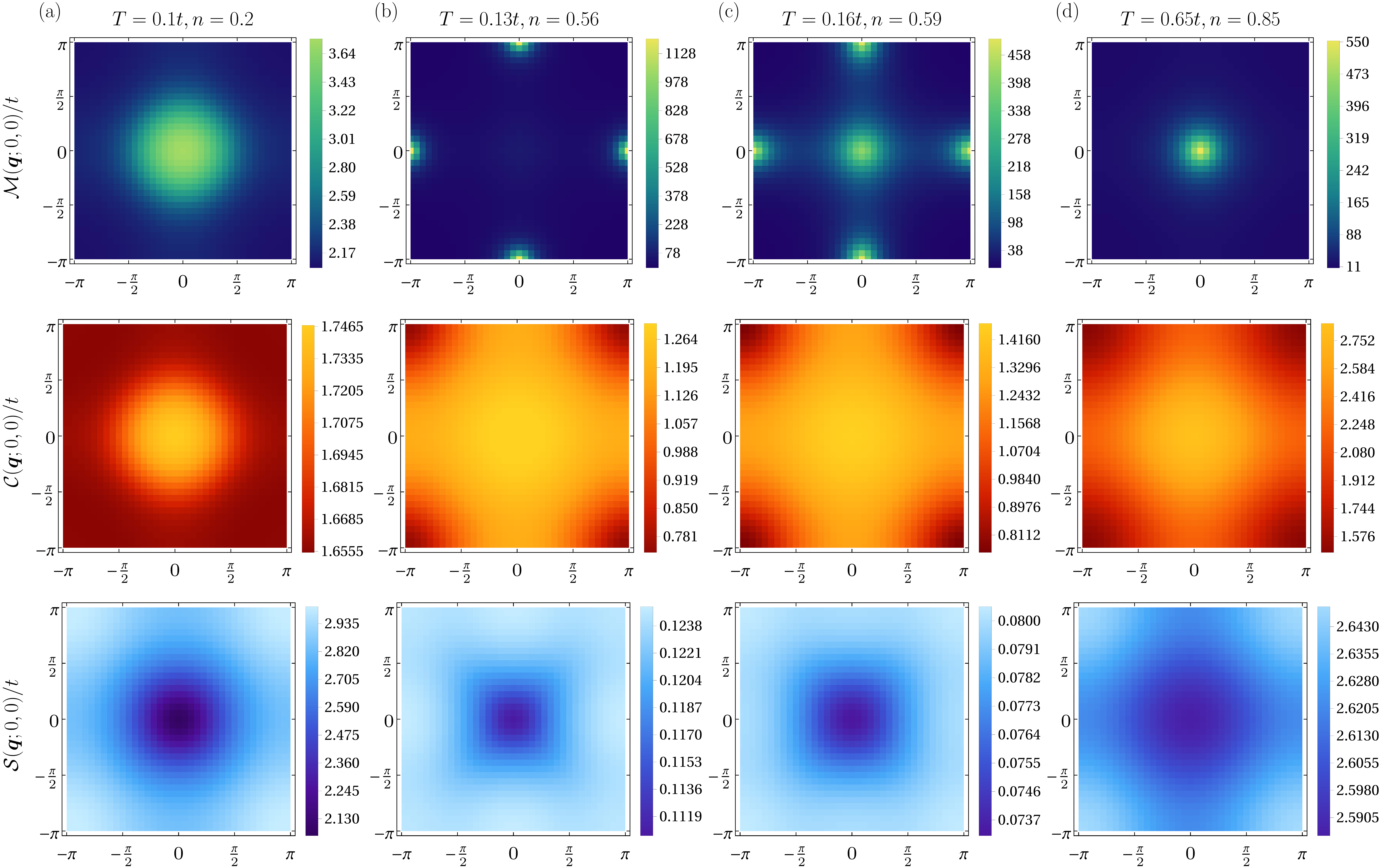}
\caption{Momentum dependence of  the static interaction channels within the first Brillouin zone, evaluated for four selected densities and corresponding lowest temperature accessible via the truncated flow equations. First row: magnetic channel $ \mathcal{M}_\Lambda(\bm{q};0,0)$; second row: charge channel $ \mathcal{C}_\Lambda(\bm{q};0,0)$; third row: superconducting channel $\mathcal{S}_\Lambda(\bm{q};0,0)$.}
\label{fig:channels3x4}
\end{figure*}

To map out the phase diagram, we begin by scanning the three static contributions of the interaction channels $ \mathcal{M}_\Lambda(\bm{q};0,0)$, $ \mathcal{C}_\Lambda(\bm{q};0,0)$ and $\mathcal{S}_\Lambda(\bm{q};0,0)$ for divergences at potential ordering vectors.
In Fig.~\ref{fig:channels3x4} we show the momentum dependencies of the channels within the first Brillouin zone for four representative density values. The temperature is chosen to be the lowest possible one which our truncated flow equations can safely reach for these densities. 
We do not find any instabilities within either the charge or superconducting channel, 
though the latter decreases by two orders of magnitude for intermediate densities $n \approx 0.59$. 
The magnetic channel, on the other hand, 
exhibits pronounced ordering tendencies.
For small densities,
it is likewise featureless,
suggesting a paramagnetic ground state.
However,
at $n = 0.56$ an instability towards an antiferromagnetic stripe state with ordering vector 
$ \bm{q}_{ \textrm{stripe} } = ( \pi , 0 ) $ 
can be clearly identified. 
Upon increasing the density, 
an additional peak at the ferromagnetic (FM) ordering vector 
$ \bm{q}_{ \textrm{FM} } = ( 0 , 0 ) $ 
appears. 
For even higher densities,
the stripe peaks decrease until only a single, pronounced instability towards the Nagaoka ferromagnet remains.

\begin{figure*}
\includegraphics[width=\linewidth]{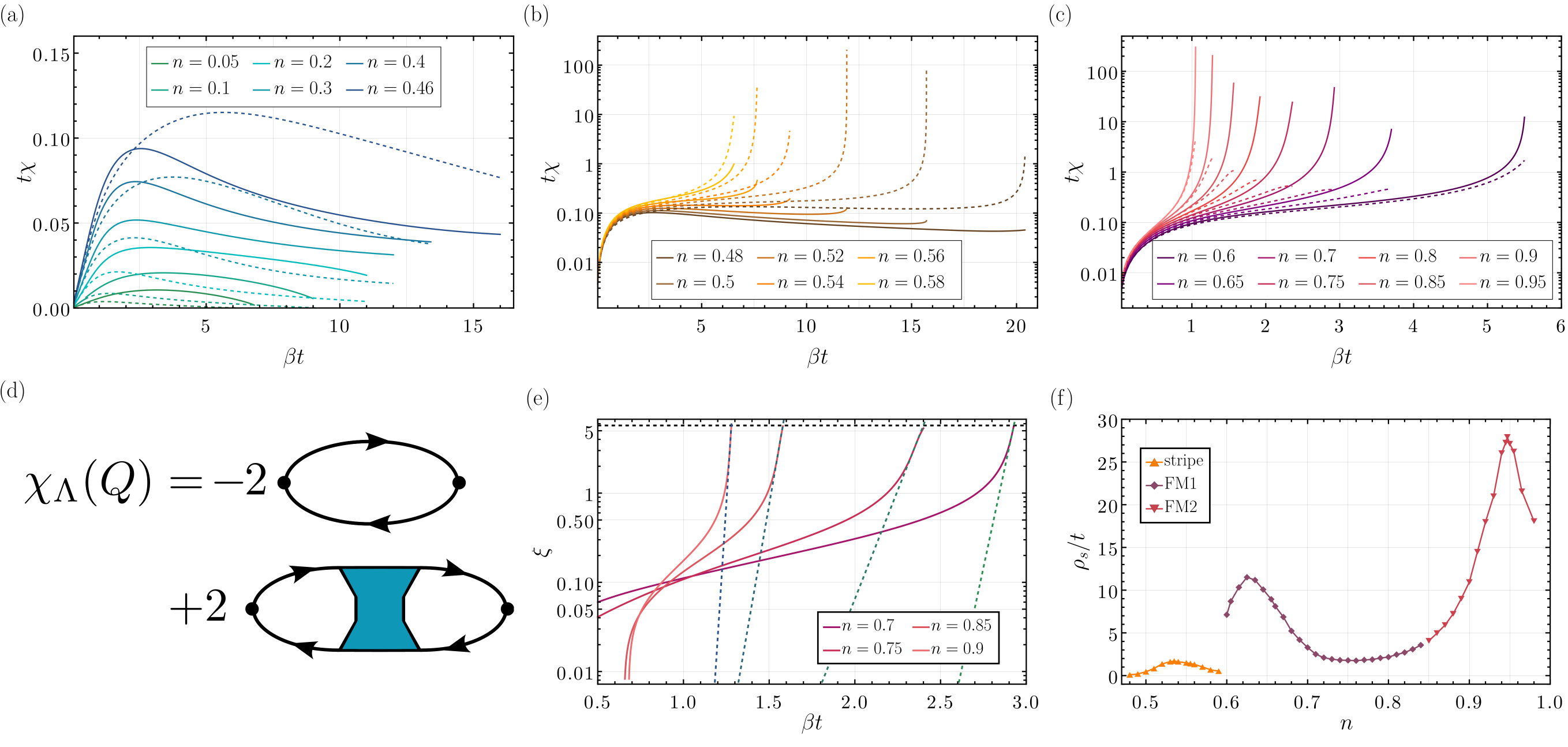}
\caption{(a)-(c) Static magnetic susceptibility 
$ \chi ( \bm{q} , 0 ) $
at momenta 
$ \bm{q} = \bm{q}_{ \textrm{FM} } = ( 0 , 0) $
(solid lines) and 
$ \bm{q} = \bm{q}_{ \textrm{stripe} } = ( \pi , 0) $
(dashed lines)
in the low-density paramagnetic,
intermediate stripe, 
and large-density ferromagnetic regimes,
respectively.
(d) Diagrammatic representation of the dynamical spin susceptibility 
$ \chi ( Q ) $.
(e) Magnetic correlation length $ \xi ( T ) $ for different densities in the Nagaoka regime calculated from Eq.~\eqref{eq:xi}.
Dashed lines are exponential fits of the magnetic instabilities.
The dashed horizontal line corresponds to the limit 
$ 1 / \Delta q = 5.73 $ imposed by our finite momentum resolution.
(f) Spin stiffness $ \rho_s ( n ) $ extracted from the exponential fit of the divergence of $ \xi ( T ) $.
}
\label{fig:magsuz}
\end{figure*}
To further investigate the magnetic instabilities, 
we next calculate the dynamic spin susceptibility 
$\chi_\Lambda(Q) $, 
which is the 4-holon correlation function shown diagrammatically in Fig.~\ref{fig:magsuz} (d);
see also Appendix~\ref{app:xopsidentity}. 
In Figs.~\ref{fig:magsuz} (a)-(c),
we show the temperature dependence of the static susceptibilities $ \chi ( \bm{q} , 0 ) $
at the two relevant ordering vectors
$ \bm{q}_{ \textrm{stripe} } $ and
$ \bm{q}_{ \textrm{FM} } $,
for small,
intermediate,
and large densities $ n $.
Figures~\ref{fig:magsuz} (b) and (c) clearly exhibit exponential growth of the stripe and FM susceptibilities, respectively, 
for $ T \to 0 $.
Shortly after encountering these instabilities,
our flow generically breaks down.
Physically,
the system then enters the renormalized classical 
regime \cite{Chakravarty1989, Vilk1997, Sachdev2011, Schafer2020} characterized by exponentially large magnetic correlation lengths,
which is beyond the scope of our truncation.
Note also that the crossover temperature where our flow breaks down continually increases with density.
This reflects the increasing relevance of the kinematic interactions due to the constraint of no double occupancy with increasing density in the $ t $ model.
This is to be contrasted with the low-density paramagnetic regime,
where our flow eventually breaks down or becomes unreliable at much lower temperatures because of the finite Matsubara frequency cutoff,
without exhibiting any kind of instability.
At densities $ n \gtrsim 0.99 $,
we face the twin difficulties of a large chemical potential (requiring many Matsubara frequencies) and strong kinematic interactions.
In particular,
the exact Nagaoka \cite{Nagaoka1965, Nagaoka1966, Kollar1996, Tasaki1998} and Lieb \cite{Lieb1989} theorems imply that for $ n \to 1 $ there must be a competition between the ferro- and antiferromagnetic ground states.
In our current truncation,
we are however not able to reach the low-temperature regime for $ n \sim 1 $;
the flow breaks down already at high temperatures $ T > t $.
This is accompanied by a growth of the static magnetic fluctuations at all momenta,
which may be a precursor of the competition between the two different ground states.

Because the Hohenberg-Mermin-Wagner theorem \cite{Mermin1966, Hohenberg1967} forbids the spontaneous breaking of continuous symmetries at any finite temperature in two dimensions,
we expect the magnetic instabilities to signal the existence of zero-temperature phase transitions to the corresponding magnetic states.
This allows us to construct the magnetic ground state phase diagram of the $ t $ model shown in Fig.~\ref{fig:PhaseDiag}. 
Explicitly, 
we find that the Nagaoka ferromagnet is stable at large densities, 
down to $n_{ c , 2 } = 0.6$, 
in good agreement with previous studies \cite{Elser1990,Coleman2002,Wurth1995, Hanisch1993,Obermeier1997,Newby2025,Blesio2019,Becca2001,vonderLinden1991}. 
This is followed by an antiferromagnetic stripe regime, 
which exists until $n_{ c , 1 } = 0.48$. 
Various (in-)commensurate Néel type orders have been observed for intermediate densities \cite{Liu2012, Sharma2025, Blesio2019}. 
A definite agreement regarding the nature and even existence of this phase has, 
to our knowledge, 
not been reached.
However,
consistent with our findings,
the best variational wavefunctions for the Nagaoka ferromagnet \cite{vonderLinden1991, Wurth1995}
likewise become unstable at a critical hole doping because the gap of a spin wave with momentum $ \bm{q}_{ \textrm{stripe} } $ closes.
For densities lower than $n_{ c , 1 }$ 
no order is observed and the system remains  paramagnetic.

Finally,
to estimate the strength of magnetic fluctuations in the renormalized classical regime \cite{Chakravarty1989, Vilk1997, Sachdev2011, Schafer2020} at low temperatures,
we extract the correlation length 
$ \xi $
from the exponential growth of the static susceptibilities 
just before the breakdown of the flow.
For a two-dimensional lattice,
the correlation length can be defined as \cite{Sandvik2010}
\begin{equation}
\label{eq:xi}
\xi = \sqrt{ \frac{16}{15} }
\frac{ 1 }{ \Delta q }
\sqrt{
\frac{ 
\chi ( \bm{q}_x , 0 ) 
}{ 
\chi ( \bm{q}_x + \Delta \bm{q} , 0 ) 
}
- 1 
} ,
\end{equation}
where $ x = \textrm{FM} $ or $ \textrm{stripe} $ depending on which correlation dominates,
and $ \bm{q}_x + \Delta \bm{q} $ is the wave vector closest to the ordering vector $ \bm{q}_x $.
With our momentum resolution,
$ \Delta q = \pi / 18  $.
The correlation length computed in this manner is shown in Fig.~\ref{fig:magsuz}(e) as a function of inverse temperature $ \beta t $ for several densities in the Nagaoka regime.
Overall,
the behavior of $ \xi ( T ) $ is qualitatively similar to the susceptibilities shown in Fig.~\ref{fig:magsuz}(c),
with a rather sharp transition from the high-temperature Curie-law behavior to the low-temperature exponential rise for 
$ T \lesssim t $;
the same qualitative behavior is also observed in classical simulations of two-dimensional ferromagnets~\cite{Shenker1980}.
Since a growing correlation length corresponds to a sharply peaked susceptibility in momentum space that we eventually cannot resolve properly any more,
the finite resolution also limits us to the regime where
$ \xi \lesssim 1 / \Delta q = 5.73 $.
This limit is indicated as dashed horizontal line in Fig.~\ref{fig:magsuz}(e).
Remarkably,
it agrees very well with the temperatures where our flow breaks down,
confirming that the strong magnetic fluctuations are the underlying reason.

Because the incipient magnetic ordering breaks the $ O ( 3 ) $ spin-rotation invariance of the $ t $ model \eqref{eq:tModel} at $ T = 0 $,
we expect that the magnetic fluctuations at sufficiently low temperatures are described by an appropriate classical field theory.
This classical theory is characterized by the spin stiffness $ \rho_s $,
which measures the resistance of the spins to a small local twist of the magnetization pattern
and determines the divergence of the correlation length, 
$ \xi \sim e^{ 2 \pi \rho_s / T } $ 
\cite{Shenker1980, Kopietz1989, Sandvik2010}.
By fitting the divergence of $ \xi ( T ) $ to this form,
we obtain the estimate for the spin stiffness
$ \rho_s ( n ) $
that is shown in Fig.~\ref{fig:magsuz}(f).
Besides the strong stripe correlations centered around $ n \approx 0.54 $,
the unexpected behavior exhibited in the ferromagnetic regime is of particular interest:
There,
the spin stiffness has not one but two distinct maxima,
a smaller one centered around $ n \approx 0.64 $,
and a larger one at $ n \approx 0.95 $.
This suggests that the ferromagnetic phase actually comprises of two distinct phases,
with the true Nagaoka ferromagnet only appearing at 
$ n_{ c , 3 } \approx 0.85 $.
This is close to the most recent variational \cite{vonderLinden1991, Hanisch1993, Wurth1995}, Monte Carlo \cite{Becca2001, Baroni2011}, and density-matrix renormalization group \cite{Liu2012, Blesio2019} estimates.
We can only speculate about the nature of the second ferromagnetic phase.
However,
it appears consistent with the partially polarized state observed in the same density range $ n_{ c , 2 } \le n < n_{ c , 3 } $
in the Monte Carlo study \cite{Becca2001}.
Additional support for a partially polarized ferromagnet at densities below the Nagaoka state comes from exact diagonalization \cite{Chiappe1993},
density-matrix renormalization group~\cite{Liu2012, Blesio2019}, 
and a supersymmetric slave-particle approach~\cite{Coleman2002}.
Moreover,
we show in Sec.~\ref{sec:FM} below 
that the transition at $ n_{ c , 3 } $ is related to a Lifshitz transition of the electronic Fermi surface. 
Last,
we note that the apparent decrease of ferromagnetic correlations visible in the downturn of $ \rho_s ( n ) $ in Fig.~\ref{fig:magsuz} (f) for densities $ n \gtrsim 0.95 $ is not unexpected and has also been observed in dynamical mean field theory \cite{Park2008}.
Older works based on variational wavefunctions \cite{Shastry1990, vonderLinden1991} similarly found that the spin stiffness has a maximum at a finite hole concentration and vanishes linearly for $ n \to 1 $.
This accounts for the fact that at half-filling,
$ n = 1 $,
the ground state of the square lattice Hubbard model is actually antiferromagnetic at any value of $ U $ \cite{Lieb1989},
and reflects the extreme sensitivity of the Nagaoka state to changes in the boundary conditions for a small number of holes \cite{Aruiac1990, Zotos1993, White2001, Kochetov2017}.

\subsection{Equation of state}
\begin{figure*}%
\includegraphics[width=\linewidth]{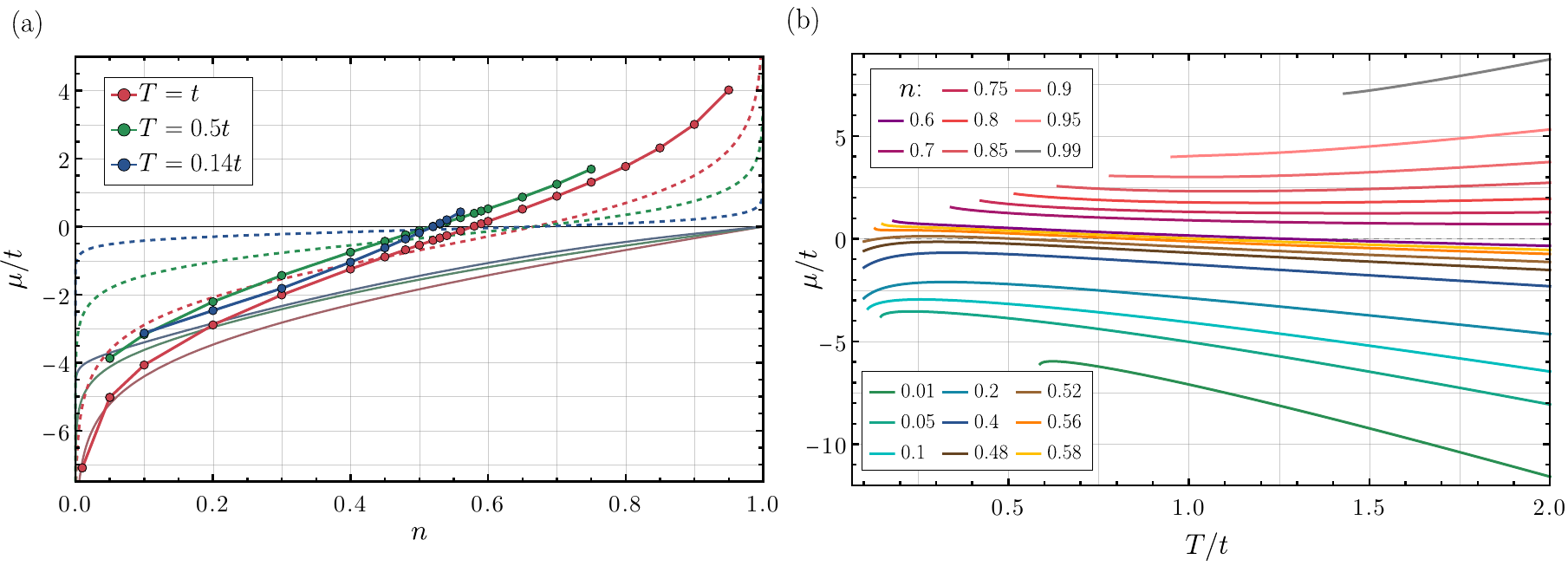}
\caption{(a) Chemical potential $ \mu $ as function of density $ n $ for various temperatures. 
Connecting lines serve as guide for the eye. 
Dashed lines show the corresponding chemical potential $ \mu_0 $ in the atomic limit,
given in Eq.~\eqref{eq:mu0}. 
The solid lines with increased gray level correspond to a free electron gas.
(b) Chemical potential $ \mu $ as function of temperature $ T $ for various densities.
The coloring is based on the magnetic ground state of the system at the respective density; 
green-blue: paramagnetic regime; 
brown-orange: stripe regime; 
purple-pink: ferromagnetic regime.
}
\label{fig:muofn}
\end{figure*}%
The electronic equation of state $\mu = \mu (n, T)$ for selected temperatures is shown in Fig.~\ref{fig:muofn} (a), 
with the equations of state for free electrons and in the atomic limit for comparison.
Here,
it is notable that as the temperature is lowered,
the chemical potential is renormalized ever stronger away from the atomic limit $ \mu_0 $,
which is our initial condition,
for almost all densities.
At small densities, 
the chemical potential of the $ t $ model converges to the free-electron equation of state for all temperatures. 
However,
upon increasing the density, 
deviations from the free electron gas set in quickly, 
culminating in a crossing of zero. 
This change of sign already happens in the atomic limit at $ n = 2/3 $ 
and is a direct consequence of the holon algebra.  
Upon lowering the temperature and incorporating the hopping, 
the crossing is shifted towards 
$ n \approx 0.529 $ at $ T = 0.14 t $.
Note that this density roughly corresponds to the region with the strongest antiferromagnetic stripe fluctuations; 
see Fig.~\ref{fig:magsuz} (f). 
A similar value has been obtained in Ref.~\cite{Khatami2014}.
Likewise, 
a parton mean-field approximation \cite{Kumar2008} predicts that the critical point for the Nagaoka state is at $ n = 1 / 2 $ and $ \mu = 0 $.

Figure~\ref{fig:muofn} (b) shows the chemical potential as function of temperature for various densities. 
The values of the chemical potential can be associated with the nature of the ground state into which the system evolves as we lower the temperature. 
The greenish-blue lines at negative values of the chemical potential
flow into the paramagnetic ground state. 
The purple-pink lines on the other hand are situated at positive values and belong to densities where ferromagnetic order eventually develops. 
Stripe ordering occurs near $ \mu = 0 $. 
Going from high to low temperatures,
$ | \mu ( T ) | $ initially decreases for all densities.
We attribute this to the density-dependent reduction of the electronic bandwidth $ W $ with temperature 
that we discuss in Sec.~\ref{sec:Z_and_W}.
It is also noteworthy that for all densities, 
the slope of $ \mu ( T ) $ changes at low temperature. 
At small densities,
in the paramagnetic regime,
the downturn merely signifies the approach to the free electron limit;
compare Fig.~\ref{fig:muofn} (a).
Here,
we expect Fermi liquid physics;
see also Sec.~\ref{sec:PM}.
Then the band structure freezes out at sufficiently low temperature 
(compare the bandwidth in Fig.~\ref{fig:bandwidthofn} below),
so that the chemical potential has to decrease with temperature to maintain constant density.
On the other hand,
the upturn at large, ferromagnetic densities is likewise to be expected.
This is because at half-filling, 
$ n = 1 $,
the particle-hole symmetry of the Hubbard model implies
$ \mu = U / 2 $,
which diverges at $ U = \infty $. 
Hence,
we expect the chemical potential of the $ t $ model to flow to large positive values at large densities,
which is also visible also in Fig.~\ref{fig:muofn} (a).

\subsection{Electronic spectral properties}
%
%
We now turn the electronic spectral properties of
the $t$ model. 
To this end we evaluate the
electronic spectral function, 
given in terms of the  holon propagator \eqref{eq:holon_propagator} as
\begin{equation} \label{eq:spectral_function}
A(\bm{k},\omega) = -\frac{1}{\pi}\,  \textrm{Im}\big[\left.G_{\Lambda = 1}(\bm{k},\omega)\right|_{i\omega \rightarrow \omega + i \eta}\big].
\end{equation}
Because for $ U = \infty $ the upper Hubbard band is entirely removed from the spectrum,
the spectral function of the $ t $ model carries less weight than the Hubbard model
at finite $U$ \cite{Shastry2010}.
Explicitly, the spectral function  of the $ t $ model
it is normalized to
\begin{equation}
\label{eq:A_normalization}
\int_{ - \infty }^\infty \textrm{d} \omega A ( \bm{k} , \omega )
= Z 
= 1 - \frac{ n }{ 2 } ,
\end{equation}
which follows immediately from the holon algebra \eqref{eq:holon_algebra}.
In order to quantify the quasi-particle weight $Z_{\bm{k}}$, dispersion $\xi_{\bm{k}}$ and damping $\gamma_{\bm{k}}$, 
we fit our data for each $ \bm{k} $ to a Lorentzian distribution,
\begin{equation}
A(\bm{k},\omega) = \frac{Z_{\bm{k}}}{\pi}\frac{\gamma_{\bm{k}}}{\left(\omega - \xi_{\bm{k}}\right)^2 + \gamma_{\bm{k}}^2} ,
\label{eq:Lorentzian}
\end{equation}
such that
\begin{subequations}
	\begin{align}
		\gamma_{\bm{k}} 	    ={}& \frac{1}{2}\textrm{FWHM}\left[A(\bm{k},\omega)\right],\\
		A(\bm{k}, \xi_{\bm{k}}) ={}& \textrm{max}\left[A(\bm{k},\omega)\right],\\
		Z_{\bm{k}}				={}& \pi \gamma_{\bm{k}} \textrm{max}\left[A(\bm{k},\omega)\right]. \label{eq:Zk}
	\end{align}
\end{subequations}
Here,
$\textmd{FWHM}\left[A(\bm{k},\omega)\right]$ is the full width at half maximum of the distribution.
Even when the quasi-particle picture breaks down and the Lorentzian shape is not adequate anymore, $ \gamma_{\bm{k}} $ still provides a measure for the width of the distribution.
In the following, 
we furthermore define the Fermi surface at finite temperature as the locus of $\bm{k}$-points where 
$ A(\bm{k},\omega = 0)$ is maximal. 
We review alternative approaches to find the Fermi surface in Sec.~\ref{sec:LTviolation} and Appendix~\ref{app:firstmatsubara}.

We begin by discussing the features of the electronic spectral function within the three distinct magnetic regimes. 
Afterward, we turn to more general considerations spanning the entire density range.

\subsubsection{Paramagnetic regime}
\label{sec:PM}
\begin{figure*}%
\includegraphics[width=\textwidth]{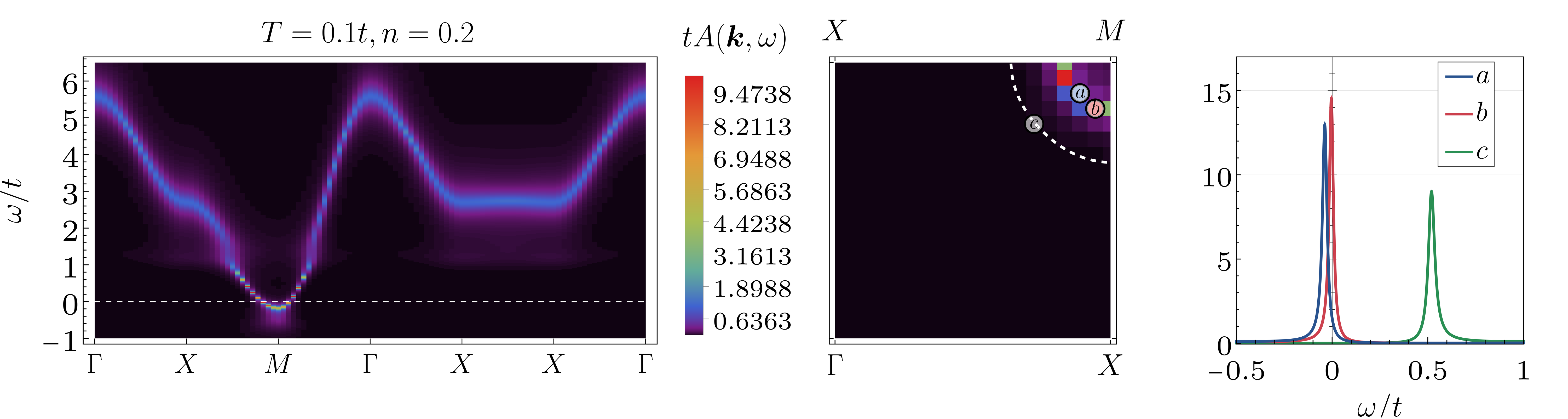}
\caption{Electronic spectral function $A(\bm{k},\omega)$ at $n = 0.2$, $T = 0.1t$.
From left to right: $A(\bm{k},\omega)$ along the high symmetry path of the Brillouin zone; $A(\bm{k},0)$ within the Brillouin zone, with the dashed red line showing the free Fermi surface; 
spectral function at momenta corresponding to the three insets in the second picture. 
Excitations sharper than our momentum resolution near the Fermi level indicate Fermi liquid behavior.}
\label{fig:tripleplotn02T1}
\end{figure*}%
For the low-density regime $n < n_{ c , 1 }$, 
we discuss $ n=0.2 $ as a representative example for the qualitative behavior of the spectral properties.
In Fig.~\ref{fig:tripleplotn02T1} we show the electronic spectral function along the high symmetry path of the first Brillouin zone, following both the $\Gamma$-$M$-diagonal, as well as the $X$-$X$ diagonal along the magnetic Brillouin zone to highlight the behavior of the Van Hove singularity.
We find a band consisting of sharp excitations near the Fermi level, 
whereas the spectrum is much broader at higher frequencies.
This behavior suggests that the ground state of the system behaves according to Fermi liquid theory. 
\begin{figure}%
\includegraphics[width=\linewidth]{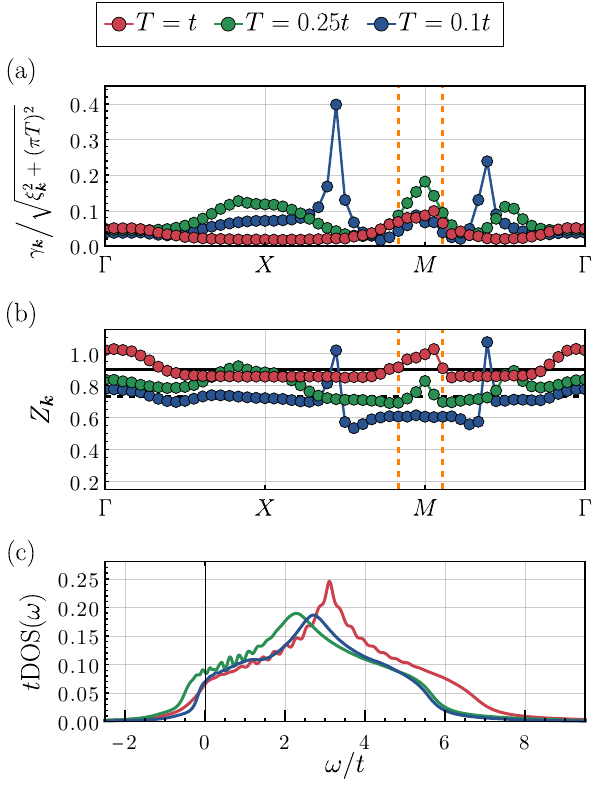}
\caption{(a) Damping $ \gamma_{\bm{k}} $ at $ n = 0.2 $ for various temperatures, 
relative to the relevant thermal energy scale $ \sqrt{ \xi_{ \bm{k} }^2+(\pi T)^2 } $ of a Fermi liquid. 
Dashed orange line: Fermi surface.
%
%
(b) Quasi-particle weight $Z_{\bm{k}}$ at $ n = 0.2 $ for various temperatures. 
Dashed orange line: Fermi surface. 
Black line: atomic limit $Z$. 
Dashed black line: $Z^3$ as predicted for the quasi-particle weight in \cite{Rueckriegel2023}.
Values $ Z_{ \bm{k} } > 1 $ signal the breakdown of the Lorentzian fit \eqref{eq:Lorentzian} for the spectral function when the damping becomes too large.
(c) Density of states $\textrm{DOS}(\omega)$ at $n = 0.2$ for various temperatures.
Since the peaks of the individual spectral functions are in some regimes sharper than our momentum resolution, 
the artificial broadening has been increased to 
$ \eta = 0.05t $ for $ T/t = 1,~0.25 $ and 
$ \eta = 0.1t $ for $ T/t = 0.1 $
for this plot.}
\label{fig:dampingn02multT}	
\end{figure}%
We can probe this further by considering the stability of the quasi-particles near the Fermi level. 
For a Fermi liquid,
we expect that 
$ \textrm{Im} [ \Sigma^R ( \bm{k} , \omega ) ] \propto
\textrm{DOS} (\omega) 
[\omega^2 + ( \pi T )^2 ] $
at low frequencies \cite{Chubukov2012},
where $ \textrm{DOS} (\omega)  $ is the density of states.
This yields a Fermi liquid damping
$ \propto 
\xi_{ \bm{k} }^2 + ( \pi T )^2 $,
which vanishes on the Fermi surface at $ T = 0 $. 
In Fig.~\ref{fig:dampingn02multT} (a) we therefore show the damping $\gamma_{\bm{k}}$ relative to the finite-temperature energy scale
$ \sqrt{ \xi_{ \bm{k} }^2 + ( \pi T )^2 } $ that is relevant for a conventional Fermi liquid. 
We find that with decreasing temperature, 
the damping develops a non-trivial momentum dependence, 
with sharp peaks at the Van Hove singularity as well as stable minima located near the Fermi surface, 
just as Fermi liquid theory predicts.
Note,
however,
that a slight particle-hole asymmetry around the Fermi energy is visible in the damping in Fig.~\ref{fig:dampingn02multT} (a),
corresponding to slightly sharper quasi-particle peaks above than below the Fermi surface in Fig.~\ref{fig:tripleplotn02T1}. 
This feature---while not accounted for by the conventional Fermi liquid phenomenology---is generic for extremely correlated electrons \cite{Zitko2013, Khatami2014, Wang2018}.
At higher densities,
the particle-hole asymmetry is much more pronounced (see, for example, Figs.~\ref{fig:tripleplotn056T13} and \ref{fig:tripleplotn085T065} below).
We argue in Sec.~\ref{sec:FM} that the microscopic origin of this asymmetry for extremely correlated electrons are (ferro-)magnetic fluctuations,
which are small but already finite in the low-density paramagnetic regime;
compare Fig.~\ref{fig:magsuz} (a).

Despite the overall Fermi liquid behavior, 
the Fermi surface, 
displayed in Fig.~\ref{fig:tripleplotn02T1} (center), 
still reveals a marked difference to the conventional metallic state, 
as it violates Luttinger's theorem \cite{Luttinger1960}. 
The volume enclosed by the Fermi surface differs from the one enclosed by the free Fermi surface, 
indicated by the dashed red line in Fig.~\ref{fig:tripleplotn02T1} (center). 
In the context of strongly correlated electrons, 
this violation has been repeatedly observed~\cite{Stephan1991, Singh1992, Putikka1998, Kokalj2007, Kozik2015, Quinn2018, Osborne2021} in the high-density/low-doping regimes. 
In fact, for decreasing density the fulfillment of Luttinger's theorem together with usual Fermi liquid traits is usually expected \cite{Singh1992},
but by no means mandatory \cite{Shastry2010, Rueckriegel2023}. 
We comment on this further in Sec.~\ref{sec:LTviolation}.

%
%
%
%
%
%
%
%
%
%
%
%
The momentum-dependent quasi-particle weight $Z_{\bm{k}}$ is displayed in Fig.~\ref{fig:dampingn02multT} (b). 
For comparison, 
we also show the initial, atomic value given by the high-frequency asymptotic $Z$ as a gray line. 
Additionally, 
the gray dashed line shows a prediction for the zero-temperature quasi-particle weight $Z_{\bm{k}} \approx Z^3$ obtained in a simple approximation in our previous work \cite{Rueckriegel2023}.
For low temperatures, 
this earlier result agrees quite well with our new data. 
However, 
$ Z_{\bm{k}} $ develops an additional momentum dependence, 
dropping in magnitude near the Fermi surface and within the hole sector. 
This behavior should be considered with care; 
following Eq.~\eqref{eq:Zk} the weight is at low temperatures given by the product 
of a very small quantity, 
the damping $ \gamma_{\bm{k}}$, 
and a very large quantity, 
the height of the spectral function, 
making the calculation of $Z_{\bm{k}}$ potentially unreliable in this region.

%
%
%
%
%
%
%
%
%
%
%
%
Last,
we compute the density of states,
\begin{equation}
\textrm{DOS}(\omega) = \frac{ 1 }{ N } \sum_{\bm{k}} A(\bm{k},\omega)
\end{equation}
in Fig.~\ref{fig:dampingn02multT} (c).
Consistent with the Fermi liquid behavior and overall weak renormalization of the band,
it closely resembles the standard nearest-neighbor tight-binding density of states.
Note,
however,
that even so,
the bandwidth $ W $ converges at low temperatures to a value lower than the tight-binding value of $ 8 t $.  

Summarizing,
the overall behavior of the $t$ model at low densities is well described by a renormalized free electron theory,
and in particular follows the Fermi liquid phenomenology (albeit with an additional slight particle-hole asymmetry). 
This is unsurprising as the constraint of no double occupancy is rarely noticed by the electrons in a sparsely filled lattice. 
Accordingly,
the chemical potential shown in Fig.~\ref{fig:muofn} approaches the free electron value for vanishing density already at intermediate temperatures $ T = t $. 
Despite this,
a fundamental difference to free electrons remains even at low densities:
the violation of Luttinger's theorem,
which can ultimately be traced back to the missing spectral weight of the upper Hubbard band in the $ U = \infty $ limit \cite{Shastry2010}.

\subsubsection{Stripe regime}
\begin{figure*}%
\includegraphics[width=\textwidth]{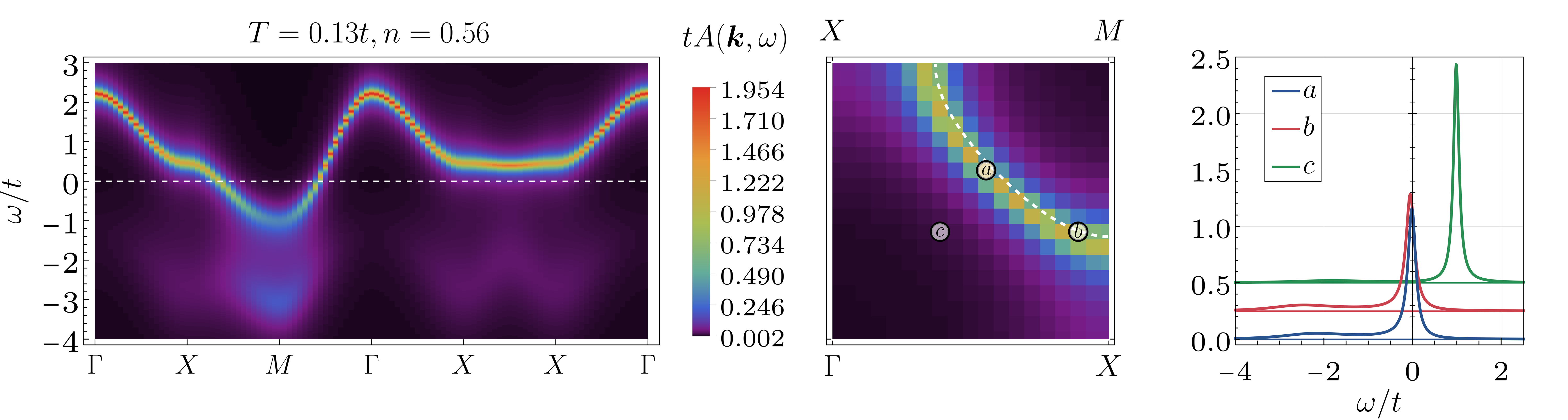}
\caption{Electronic spectral function $A(\bm{k},\omega)$ at $n = 0.56$, $T = 0.13t$.
From left to right: $A(\bm{k},\omega)$ along the high symmetry path of the Brillouin zone; $A(\bm{k},0)$ within the Brillouin zone, with the dashed red line showing the free Fermi surface; spectral function at momenta corresponding to the three insets in the second picture.  Notably, the sharpest excitations are now found away from the Fermi level. Additionally a continuum has formed at negative frequencies.}
\label{fig:tripleplotn056T13}
\end{figure*}
\begin{figure}%
\includegraphics[width=\linewidth]{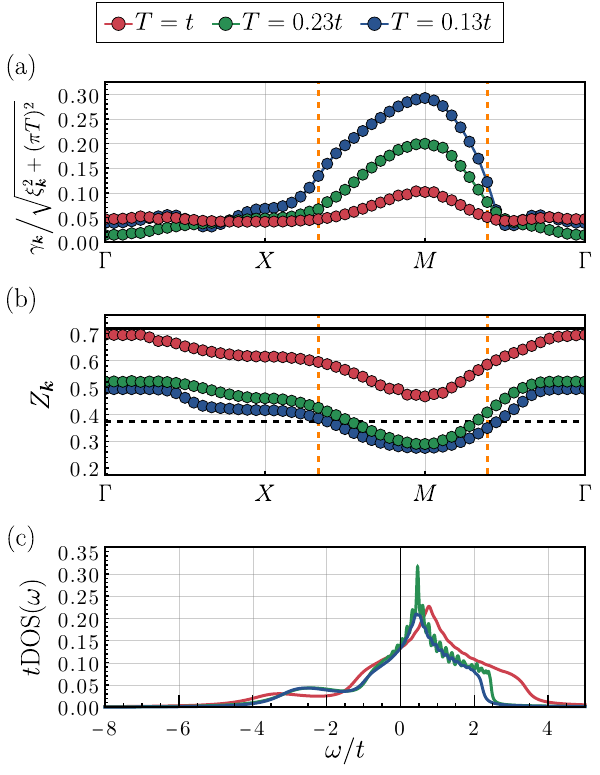}
\caption{(a) Damping $ \gamma_{\bm{k}} $ at $ n = 0.56 $ for various temperatures, 
relative to the relevant thermal energy scale $ \sqrt{ \xi_{ \bm{k} }^2+(\pi T)^2 } $ of a Fermi liquid. 
Dashed orange line: Fermi surface.
The relative damping remains roughly constant above the Fermi energy,
which is indicated by the dashed orange line. 
Below it, 
around the $M$-point, 
the damping increases upon lowering the temperature
due to the formation of a second, 
incoherent spectral peak below the single-particle band.
It also does not decrease near the Fermi surface, 
signaling non-Fermi liquid behavior.
(b) Quasi-particle weight $Z_{\bm{k}}$ at $ n = 0.56 $ for various temperatures. 
Dashed orange line: Fermi surface. 
Black line: atomic limit $Z$. 
Dashed black line: $Z^3$ \cite{Rueckriegel2023}.
(c) Density of states $\textmd{DOS}(\omega)$ at $n = 0.56$ for various temperatures.
At negative frequencies a broad secondary hump emerges from a continuum below the single-particle band edge at low temperature.}
\label{fig:dampingn056multT}	
\end{figure}

%
Moving into the intermediate density regime $n_{c,1} \leq n = 0.56 < n_{c,2}$,
the band  narrows significantly
and the Fermi surface becomes less defined;
see Fig.~\ref{fig:tripleplotn056T13}. 
The spectral weight is shifted to the $\Gamma$-point at finite positive frequencies. 
In the vicinity of the Fermi energy
the excitations broaden and leak spectral weight into an incoherent continuum at negative frequencies. 
This is reflected both in the density of states, Fig.~\ref{fig:dampingn056multT} (c), 
and in the spectral function, Fig.~\ref{fig:tripleplotn056T13}, 
in the form of a shoulder at negative frequencies, 
below the single-particle band. 
We attribute this incoherent spectral weight at negative frequencies
and the resulting particle-hole asymmetry
to the existence of ferromagnetic spin polaron states \cite{Brinkman1970, White2001, Maska2012, Lebrat2024, Prichard2024, Samajdar2024a}.
These are also responsible for the strong ferromagnetic fluctuations observable in the static spin susceptibility in Fig.~\ref{fig:magsuz} (b),
even though antiferromagnetic stripe correlations always dominate in this density range.
We comment on the underlying physical mechanism for the spin polaron formation in Sec.~\ref{sec:FM} below, 
when we consider the high-density regime, 
where the effect is much more pronounced.
The damping, 
displayed in Fig.~\ref{fig:dampingn056multT} (a), 
clearly exhibits non-Fermi liquid characteristics,
since it does not vanish nor shrink near the Fermi surface upon decreasing temperature. 
Thus,
the $ t $ model is better described as a bad metal \cite{Emery1995,Deng2013} in this density regime.
The largest damping is found near the $M$-point, 
consistent with the previous remarks. 
There, 
the relative damping additionally increases upon lowering the temperature. 
This is however an artifact of our fitting method. 
We extract the full width at half maximum, 
which assumes a single, 
Lorentzian quasi-particle peak. 
However, 
in this intermediate density region, 
the spectral function develops a broad second peak associated with the bound polaronic states below the Fermi level; 
see Fig.~\ref{fig:tripleplotn056T13}. 
Around the $ M $-point,
the two peaks are sufficiently close to each other that our fit only measures the width of the two-peak structure instead of the proper quasi-particle damping.  

%
%
%
%
%
%
%
%
%
%
%
%
Like the damping,
the quasi-particle weight shown in Fig.~\ref{fig:dampingn056multT} (b)
does not attain any strong momentum dependence 
and converges to a finite value at low temperatures. 
The qualitative behavior of $ Z_{ \bm{k} } $ reflects the shape of the single-particle band,
so that it remains smallest below the Fermi energy.
%
%
%
%
%
%
%
%
%
%
%
%
Finally, 
note that the violation of Luttinger's theorem is minimal;
see Fig.~\ref{fig:tripleplotn056T13} (center). 
This remains true in the entire density range $n_{c,1} \leq n < n_{c,2}$. 
For a more in-depth discussion see  Sec.~\ref{sec:LTviolation} below.

\subsubsection{Ferromagnetic regime}
\label{sec:FM}
%
%
%
%
%
%
%
%
%
%
%
%
%
%
\begin{figure*}%
\includegraphics[width=\textwidth]{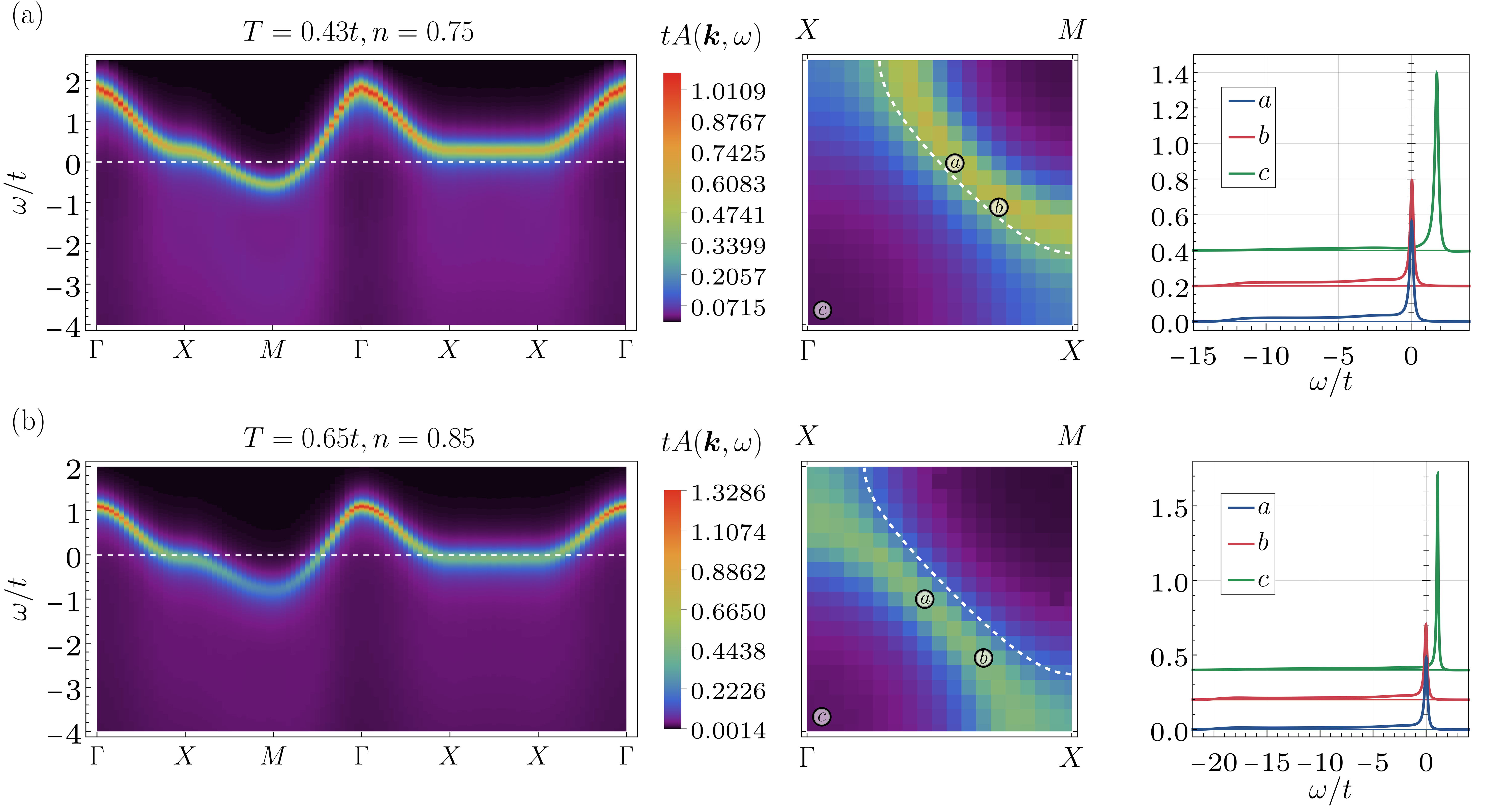}
\caption{Electronic spectral function $A(\bm{k},\omega)$ at
(upper panel) $n = 0.75$, $T = 0.43t$ and
(lower panel) $n = 0.85$, $T = 0.65t$.
From left to right: $A(\bm{k},\omega)$ along the high symmetry path of the Brillouin zone; $A(\bm{k},0)$ within the Brillouin zone, with the dashed red line showing the free Fermi surface; 
spectral function at momenta corresponding to the three insets in the second picture.}
\label{fig:tripleplotn085T065}
\end{figure*}%
\begin{figure}%
\includegraphics[width=\linewidth]{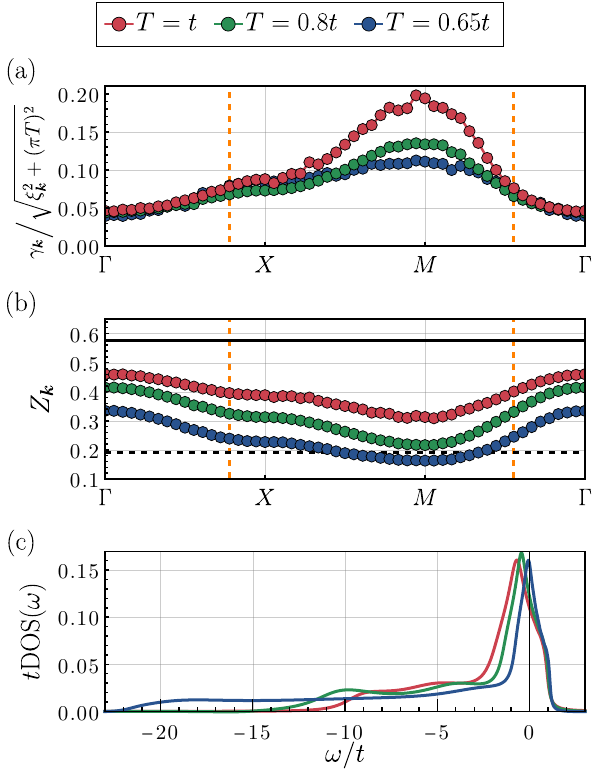}
\caption{(a) Damping $ \gamma_{\bm{k}} $ at $ n = 0.85 $ for various temperatures, 
relative to the relevant thermal energy scale $ \sqrt{ \xi_{ \bm{k} }^2+(\pi T)^2 } $ of a Fermi liquid. 
Dashed orange line: Fermi surface.
There is only a weak momentum dependence and no Fermi liquid behavior.
The Fermi energy only separates the incoherent hole from the more coherent particle sector.
(b) Quasi-particle weight $Z_{\bm{k}}$ at $ n = 0.85 $ for various temperatures. 
Dashed orange line: Fermi surface. 
Black line: atomic limit $Z$. 
Dashed black line: $Z^3$ \cite{Rueckriegel2023}.
(c) Density of states $\textmd{DOS}(\omega)$ at $n = 0.85$ for various temperatures.
Note the extended negative-energy continuum,
and that the Van Hove singularity is located just below the Fermi level in the hole sector.}
\label{fig:dampingn085multT}	
\end{figure}%
%
%
%
%
%
%
%
%
%
%
%
%
%
%
%
%
%
%
%
%
%
%
%
%
For densities $n \geq n_{c,2}$ the static susceptibility signals  an instability towards a ferromagnetic ground state.
Here the Van Hove singularity remains close to the Fermi level, 
giving rise to a large number of states at and near $\omega = 0$, 
visible in Fig.~\ref{fig:dampingn085multT} (c).
In combination with the drastically decreasing bandwidth, 
we thus arrive at a scenario closely resembling the idealized flat-band correlated ferromagnetism \cite{Kanamori1963, vonderLinden1991, Mielke1991a, Mielke1991b, Tasaki2003, Hu2025}.
However, 
we already argued in Sec.~\ref{sec:phase} that there is an additional qualitative change at density $ n_{c,3} = 0.85 $ 
that separates two distinct ferromagnetic phases.
As representative examples for these two regimes,
we therefore display the electronic spectra for 
$ n = 0.75 $ and 
$ n = 0.85 $ 
in the upper and lower panels of Fig.~\ref{fig:tripleplotn085T065},
respectively.
While the overall shape of the electronic spectral function is similar in both cases,
there is a notable difference to distinguish the two phases:
At $ n_{ c , 3 } $,
the Fermi surface undergoes a Lifshitz transition from electron- to hole-like \cite{Volovik2016,Simkovic2024}. 
This transition is associated with the Van Hove singularity moving into the hole sector below the Fermi level for densities $n \geq n_{ c , 3 }$.
While Luttinger's theorem is violated both for densities smaller and larger than $ n_{ c , 3 } $,
the violation also changes its sign,
further substantiating the existence of two distinct phases.
Two qualitatively different ferromagnetic regimes have also been found in the literature, 
where commonly an unsaturated ferromagnetic phase leading up the fully polarized ferromagnetic ground state at higher densities has been predicted \cite{vonderLinden1991, Chiappe1993, Hanisch1993, Wurth1995, Zitzler2002, Coleman2002, Becca2001}. 
Though we cannot distinguish these two cases, 
the phase boundaries obtained in our calculation are in quantitative agreement with earlier findings.

%
%
%
%
%
%
%
%
%
%

From the spectra in
Fig.~\ref{fig:tripleplotn085T065},
it is clear that the trends observed in the intermediate stripe regime continue also for $ n \ge n_{ c , 2 } $:
With increasing density,
the electronic single-particle band becomes ever narrower and more incoherent \cite{Wang2018},
without well-defined quasi-particles at the Fermi level.
Thus,
the $ t $ model exhibits non-Fermi liquid, bad metal behavior for all densities $ n \ge n_{ c , 1 } $.
This is further substantiated by the 
damping and quasi-particle residue shown in Figs.~\ref{fig:dampingn085multT} (a) and (b),
respectively.
The particle-hole asymmetry, 
already present at densities exhibiting instabilities towards stripe order,
likewise becomes even more pronounced with increasing density \cite{Wang2018}.
The high-energy particle excitations at the $\Gamma$-point remain the sharpest,
whereas the band in the negative-energy hole sector is significantly broadened. 
Moreover,
the density of states in Fig.~\ref{fig:dampingn085multT} (c) reveals 
that a significant amount of the spectral weight is transferred to a broad negative-energy continuum extending far below the Fermi energy.
In the momentum-resolved spectral functions 
displayed in Fig.~\ref{fig:tripleplotn085T065} (right),
this continuum is visible as the incoherent, low-weight band tails extending to negative frequencies.
The existence of such negative-frequency band tails in the $ t $ model at high densities has been conjectured already by Brinkmann and Rice in 1970 \cite{Brinkman1970}.
The underlying mechanism is the formation of spin polarons,
consisting of a bound state of a single hole and a local ferromagnetic environment.
These so-called ``Nagaoka polarons'' \cite{White2001, Maska2012, Lebrat2024, Samajdar2024a} can form because constructive interference of paths for the motion of a hole in a ferromagnetic background lowers its kinetic energy.
They can be viewed as
finite-size precursor of the Nagaoka ferromagnet and have not only been predicted in cluster simulations \cite{White2001, Maska2012, Samajdar2024a} but also recently observed in cold-atom experiments \cite{Lebrat2024, Prichard2024}.
Thus,
provided that the ferromagnetic correlation length is large enough,
we expect the existence of ferromagnetic polaron states in the hole sector
and below the single-particle band.
By increasing the size of the ferromagnetic bubble,
one can lower the energy of these polaron states, 
until the size reaches the finite ferromagnetic correlation length.
Hence,
there is a large low-energy continuum of polaronic states that individually carry only very little spectral weight.
We have already seen in Fig.~\ref{fig:magsuz} 
that there are strong ferromagnetic fluctuations,
and hence large correlation lengths, 
at low temperatures for all densities $ n \ge n_{ c , 1 } $.
They also drastically increase with the density,
especially for $ n \ge n_{ c , 2 } $ where the dominant instability is ferromagnetic.
Therefore,
we expect that this polaronic mechanism drives 
both the low coherence of hole excitations and the formation of the large tails below the single-particle band in the magnetic regimes.
This is further substantiated by the magnetic spectral functions shown in the companion paper~\cite{Arnold2025} and Appendix~\ref{app:xopsidentity},
which similarly show a polaronic continuum that grows with $ n $.

\subsubsection{Quasi-particle weight and bandwidth}
\label{sec:Z_and_W}
\begin{figure}%
\includegraphics[width=\linewidth]{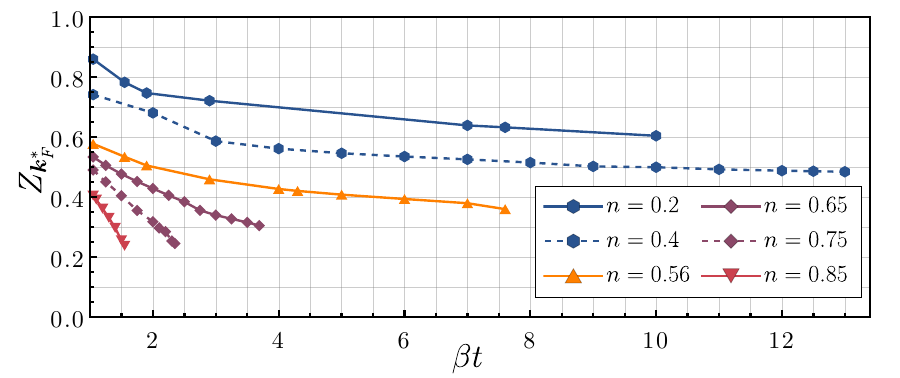}
\caption{Temperature dependence of the quasi-particle weight $Z_{\bm{k}^*_F}$ at the nodal Fermi momentum. 
Connecting lines are guides to the eye.
}
\label{fig:ZofT}
\end{figure}

%
%
%
%
Figure~\ref{fig:ZofT} shows the temperature dependence of the quasi-particle weight $Z_{\bm{k}}$ evaluated at the nodal Fermi momentum $\bm{k}^*_F$ (i.e., the Fermi momentum on the high symmetry path $\Gamma \rightarrow M$  in the Brillouin zone)
in the four different density regimes. 
For all densities,
$Z_{\bm{k}}$ decreases significantly with decreasing temperature.
For $ n < n_{ c , 2 } $,
it however appears to converge to a finite value in the zero-temperature limit $ \beta t \to \infty $.
In the ferromagnetic regimes for $ n \ge n_{ c , 2 } $ on the other hand, 
the quasi-particle weight drops drastically when the ferromagnetic susceptibility starts to grow exponentially;
compare Fig.~\ref{fig:magsuz} (c).

We associate this with the transfer of spectral weight from the Fermi surface to the emerging negative-energy polaronic continuum
that is visible in Figs.~\ref{fig:dampingn056multT} (c) and \ref{fig:dampingn085multT} (c).
Extrapolating from this behavior,
we speculate that at zero temperature the quasi-particle weight becomes very small \cite{Chiappe1993, Galan1992, Louis1993, Park2008, Dagotto1994, Deng2013} in the high-density FM2 state and may in fact vanish at half filling \cite{Galan1992, Louis1993, Dagotto1994},
while it remains finite but reduced in the FM1, stripe and paramagnetic states.
At elevated temperatures $ T \gtrsim 0.5 t $,
the reduction of the nodal quasi-particle weight with increasing density moreover agrees quantitatively with extrapolations based on high-order numerical linked-cluster expansion \cite{Khatami2014}.

\begin{figure*}%
\includegraphics[width=\linewidth]{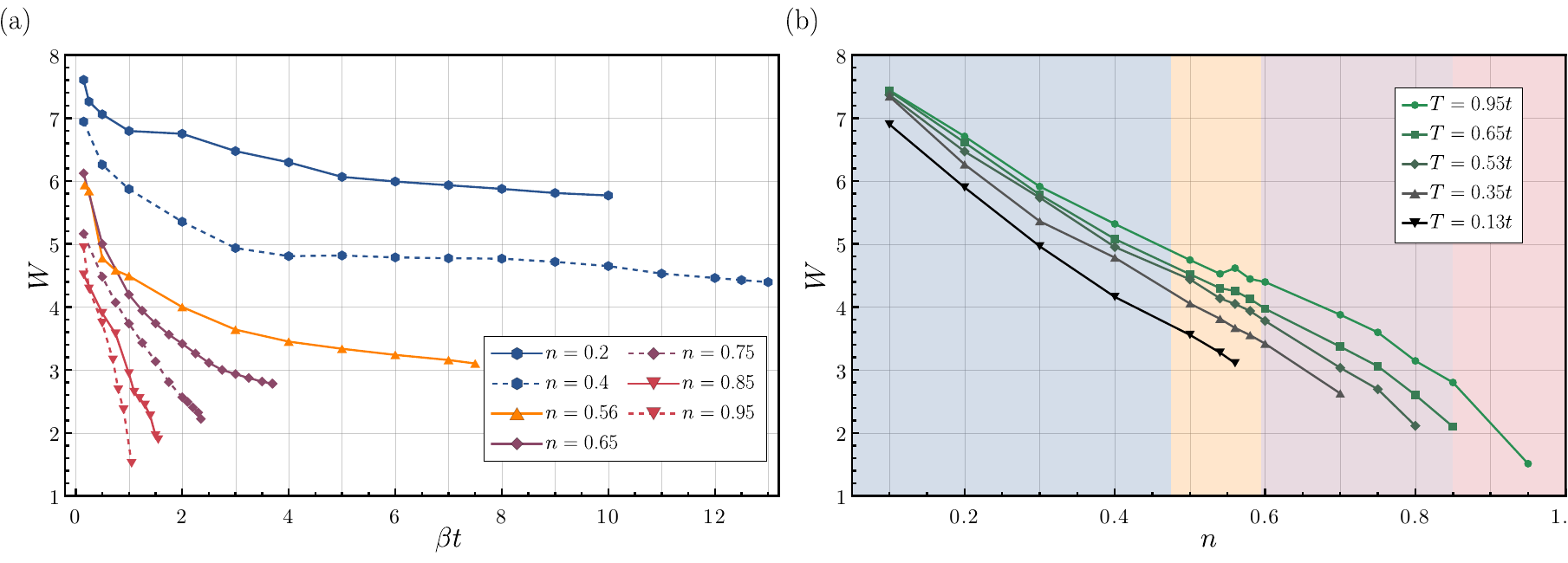}
\caption{Single-particle electronic bandwidth $ W $ 
as 
(a) function of inverse temperature for densities representative for the four distinct magnetic phases
and
(b) function of density for various temperatures. 
The colored backgrounds in (b) distinguish the magnetic ground states of the system, 
following the color scheme of Fig.~\ref{fig:PhaseDiag}.
Connecting lines are guides to the eye.}
\label{fig:bandwidthofn}
\end{figure*}

The bandwidth $ W $ of single-particle electronic excitations is shown
in Figs.~\ref{fig:bandwidthofn} (a) and (b) 
as function of temperature and density, 
respectively. 
We extract $ W $ via the positions of the maxima of the spectral functions $A(\bm{k},\omega)$ at the $\Gamma$- and $M$-points. 
For the latter,
the significant broadening and the nearby negative-energy continuum introduce a noticeable dependence  of the position of the maximum
on the Pad\'{e} approximant,
which somewhat reduces our numerical accuracy.
%
%
%
%
%
%
%
%
%
%
%
%
The bandwidth $ W $ is renormalized at all densities and always decreases with temperature. 
The behavior is overall very similar to the nodal quasi-particle weight $ Z_{ \bm{k}^*_F } $ shown in Fig.~\ref{fig:ZofT}.
Intuitively,
the narrowing of the single-particle band with increasing density is related to the decreasing number of sites accessible to an added electron because of the no-double occupancy constraint.  
In the paramagnetic and stripe regime, 
Fig.~\ref{fig:bandwidthofn} (a) suggest that $ W $ converges to a finite value at zero temperature. 
For higher densities $ n \ge n_{ c , 2 }$,
this becomes much less clear. 
At both $n = 0.65$ and 
$n = 0.75$ a decrease in slope of the temperature dependence of bandwidth can be argued for, 
possibly allowing an extrapolation to a finite width. 
At densities $n = 0.85$ and greater, towards the FM2 state,
our temperature ranges are insufficient to extrapolate any behavior beyond $ W \sim t $ \cite{Sherman2003}. 
The overall drastic decrease of the bandwidth with increasing density and decreasing temperature seems,
however,
consistent with the vanishing of the bandwidth at half filling \cite{Dagotto1994},
and points towards the formation of a flat band as a defining feature of the FM2 state.
Reference~[\onlinecite{Wang2018}] on the other hand reports a much larger bandwidth 
$ W \simeq 0.83 \times 8 t = 6.64 t $
at $ n =0.875 $,
using cluster perturbation theory based on $ 4 \times 4 $ clusters.
However,
in this study the band is not continuous but fractured in the hole sector.
We therefore believe that the discrepancy to our results is due to the finite cluster size,
which cannot capture the long-range physics of the polaronic band tails correctly,
yielding a much larger hole band.
At elevated temperatures $ T \gtrsim 0.5 t $,
our bandwidths also have a similar magnitude as the estimates obtained in Ref.~[\onlinecite{Khatami2013}] from the leading moments of the linked cluster expansion,
although the slight curvature of the $ W (n ) $ curve is different in both approaches.

\subsection{Violation of Luttinger's theorem}
\label{sec:LTviolation}
A central question that arises perpetually in strongly correlated electron systems is the validity of Luttinger's theorem \cite{Luttinger1960, Oshikawa2000, Seki2017}.
It states that the volume $ n_L $ enclosed by all Fermi surfaces equals the density
and has been proven to all orders in perturbation theory by Luttinger \cite{Luttinger1960} 
and much later also non-perturbatively with topological arguments \cite{Oshikawa2000, Seki2017}. 
Luttinger's theorem may fail at strong coupling $ U \gg W $
because the Hubbard model is no longer adiabatically connected to the $ U = 0 $ limit of free fermions \cite{Chiappe1993, Anderson2008, Anderson2009, Shastry2010, Casey2011}.
This disconnection is embodied by the Hilbert space projection in the $ t $-$ J $ and $ t $ models,
which ejects doubly occupied states and thus the upper Hubbard band from the spectrum entirely.
In consequence, 
the low-energy excitations of the weak- and strong-coupling Hubbard models live in Hilbert spaces of different dimensionality,
such that they cannot be connected by a convergent perturbation series.
The non-perturbative proofs of the validity of Luttinger's theorem on the other hand rely crucially on the algebra of canonical fermions \cite{Oshikawa2000, Seki2017}.
Thus, they are not directly applicable to the non-canonical operators acting on the projected Hilbert space.
For the Hubbard, $ t $-$ J $, and $ t $ models,
this issue has attracted much attention and remains, 
to our knowledge, 
controversial to this day \cite{Stephan1991, Singh1992, Putikka1998, Kokalj2007, Shastry2010, Kozik2015, Quinn2018, Osborne2021, Shastry2013, Perepelitsky2015, Shastry2019}.
In fact,
Shastry has argued in Ref.~[\onlinecite{Shastry2010}] that when Fermi liquid phenomenology applies,
extremely correlated electrons satisfy the following modified version of Luttinger's theorem:
\begin{equation}
\label{eq:Luttinger_Shastry}
n_L 
= \frac{ n }{ Z } 
= \frac{ n }{ 1 - n / 2 } 
\ge n . 
\end{equation}
This takes into account the density-dependent reduction of the spectral weight to $ Z = 1 - n / 2 $ in the projected Hilbert space;
see Eq.~\eqref{eq:A_normalization}.

\begin{figure}%
\includegraphics[width=\linewidth]{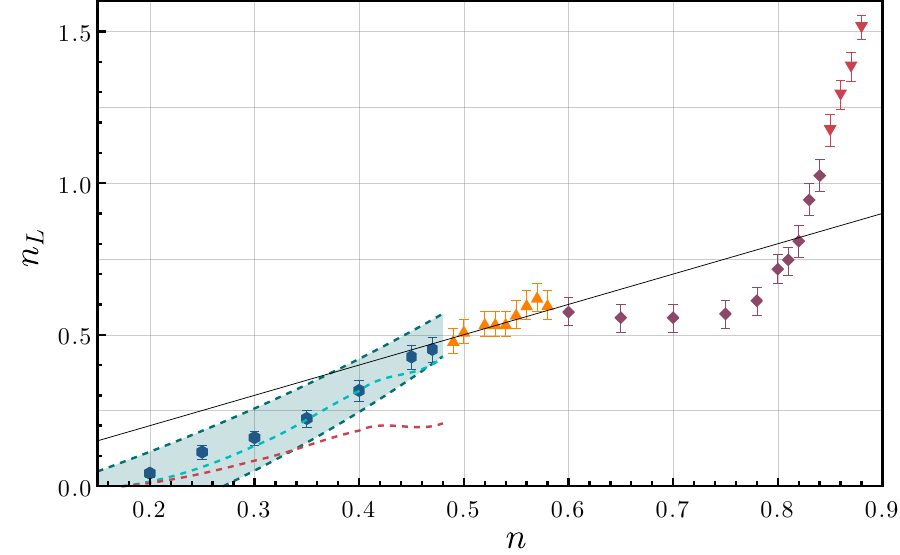}
\caption{Volume $n_L$ enclosed by the interacting Fermi surfaces, 
evaluated at the lowest temperature available at each density. 
The colors represent the magnetic ground state of the system based on Fig.~\ref{fig:PhaseDiag}. 
Error bars indicate the finite momentum resolution.
The black line marks the volume predicted by Luttinger's theorem. 
The dashed red line show the Luttinger's theorem under consideration of the 
systematic error $ \delta $ within our data.
The dashed cyan line follows from the modified Luttinger's theorem \eqref{eq:Luttinger_numerics} that accounts both for the reduced spectral weight and the actual systematic error $ \delta $.
The shaded area is a measure of the numerical uncertainty of $ \delta $;
it is bounded by Eq.~\eqref{eq:Luttinger_numerics} evaluated for the minimal ($ 0.11 $) 
and maximal ($ 0.27 $)
values of $ \delta $;
compare Fig.~\ref{fig:ZDeltaError}.}
\label{fig:LVviolation}
\end{figure}

Clearly our results violate Luttinger's theorem, 
as seen in the center panels of Figs.~\ref{fig:tripleplotn02T1}, \ref{fig:tripleplotn056T13}, and \ref{fig:tripleplotn085T065}. 
Our findings are summarized in Fig.~\ref{fig:LVviolation},
where we show the volume $ n_L $ enclosed by the spin-up and spin-down Fermi surfaces,
with the prediction
$ n_L = n $ 
of Luttinger's theorem for comparison. 
The error bars are a consequence of the finite momentum grid, 
giving rise to some uncertainty on the exact position of the Fermi surface. 
Corresponding to the four different magnetic regimes, 
we observe qualitative differences in the Luttinger count $ n_L $. 
In the low-density paramagnetic regime, 
the volume is consistently smaller than predicted by Luttinger's theorem,
and hence also by the modified rule \eqref{eq:Luttinger_Shastry}.
This is contrary to the usual expectation that the Hilbert space projection should not matter at sufficiently low densities,
where the $ t $ model behaves according to conventional Fermi liquid theory.
To further illuminate this issue,
we consider the occupation number
\begin{subequations} \label{eq:nk}
\begin{align}
n( \bm{k} ) 
= {} & \frac{ 2 }{ \beta } \sum_\omega e^{ i \omega 0^+ } G_{ \Lambda = 1 } ( \bm{k} , \omega )
\label{eq:n_via_G_normal}
\\
= {} & Z + \frac{ 2 }{ \beta } \sum_\omega \cos ( \omega 0^+ ) G_{ \Lambda = 1 } ( \bm{k} , \omega )
\label{eq:n_via_G_sym}
\\
= {} & 2 \int_{-\infty}^{\infty} \textrm{d} \omega \frac{ 1 }{ 1 + e^{ \beta \omega } } A( \bm{k} , \omega ) .
\label{eq:n_via_A}
\end{align}
\end{subequations}
Note that in the above, 
to obtain from the normal ordered expression \eqref{eq:n_via_G_normal}
the symmetrically ordered one \eqref{eq:n_via_G_sym}
and the spectral representation \eqref{eq:n_via_A},
it is crucial that the holon algebra \eqref{eq:holon_algebra} is exactly preserved.
In the Matsubara propagator \eqref{eq:holon_propagator},
this means that the high-frequency asymptotic is 
$ G_{ \Lambda = 1 } ( \bm{k} , \omega )
\sim Z / i \omega  $
as in the atomic limit.
In the spectral function,
this is reflected in the normalization to $ Z $,
Eq.~\eqref{eq:A_normalization}.
Unfortunately,
our truncation of the X-FRG flow does not perfectly preserve these properties.
Both the coefficient of the high-frequency asymptotic of 
$ G_{ \Lambda = 1 } ( \bm{k} , \omega ) $
and the normalization of 
$ A ( \bm{k} , \omega ) $
at low temperatures slightly reduce from
$ Z $ to $ Z - \delta $;
compare Fig.~\ref{fig:ZDeltaError} in Appendix~\ref{app:Merror}.
Since our numerics furthermore employs a finite Matsubara frequency cutoff,
we naturally determine the density $ n $ from the symmetric regularization \eqref{eq:n_via_G_sym}.
Taking into account the finite systematic error $ \delta $ then yields 
\begin{equation} \label{eq:n_error}
n 
= \frac{ 1 }{ N } \sum_{ \bm{k} } n ( \bm{k} )
=
\delta + 2 \int_{-\infty}^{\infty} \textrm{d} \omega \frac{ 1 }{ 1 + e^{ \beta \omega } } 
\textmd{DOS}( \omega ).
\end{equation}
Provided that we are in the Fermi liquid regime,
the frequency integration in the presence of the normalization error $ \delta $ equals 
$ ( Z - \delta ) n_L $ 
\cite{Shastry2010, Luttinger1960}.
Solving for the Luttinger volume,
we obtain the modified version of Luttinger's theorem appropriate for our truncation,
\begin{equation}
\label{eq:Luttinger_numerics}
n_L = \frac{ n - \delta }{ 1 - n/2 - \delta } .
\end{equation}
This corrected form in fact agrees quite well with our data in the low-density Fermi liquid regime. On the other hand, considering the unmodified Luttinger's theorem together with our normalization error, we find significant disagreement with our data;
see Fig.~\ref{fig:LVviolation}.
Comparing both scenarios,
we therefore
argue that in this regime,
the physical behavior should actually follow the modified version \eqref{eq:Luttinger_Shastry} of Luttinger's theorem derived by Shastry \cite{Shastry2010}.
Especially at low densities,
the systematic error $ \delta $ can be almost as large as the density itself and therefore cannot be neglected.
While suggestive, our conclusion should be understood with this in mind.
In contrast,
we show in Fig.~\ref{fig:ZDeltaError} in Appendix~\ref{app:Merror}
that at higher densities
both the absolute and the relative magnitude of the error decrease,
so that the lack of numerical accuracy  alone cannot account fo the violation of Luttinger's theorem revealed by our calculation.

With the transition to the intermediate density regime
where $ \mu \approx 0 $ (see Fig.~\ref{fig:muofn})
and antiferromagnetic stripe fluctuations dominate,
Luttinger's theorem is surprisingly restored within our numerical accuracy.
However,
in the high-density ferromagnetic regimes,
Luttinger's theorem is again broken,
with the behavior changing qualitatively
between the FM1 and FM2 regions, as already anticipated in Sec.~\ref{sec:FM}.
For densities between $ n_{ c , 2 } $ and $ n_{ c , 3 } $,
the Luttinger volume $ n_L $ remains roughly stationary 
and thus stays again below the value demanded by Luttinger's theorem.
Around $ n_{ c , 3 } $ on the other hand,
$ n_L $ rises drastically,
so that the Fermi surface becomes much larger than the corresponding non-interacting one.
This feature has also been inferred previously from exact diagonalization \cite{Chiappe1993, Kokalj2007}, the high temperature series expansion \cite{Putikka1998} and dynamical mean field theory \cite{Deng2013}.  
It gives yet another indication that the nature of the two ferromagnetic regimes below and above the critical density $ n_{ c , 3 } $ is qualitatively different.
Of course,
these results come with their own caveats:
First,
with increasing density,
the Fermi surfaces are evaluated at ever higher temperatures because of the incipient magnetic instabilities.
In the density range 
$ n = 0.6 $ to $ 0.9 $,
the lowest available temperatures lie between 
$ T \approx 0.18t $ and $ t $.
Second,
one has to keep in mind that for densities beyond $ n_{ c , 2 } $,
the electronic spectrum becomes ever more incoherent;
see Figs.~\ref{fig:tripleplotn056T13} and \ref{fig:tripleplotn085T065}.
Thus,
there is no proper Fermi surface that hosts sharp excitations in this regime in any case.

\begin{figure}%
\includegraphics[width=\linewidth]{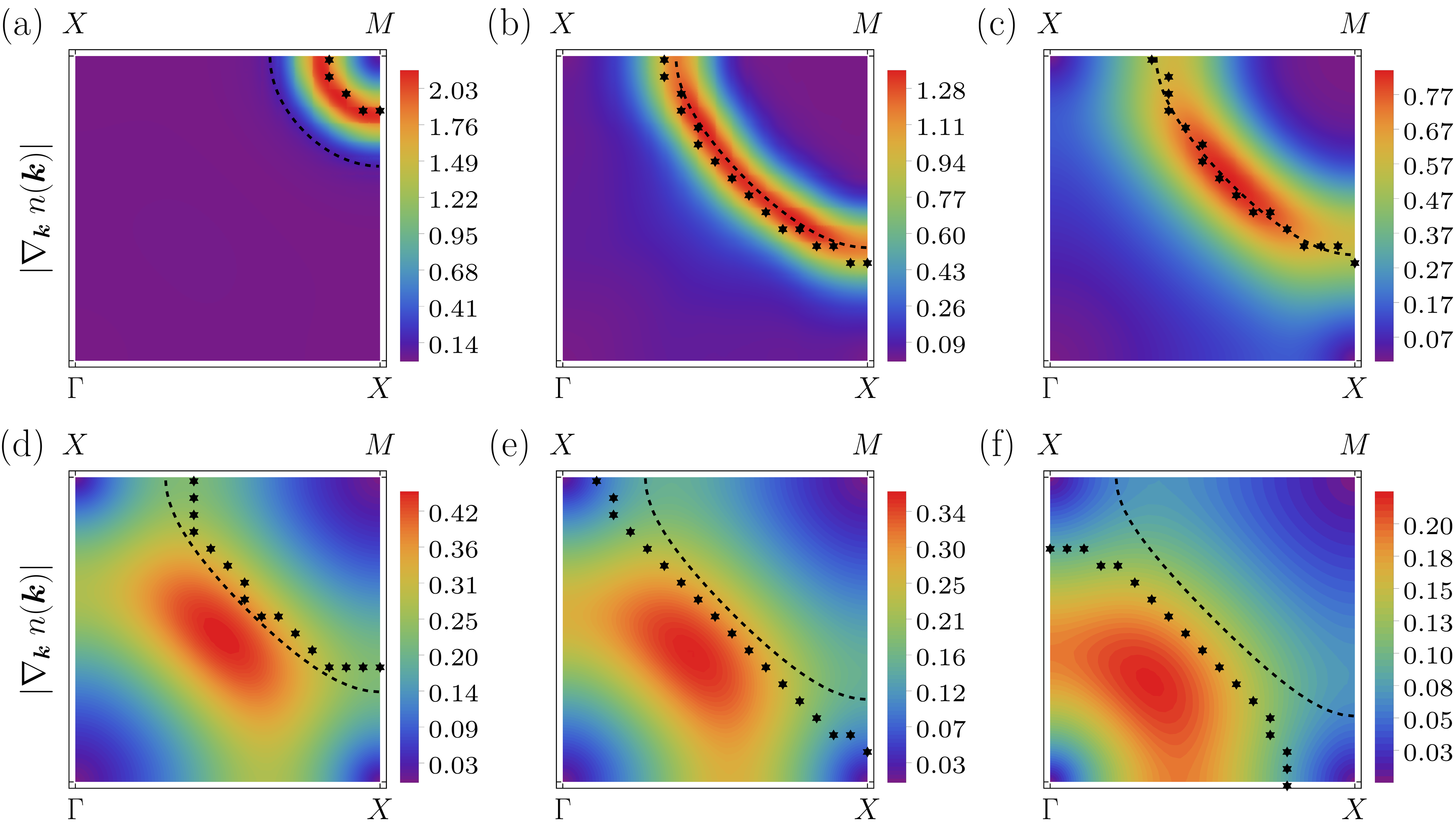}
\caption{Contour plot of the gradient of the occupation number, 
$|\nabla_{\bm{k}}n(\bm{k})|$. 
The dashed black line marks the free Fermi surface 
and the stars the interacting one determined via the spectral function $A(\bm{k},0)$. 
(a) $n = 0.2$,  $ T = 0.1 t $;
(b) $n = 0.56$, $ T = 0.13 t $;
(c) $n = 0.6$, $ T = 0.18 t $;
(d) $n = 0.7$, $ T = 0.34 t $;
(e) $n = 0.75$, $ T = 0.43 t $;
(f) $n = 0.85$, $ T = 0.65 t $. 
At low and intermediate densities the interacting Fermi surface coincides with 
$\textrm{max}|\nabla_{\bm{k}}n(\bm{k})|$. 
Beyond $ n_{ c , 2 } = 0.6 $ however, 
a second hole-like contour forms, 
dominating the gradient at high density. 
From this point onward the Luttinger theorem is broken again,
and the interacting Fermi surface lies between the two 
$\textrm{max}|\nabla_{\bm{k}}n(\bm{k})|$-contours.}
\label{fig:DNsurfaces}
\end{figure}

Since our determination of the Fermi surface depends on an admittedly ill-defined analytical continuation,
we here also discuss an alternative method of obtaining the Fermi surface that does not require analytical continuation \cite{Singh1992,Putikka1998}.
This strategy is based on the occupation number \eqref{eq:nk}.
At least for a conventional Fermi liquid with sharp excitations at the Fermi level,
one can estimate the position of the Fermi surface from the maximum of the slope of $ n ( \bm{k} ) $ in the Brillouin zone.
In Fig.~\ref{fig:DNsurfaces},
we therefore show the gradient $|\nabla_{\bm{k}} n(\bm{k})|$
for the six densities $n = 0.2$, $0.56$, $0.6$, $0.7$, $0.75$ and $0.85$.
For the first three densities, we indeed find good agreement between the interacting Fermi surface (black stars) and 
$ \textrm{max} | \nabla_{\bm{k}} n(\bm{k}) | $. 
This changes for higher densities,
in the ferromagnetic regimes beyond $ n_{ c , 2 } $. 
Here, 
the occupation number develops a strong momentum dependence even away from the Fermi surface \cite{Kuzmin1997, Singh1992, Dagotto1994}.
For such high densities we can identify two contours of local maxima. 
The first one is weaker and positioned near the non-interacting Fermi surface (dashed black line). 
It appears to be a remnant of the low-density contour line. 
The second one has newly formed and traces a hole-like surface.
According to the spectral function,
the actual interacting Fermi surface lies in between these two contours. 
In Fig.~\ref{fig:DNsurfaces} (d) the Fermi surface is still close to the electron-like remnant contour. 
Upon increasing the density further the electron-like slope decreases while the hole-like slope increases, 
which is accompanied by the interacting Fermi surface changing its topology from electron- to hole-like in Fig.~\ref{fig:DNsurfaces} (f).

\begin{figure}%
\includegraphics[width=\linewidth]{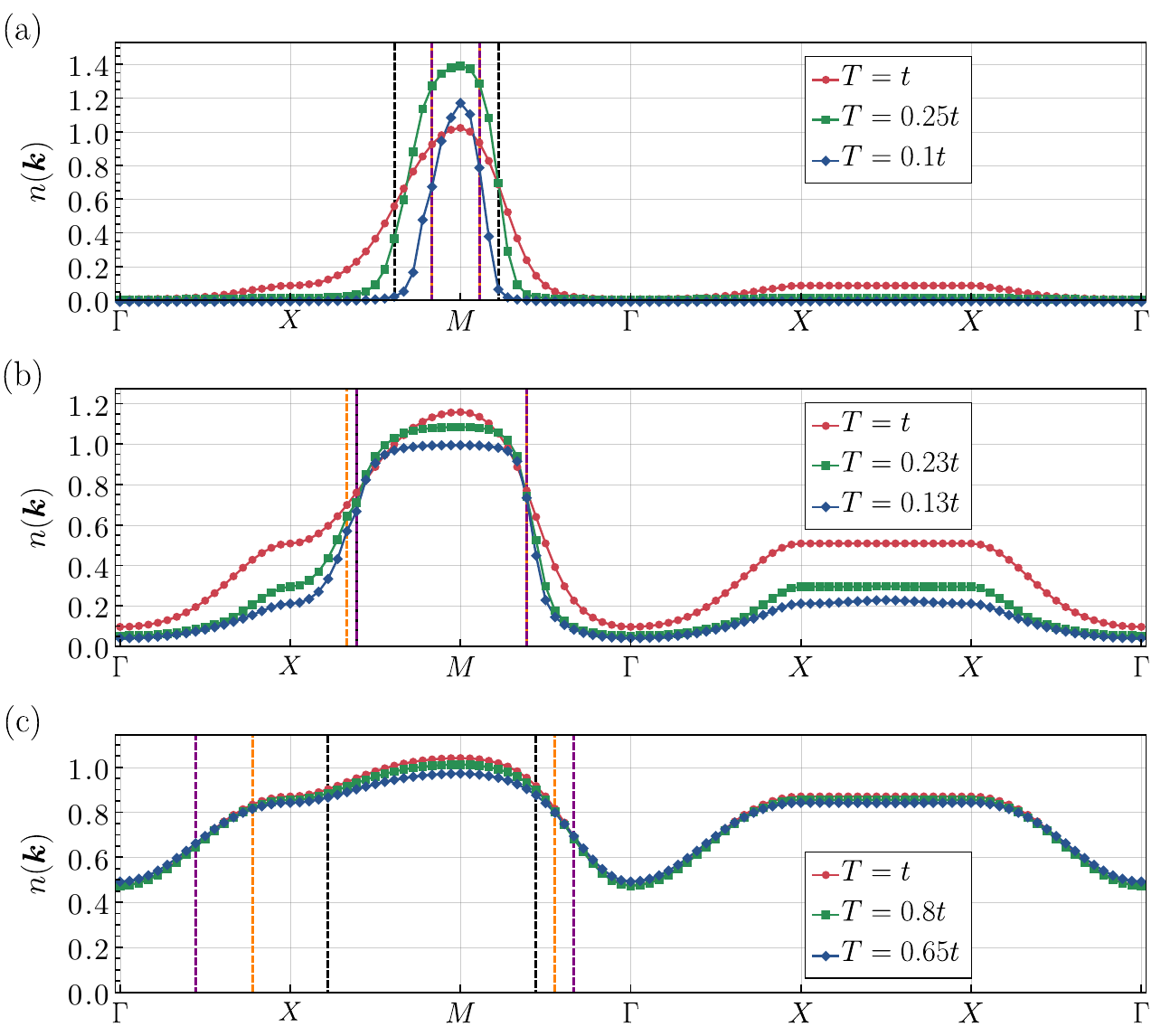}
\caption{Occupation number $n(\bm{k})$ along the high symmetry path through the Brillouin zone, 
for various temperatures, 
at densities
(a) $ n = 0.2 $,
(b) $ n = 0.56 $, and
(c) $ n = 0.85 $.
Dashed orange lines mark the Fermi surface obtained from $ A ( \bm{k} , 0 ) $,
whereas dashed purple lines mark 
$ \textrm{max} | \nabla_{ \bm{k} } n_{ \bm{k} } | $.
The dashed black lines are the corresponding non-interacting Fermi surface.}
\label{fig:nwalk}
\end{figure}%
Given the low coherence and the non-Fermi liquid behavior of the spectral function in the ferromagnetic regime,
the disagreement between the two methods of determining the Fermi surface is not surprising.
The root cause is the lack of a true Fermi edge in the occupation number.
To illustrate this,
we show the momentum dependence of the occupation number $ n ( \bm{k} ) $ in Figs.~\ref{fig:nwalk} (a)-(c) for low, intermediate, and high densities \cite{footnote_occupation}.
At density $n = 0.2$, 
displayed in Fig.~\ref{fig:nwalk} (a), 
we find standard Fermi liquid behavior towards zero temperature, 
where only states within the Fermi surface are occupied.
%
%
%
%
%
%
%
%
%
%
%
The behavior is similar,
but already appreciably smeared out in the stripe regime at $n = 0.56$, 
shown in Fig.~\ref{fig:nwalk} (b).
States outside the Fermi surface are emptied upon decreasing temperature. 
However a sharp edge never truly forms, at least not in within the temperature range accessible to us.
Likewise,
a finite but small occupation remains away from the Fermi edge along the Van Hove singularity,
reflecting the existence of the continuum below the single-particle band;
compare Fig.~\ref{fig:tripleplotn056T13}.
%
%
%
%
%
%
%
%
%
%
%
%
At high density $n = 0.85$ the picture has qualitatively changed. 
Figure~\ref{fig:nwalk} (c) shows occupation of all states. 
The momentum dependence follows a much more generic cosine behavior 
and clear-cut features of a Fermi edge are completely lost \cite{Singh1992, Dagotto1994}. 
Lowering the temperature has surprisingly little effect on the allocation of states. Of course one must keep in mind that the onset of the ferromagnetic instability limits our temperature range notably.
We conclude that the spectral function is the only reliable means of determining the Fermi surface 
when Fermi liquid phenomenology is absent.
Yet another way of determining $ A ( \bm{k} , 0 ) $,
and hence the Fermi surface, 
at low temperatures
without analytical continuation is discussed in Appendix~\ref{app:firstmatsubara}.

\section{Conclusions and Outlook}

\label{sec:conclusions}

Extreme correlations remain a challenge for condensed matter theory.
They arise when large or even infinite contact interactions restrict the low-energy dynamics to a projected Hilbert space.
The kinematic interactions due to this Hilbert space projection necessitate a genuinely non-perturbative treatment.
The fundamental difference to weak-coupling physics,
which can be perturbatively related to non-interacting canonical fermions,
is mathematically expressed by the non-canonical algebra of the basic operators
of the model acting on the restricted Hilbert space. 
In this work,
we have extended the applicability of the non-perturbative functional renormalization group (FRG) to extremely correlated electrons.
To that end,
we have considered the so-called $ t $ model,
that is,
the Hubbard model at $ U = \infty $,
where only the correlated hopping in the projected Hilbert space remains.
This has allowed us to study the kinematic interactions due to the Hilbert space projection at its purest,
without any other competing interactions.

Because of the long-range character of the low-energy physics,
which is strikingly illustrated by the exact Nagaoka theorem,
the $ t $ model is notoriously difficult to treat with existing numerical techniques that rely on finite clusters.
Using the X-FRG approach developed in Ref.~[\onlinecite{Rueckriegel2023}],
we have obtained both the magnetic ground state phase diagram as well as the finite-temperature, momentum-resolved electronic spectrum of the $ t $ model for the full range of electronic densities $ n $, 
directly in the thermodynamic limit.
In particular,
we have confirmed the existence of Nagaoka ferromagnetism in the thermodynamic limit for electronic densities $ 0.85 \le n < 0.99 $.
For lower densities $ 0.6 \le n \le 0.84 $,
where some previous studies indicate likewise strong ferromagnetic fluctuations but without a fully saturated ground state,
we have also found a second,
qualitatively different ferromagnetic regime.
Our analysis of the electronic spectrum reveals that these two ferromagnetic states are separated by a Lifshitz transition of the Fermi surface.
At even lower densities $ 0.48 \le n \le 0.59 $,
the low-temperature state of the $ t $ model exhibits antiferromagnetic stripe order,
although subdominant ferromagnetic fluctuations are still present.
In the low-density regime $ 0 < n \le 0.47 $,
where $ \mu < 0 $,
the system remains paramagnetic.
For these densities,
the low-energy electronic spectrum is described by the Fermi liquid phenomenology.
The electronic spectrum in the magnetically correlated regimes,
where $ \mu \gtrsim 0 $, 
is on the other hand best characterized as an incoherent bad metal \cite{Emery1995, Deng2013}.
Most notably, in this regime the system
develops a pronounced particle-hole asymmetry with increasing density \cite{Wang2018},
with extended band tails below the single-particle band in the hole sector.
We attribute these to the formation of Nagaoka polarons.
Moreover,
the single-particle bandwidth is significantly reduced by the kinematic interactions at high densities,
while the Van Hove singularity moves close to the Fermi level.
Hence,
the Nagaoka state of the $ t $ model shares some aspects of the idealized scenario of flat-band correlated ferromagnetism \cite{Mielke1991a,Mielke1991b,Tasaki2003,Hu2025}.
Last,
we observe distinct deviations from Luttinger's theorem for the Fermi surface volume
for small and large densities, but not for intermediate densities where 
the ground state exhibits  stripe order.
This again highlights the fundamental disconnection of the physics emerging 
in the extremely correlated $ U = \infty $ limit from the weak-coupling regime where $ U $ is smaller than the hopping energy.

The X-FRG approach to extremely correlated electrons
developed in this work
complements the established FRG for canonical fermions \cite{Kopietz2010,Metzner2012,Salmhofer2001,Honerkamp2001,Kopietz2001,
Halboth2000,Halboth2000b,Salmhofer2004,Ossadnik2008,Husemann2009,Husemann2012,Taranto2014,
Vilardi2017,Vilardi2019,Lichtenstein2017,Honerkamp2018,Ehrlich2020,
Hille2020,Honerkamp2022,Profe2022}:
While the latter sets out from the $ U = 0 $ limit of non-interacting electrons and can deal with weak to moderate interactions,
the former is designed for the extremely correlated regime
where $ U $ is very large or even infinite.
This is achieved by splitting the problem into two separate parts.
First,
the local physics is solved analytically.
This takes care of the non-canonical algebra associated with the Hilbert space projection,
which is treated exactly.
Second,
non-local hopping is included via a diagrammatic resummation of the strong-coupling expansion.
This is the actual FRG step,
which is formally identical to the weak-coupling,
canonical fermion case.
The only difference is that the initial vertices carry the non-trivial frequency dependence of the local problem,
which also determines the high-frequency asymptotics.
Since this dependence is however known analytically,
it can be factored out and treated exactly. 
This strategy is similar in spirit to the extremely correlated 
Fermi liquid theory developed by Shastry and co-authors \cite{Shastry2010,Perepelitsky2015,Mai2018},
where a local scale transformation is applied to eliminate the non-canonical high-frequency behavior of correlation functions of the $ t $-$ J $ model.

Because of its formal similarity to the FRG for canonical fermions,
we expect that many well-established methods and approximation schemes can be adapted to the strong-coupling X-FRG with minor modifications.
A case in point is the channel decomposition devised in this work.
Thus,
the X-FRG provides a promising avenue for an unbiased investigation of extremely correlated electron systems,
directly in the thermodynamic limit and with modest numerical effort.
Extensions to different lattice and hopping geometries are straightforward and currently underway.
In the future,
we plan to address the full $ t $-$ J $ model as well.
In the X-FRG framework \cite{Rueckriegel2023},
this requires bosonic fields for density and spin variables.
Thus,
the X-FRG allows one to treat
spin and density dynamics on equal footing with the projected electrons.
This may also open up new ways to incorporate the exact normalization of the spectral function via Ward identities.
Its breaking at low temperatures appears to be the most serious issue that the current X-FRG implementation faces at present.

\begin{acknowledgments}
This work was financially supported by the Deutsche
Forschungsgemeinschaft (DFG, German Research Foundation) 
through Project No. 431190042.
\end{acknowledgments}
\section*{Data availability}

The data that support the findings of this study were generated and analyzed with \textit{Wolfram Mathematica}. A custom \textit{Mathematica} notebook containing all scripts is publicly available~\cite{gude}.

\appendix

\setcounter{equation}{0}

\renewcommand{\theequation}{A\arabic{equation}}

\renewcommand{\appendixname}{APPENDIX}

\renewcommand{\thesection}{\Alph{section}}

\section{Temperature rescaling} 

\label{app:temperature}

With the multiplicative deformation 
$ t_{ \Lambda , i j } 
= \Lambda t_{ i j } $,
the temperature $ T = 1 / \beta $ can be scaled out of the generating functionals
\eqref{eq:generating_functional_G} and
\eqref{eq:generating_functional_vertices}.
To that end,
we introduce rescaled quantities 
(denoted with a bar) via
$ \Lambda 
= T \bar{ \Lambda } $,
$ \tau 
= \beta \bar{ \tau } $,
$ \omega 
= T \bar{ \omega } $,
$ \mu_\Lambda 
= T \bar{ \mu }_{ \bar{ \Lambda } } $,
$ \tilde{f}_\Lambda 
= T \bar{ \tilde{f} }_{ \bar{ \Lambda } } $,
$ G_\Lambda 
= \beta \bar{ G }_{ \bar{ \Lambda } } $,
$ \Sigma_\Lambda 
= T \bar{ \Sigma }_{ \bar{ \Lambda } } $,
$ U_\Lambda 
= T \bar{ U }_{ \bar{ \Lambda } } $,
$ \ldots $.
Note that the only dimensionful quantity remaining in this scheme is the deformation parameter 
$ \bar{ \Lambda } $,
which flows from zero to $ \beta $.
The starting point of the X-FRG flow can thus be interpreted as the infinite temperature limit 
$ \beta = 0 $,
where the correlation length vanishes and only isolated Hubbard atoms remain.
The X-FRG flow then gradually lowers the temperature down to a finite temperature 
$ \bar{ \Lambda } =\beta > 0 $.
To obtain the physical quantities at this temperature from the flowing rescaled ones,
one merely has to scale them again by the appropriate powers of $ T $.

\renewcommand{\theequation}{B\arabic{equation}}

\section{Vertex function symmetries} 

\label{app:symmetries}

In this Appendix,
we discuss the non-trivial constraints imposed on our matrix-valued channel decomposition \eqref{eq:U_channels}
that follow from the symmetries of the 2-body interaction 
$ U_\Lambda ( K_1' , K_2' ; K_2 , K_1 ) $.
There are two relevant symmetries for the $ t $ model:

\textit{1.~Crossing symmetry.}
The 2-body interaction must be symmetric under
\begin{equation}
Q_\textrm{pp} \to Q_\textrm{pp} \; ,
\;\;\; \;\;\;
Q_\textrm{ex} \to - Q_\textrm{ex} \; ,
\;\;\; \;\;\;
Q_\textrm{fs} \to - Q_\textrm{fs} \; ,
\end{equation}
which implies
\begin{equation}
U_\Lambda ( K_1' , K_2' ; K_2 , K_1 ) =
U_\Lambda ( K_2' , K_1' ; K_1 , K_2 ) \; .
\end{equation}
Inserting the channel decomposition \eqref{eq:U_channels} then yields the following matrix equation for the channels:
\begin{align}
0 = {} &
\frac{ 1 }{ 2 } \left[
\textbf{M}_\Lambda ( Q_\textrm{fs} ) -
\textbf{M}_\Lambda^\top ( - Q_\textrm{fs} ) -
\textbf{C}_\Lambda ( Q_\textrm{fs} ) +
\textbf{C}_\Lambda^\top ( - Q_\textrm{fs} )
\right]
\nonumber\\
& +
\textbf{M}_\Lambda ( Q_\textrm{ex} ) -
\textbf{M}_\Lambda^\top ( - Q_\textrm{ex} ) -
\textbf{S}_\Lambda ( Q_\textrm{pp} ) -
\textbf{S}_\Lambda^\top ( Q_\textrm{pp} )
\; .
\end{align}
This symmetry is satisfied by choosing the three channel matrices such that
\begin{subequations}
\label{eq:particle-hole_channels}
\begin{align}
\textbf{S}_\Lambda^\top ( Q ) 
& = \textbf{S}_\Lambda ( Q ) \; ,
\\
\textbf{M}_\Lambda^\top ( Q ) 
& = \textbf{M}_\Lambda ( - Q ) \; ,\label{eq:deltaMcrossing}
\\
\textbf{C}_\Lambda^\top ( Q ) 
& = \textbf{C}_\Lambda ( - Q ) \; .\label{eq:deltaCcrossing}
\end{align}
\end{subequations}

\textit{2.~Time-reversal symmetry.}
The 2-body interaction must be symmetric under
\begin{equation}
Q_\textrm{pp} \to Q_\textrm{pp} \; ,
\;\;\; \;\;\;
Q_\textrm{ex} \to Q_\textrm{ex} \; ,
\;\;\; \;\;\;
Q_\textrm{fs} \to - Q_\textrm{fs} \; ,
\end{equation}
which implies
\begin{equation}
U_\Lambda ( K_1' , K_2' ; K_2 , K_1 ) =
U_\Lambda ( K_1 , K_2 ; K_2' , K_1' ) \; .
\end{equation}
When inserting the channel decomposition \eqref{eq:U_channels},
this symmetry cannot be cast into a matrix form.
It explicitly yields 
\begin{align}
0 = {} &
- \Omega_\textrm{fs}
\left[
S_\Lambda^{ 2 1 } ( Q_\textrm{pp} ) -
S_\Lambda^{ 1 2 } ( Q_\textrm{pp} ) +
\frac{ i \Omega_\textrm{ex} }{ Z }
S_\Lambda^{ 2 2 } ( Q_\textrm{pp} )
\right]
\nonumber\\
&
+ \Omega_\textrm{fs}
\left[
M_\Lambda^{ 2 1 } ( Q_\textrm{ex} ) -
M_\Lambda^{ 1 2 } ( Q_\textrm{ex} ) +
\frac{ i \Omega_\textrm{ex} }{ Z }
M_\Lambda^{ 2 2 } ( Q_\textrm{ex} )
\right]
\nonumber\\
&
+  \frac{ \Omega_\textrm{ex} }{ 2 }
\left[
M_\Lambda^{ 2 1 } ( Q_\textrm{fs} ) -
M_\Lambda^{ 1 2 } ( Q_\textrm{fs} ) +
\frac{ i \Omega_\textrm{fs} }{ Z }
M_\Lambda^{ 2 2 } ( Q_\textrm{fs} )
\right]
\nonumber\\
&
+  \frac{ \Omega_\textrm{ex} }{ 2 }
\left[
C_\Lambda^{ 2 1 } ( Q_\textrm{fs} ) -
C_\Lambda^{ 1 2 } ( Q_\textrm{fs} ) -
\frac{ i \Omega_\textrm{fs} }{ Z }
C_\Lambda^{ 2 2 } ( Q_\textrm{fs} )
\right] 
\; .
\label{eq:channels_time-reversal}
\end{align}
Because the contribution of each channel depends only on the relevant momentum-frequency transfer,
each square bracket of Eq.~\eqref{eq:channels_time-reversal} above has to vanish separately.
This also immediately yields
$ S_\Lambda^{ 2 2 } ( Q ) = 0 $.
For the components of the magnetic and charge channels,
we obtain in this manner that
\begin{subequations}
\begin{align}
\frac{ i \Omega }{ Z } M_\Lambda^{ 2 2 } ( Q ) 
& =
M_\Lambda^{ 1 2 } ( Q ) - M_\Lambda^{ 2 1 } ( Q )
\; , 
\\
\frac{ i \Omega }{ Z } C_\Lambda^{ 2 2 } ( Q ) 
& =
C_\Lambda^{ 2 1 } ( Q ) - C_\Lambda^{ 1 2 } ( Q )
\; .
\end{align}
\end{subequations}
Note that in the atomic limit \eqref{eq:SMC_0}, 
these relations are satisfied non-trivially because of 
$ \Omega \delta_{ \Omega , 0 } = 0 $.
For $ \Lambda > 0 $,
they imply that the components of the matrix of particle-hole bubbles $ \textrm{P}_\Lambda ( Q ) $ satisfy
\begin{equation}
\label{eq:P_time_reversal}
\frac{ i \Omega }{ Z } P_\Lambda^{ 1 1 } ( Q ) =
P_\Lambda^{ 2 1 } ( Q ) - P_\Lambda^{ 1 2 } ( Q )
\; .
\end{equation}
One can easily verify explicitly that the single-scale version of this relation is fulfilled by Eq.~\eqref{eq:P_dot},
as well as that it is preserved by the Katanin substitution \eqref{eq:P_Katanin}.
The subsequent loop expansion \eqref{eq:P_1-loop} on the other hand satisfies it only to first order in loops.
However,
one should keep in mind that the time-reversal relation \eqref{eq:P_time_reversal} only fixes the dynamics of $ P_\Lambda^{ 1 1 } ( Q ) $.
Especially in light of the conserved initial conditions \eqref{eq:SMC_0},
one also has to account for possible $ \delta_{ \Omega , 0 } $ terms;
see for example the approximation \eqref{eq:P_1-loop}. 
Therefore one cannot use Eq.~\eqref{eq:P_time_reversal} 
to completely eliminate $ P_\Lambda^{ 1 1 } ( Q ) $.
On the other hand,
time-reversal symmetry can be kept exactly by using Eq.~\eqref{eq:P_time_reversal}
to eliminate $ P_\Lambda^{ 2 1 } ( Q ) $,
which we do in our numerics.
This does however violate the crossing symmetry, 
Eqs.~\eqref{eq:deltaMcrossing} and~\eqref{eq:deltaCcrossing}. 
Since we are ultimately not interested in the complete 2-body interaction 
$ U_\Lambda ( K_1' , K_2' ; K_2 , K_1 ) $,
but in the electronic propagator and the spin, charge, and superconducting susceptibilities,
we believe that neglecting the crossing symmetry is reasonable.
On the other hand,
we aim to investigate magnetic correlations.
To that end,
keeping time-reversal symmetry is essential,
which we implement via Eq.~\eqref{eq:P21}.

\renewcommand{\theequation}{C\arabic{equation}}

\section{Dynamic spin susceptibility}

\label{app:xopsidentity}

\begin{figure}%
\includegraphics[width=\linewidth]{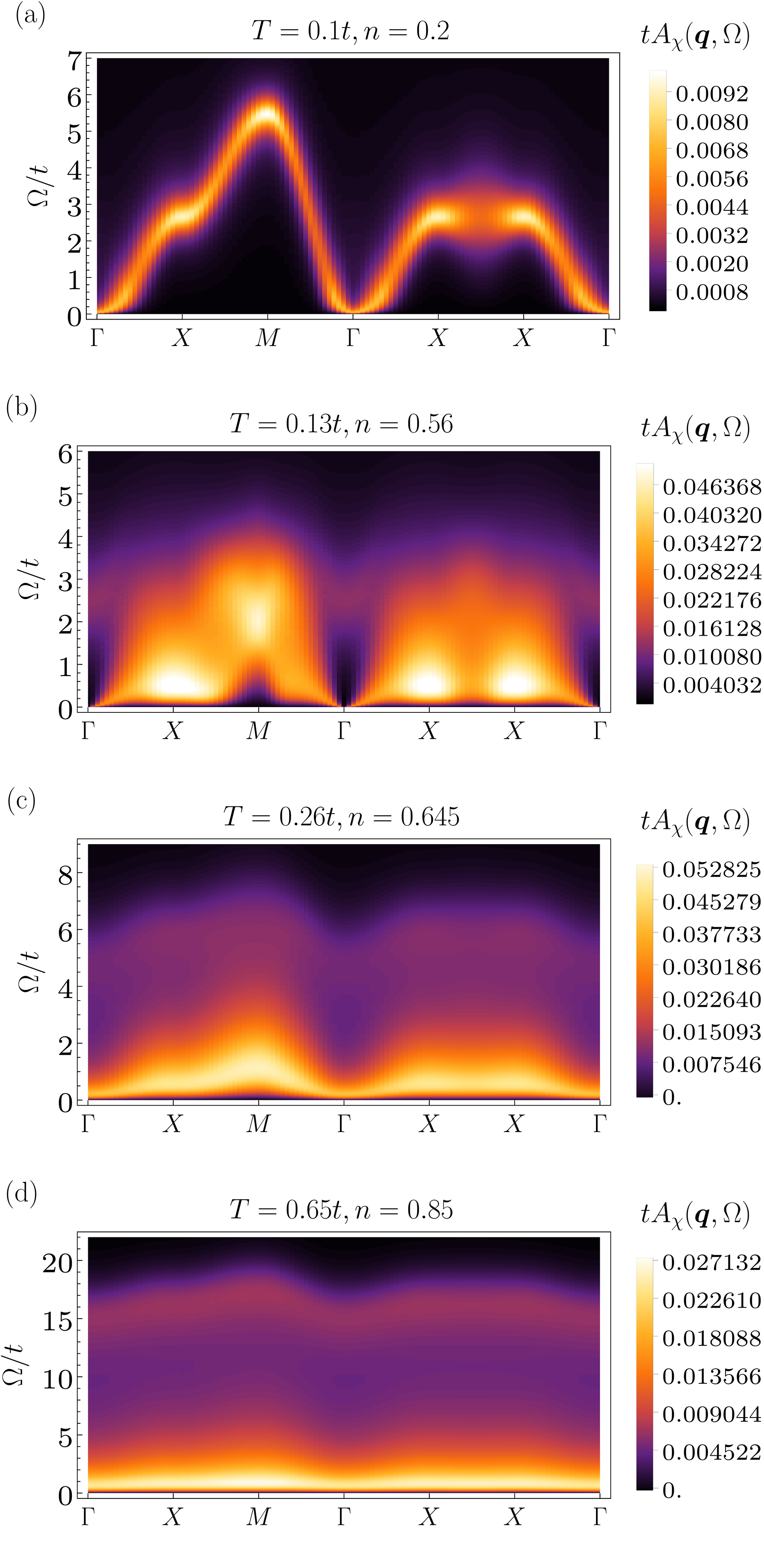}
\caption{Magnetic spectral function defined in Eq.~\eqref{eq:A_chi} in the four different density regimes corresponding to the
(a) paramagnetic,
(b) stripe antiferromagnetic,
(c) first ferromagnetic, and
(d) second ferromagnetic 
ground states;
compare the phase diagram in Fig.~\ref{fig:PhaseDiag}.}
\label{fig:SuzDyn}
\end{figure}

In terms of the holon operators,
the electronic spin operators read
\begin{subequations}
\begin{align}
s_i^+ & = ( s_i^- )^\dagger = h_{ i \uparrow }^\dagger h_{ i \downarrow } ,
\\
s_i^z & = \frac{ 1 }{ 2 } \sum_\sigma \sigma h_{ i \sigma }^\dagger h_{ i \sigma } , 
\end{align}
\end{subequations}
where 
$ s_i^\pm = s_i^x \pm i s_i^y $
are the usual spin raising and lowering operators.
For unbroken spin-rotation invariance,
we can then write the dynamic magnetic susceptibility as follows:
\begin{subequations}
\begin{align}
&
\chi_\Lambda ( Q )
= 
\sum_i \int_0^\beta \textrm{d} \tau e^{ - i \bm{q} \cdot \bm{R}_i + i \Omega \tau } 
\Braket{ \mathcal{T} s_i^z ( \tau ) s_j^z }
\\
&
= 
\sum_i \int_0^\beta \textrm{d} \tau e^{ - i \bm{q} \cdot \bm{R}_i + i \Omega \tau } 
\frac{ 1 }{ 2 }
\Braket{ \mathcal{T} s_i^+ ( \tau ) s_j^- }
\\
&
= 2
\sum_i \int_0^\beta \textrm{d} \tau e^{ - i \bm{q} \cdot \bm{R}_i + i \Omega \tau } 
\Braket{ \mathcal{T} 
h_{ i \sigma }^\dagger ( \tau ) h_{ i \sigma } ( \tau )
h_{ j \sigma }^\dagger h_{ j \sigma }
}
\\
&
=
- 2 \int_K G_\Lambda ( K + Q ) G_\Lambda ( K )
\nonumber\\
&
+ 2 \int_{ K_1 , K_2 } 
G_\Lambda ( K_1 + Q ) G_\Lambda ( K_1 ) G_\Lambda ( K_2 - Q ) G_\Lambda ( K_2 )
\nonumber\\
& \hspace{2cm} \times
U_\Lambda ( K_1 , K_2 ; K_1 + Q , K_2 - Q ) .
\end{align}
\end{subequations}
By expressing spin correlations 
in terms of holon correlation functions in the above,
we have to keep in mind that we use a symmetric equal-time regularization prescription
and apply the holon algebra \eqref{eq:holon_algebra} when appropriate. 
Next,
we insert the channel decomposition \eqref{eq:U_channels} for the 2-body interaction.
Keeping only the leading instability as in the corresponding channel flow equations \eqref{eq:MC_dot} then yields
\begin{align}
&
\chi_\Lambda ( Q )
\approx 
- 2 \int_K G_\Lambda ( K + Q ) G_\Lambda ( K )
\nonumber\\
&
+ \frac{ 2 }{ Z^2 } 
\int_{ K_1 , K_2 } G_\Lambda ( K_1 + Q ) G_\Lambda ( K_1 ) G_\Lambda ( K_2 - Q ) G_\Lambda ( K_2 )
\nonumber\\
& \hspace{1cm} \times
\left[
- G_0^{ - 1 } ( \omega_2 )
- G_0^{ - 1 } ( \omega_1 )
+ Z^2 \mathcal{M}_\Lambda ( Q ; \omega_2 , \omega_1 )
\right] .
\label{eq:chi_dyn}
\end{align}
This formula is used to determine the static magnetic susceptibilities in Sec.~\ref{sec:phase}.
However,
it also contains information about the dynamic spin response. 
To extract it,
we consider the associated magnetic spectral function
\begin{equation}
\label{eq:A_chi}
A_{ \chi } ( \bm{q} , \Omega ) 
= \frac{ 1 }{ \pi } \textrm{Im}
\big[
\left.
\chi_{ \Lambda = 1 } ( \bm{q} , \Omega )
\right|_{ i \Omega \to \Omega + i \eta } 
\big] .
\end{equation}
The required analytic continuation is again performed with Pad\'{e} approximants \cite{Beach2000}.
Note,
however,
that our solutions for the channels given in Eqs.~\eqref{eq:MC_sol} contain non-analytic 
$ \delta_{ \Omega , 0 } $ terms inherited from the atomic initial conditions \eqref{eq:SMC_0}.
These non-analytic terms do not contribute to the retarded spin propagator and should therefore be dropped before the numerical analytical continuation.
Our results are shown in Fig.~\ref{fig:SuzDyn}.
In the low-density paramagnetic state,
the dynamic spin response is dominated by the particle-hole bubble in the first line of Eq.~\eqref{eq:chi_dyn}.
This yields diffusive low-energy dynamics,
which develops into a single broad paramagnon band at higher energies;
see Fig.~\ref{fig:SuzDyn} (a).
For densities higher than $ n_{ c , 1 } $,
that is,
in the magnetically correlated phases,
this is no longer the case.
There,
the spin response is dominated by the magnetic channel $ \mathcal{ M }_\Lambda $.
In the intermediate stripe regime shown in Fig.~\ref{fig:SuzDyn} (b),
the paramagnon band has disappeared in favor of a diffusive continuum.
Moreover,
a large amount of spectral weight accumulates at very low energies around the stripe ordering wave vector 
$ \bm{q}_{ \textrm{stripe} } 
= X 
= ( \pi , 0 ) $.
Increasing the density further in the two ferromagnetic regimes displayed in Figs.~\ref{fig:SuzDyn} (c) and (d),
we observe only a diffusive low-energy continuum with extended high-frequency tails.
These consist of the Nagaoka polarons that also give rise to the hole continuum in the electronic spectra in Fig.~\ref{fig:tripleplotn085T065}.
The weak momentum dependence furthermore suggests that in the renormalized classical regime \cite{Chakravarty1989, Vilk1997, Sachdev2011, Schafer2020} above the ferromagnetic phase transition, 
the spin dynamics is almost completely local due to the dominance of long-range polaron physics.

\renewcommand{\theequation}{D\arabic{equation}}

\section{First Matsubara frequency approximation }

\label{app:firstmatsubara}

\begin{figure}%
\includegraphics[width=\linewidth]{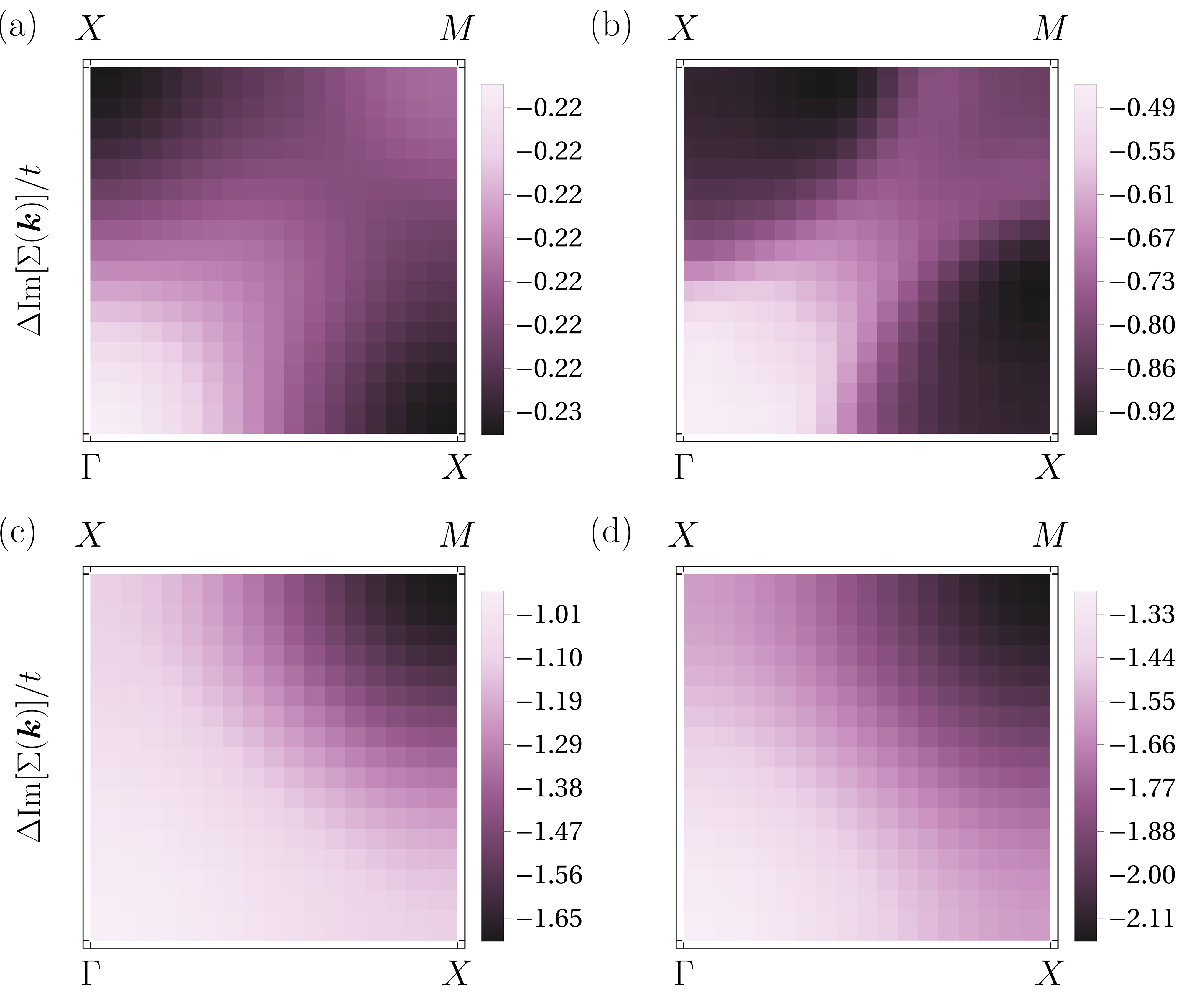}
\caption{Slope $\Delta \textrm{Im}\left[\Sigma(\bm{k})\right]$ of the imaginary part of the self-energy on the Matsubara axis. 
(a) $n = 0.2$, $T = 0.1t$; 
(b) $n = 0.56$, $T = 0.13t$; 
(c) $n = 0.75$, $T = 0.43t$; 
(d) $n = 0.85$, $T = 0.65t$.}
\label{fig:firstFsigma}
\end{figure}

\begin{figure}%
\includegraphics[width=\linewidth]{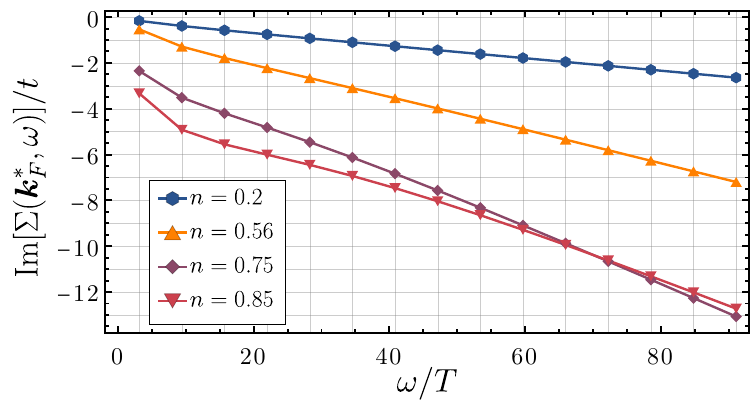}
\caption{Matsubara frequency dependence of the imaginary part of the self-energy for various densities. 
The self-energy is evaluated at the nodal Fermi momentum $\bm{k}^*_F$.}
\label{fig:SigmaSlope}
\end{figure}

Pad\'{e} analytic continuation is an ill-defined problem. 
While guidelines and requirements for the physical propagators exist, 
one can always find multiple valid Pad\'{e} approximants, 
leaving some ambiguity in the continuation procedure. 
At low enough temperatures,
it is however possible to approximate the zero-frequency retarded self-energy
$ \Sigma^R ( \bm{k} , \omega = 0 )
= \Sigma ( \bm{k} , \omega ) |_{ i \omega \to i \eta }  $
by the value of the Matsubara self-energy $ \Sigma ( \bm{k} , \pi T ) $   at the first Matsubara frequency $ \omega_0 = \pi T$ \cite{Kitatani2025,Simkovic2024, Lihm2025, Wu2017, Schafer2020}.
Effectively,
this amounts to setting $ \eta = \pi T $,
and circumvents the need for analytical continuation for obtaining 
$ \Sigma^R ( \bm{k} , \omega = 0 ) $ 
and hence also
$ A ( \bm{k} , \omega = 0 ) $
at low temperature.
In this Appendix, 
we present our results obtained in this way, 
as a validity check of our analytical continuation procedure used in the main text.
\begin{figure}%
\includegraphics[width=\linewidth]{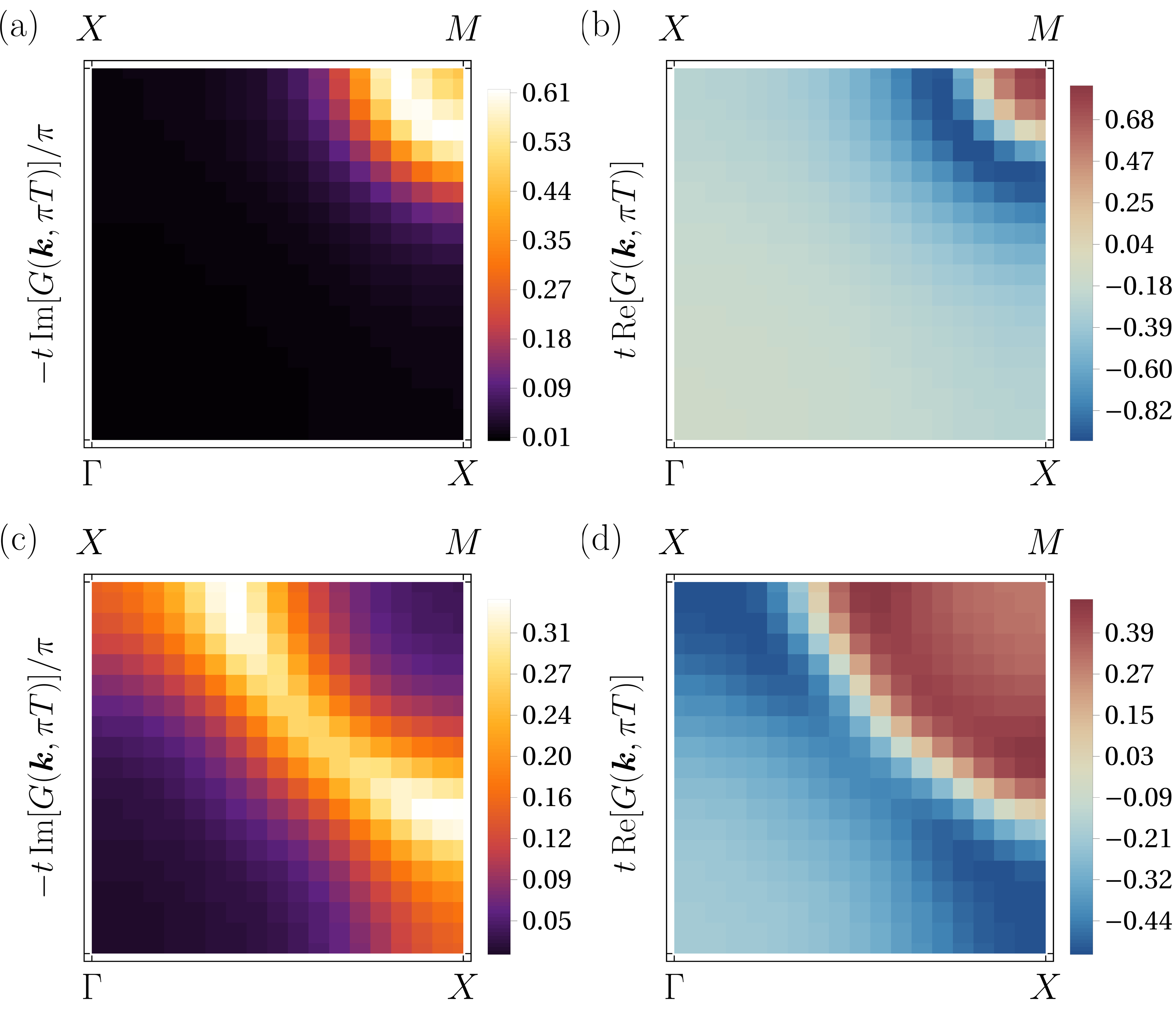}
\caption{(a) Imaginary and (b) real parts of the holon propagator evaluated at the first Matsubara frequency $ \omega_0 = \pi T$, 
in the paramagnetic regime at density $n = 0.2$ and temperature $T = 0.1t$.
(c) and (d) show the same within the stripe regime, for $n = 0.56$ and $T = 0.13t$.}
\label{fig:firstfrequency02}
\end{figure}%
Figures~\ref{fig:firstfrequency02} (a) and (b) show imaginary and real parts of the electronic Matsubara propagator at the first fermionic frequency,
respectively. 
We find excellent agreement with the Fermi surface shown in Fig.~\ref{fig:tripleplotn02T1} obtained by analytical continuation. Similarly, the real part crosses zero, as expected.
%
%
The same holds true for densities within the stripe regime; 
see Fig.~\ref{fig:firstfrequency02} (c) and (d). 
The Fermi surface found here matches again the one presented in Fig.~\ref{fig:tripleplotn056T13} (center). 
More intriguing is the clear difference in spectral weight, 
which is shifted towards the momenta at $(\pi ,\pi/2 )$. 
Such behavior is usually associated with a pseudogap, 
though in hole-doped systems a nodal arc is commonly expected \cite{Norman1998,Sherman2003,Simkovic2024,Shen2005,Kanigel2007,Yang2011,Keimer2015,Sachdev2025}. 
We also do not find this behavior within the analytically continued data, 
nor do we observe any additional behavior that can be associated with a pseudogap. 
As our Pad\'{e} approximation might wash out said behavior, 
we also investigate the slope of the self-energy 
$ \Delta \textrm{Im} \left[\Sigma(\bm{k})\right] 
= \textrm{Im} \left[\Sigma(\bm{k}, 3 \pi T)\right]
- \textrm{Im} \left[\Sigma(\bm{k}, \pi T)\right] $ 
in Fig.~\ref{fig:firstFsigma}. 
This quantity is known as a possible indicator of pseudogap behavior \cite{Simkovic2024, Lihm2025, Wu2017} 
and does not rely on analytical continuation. 
Both Figs.~\ref{fig:firstFsigma} and Fig.~\ref{fig:SigmaSlope} show only negatives slopes, 
which suggest that the varying spectral weight observed in Figs.~\ref{fig:firstfrequency02} (c) is an artifact of the high broadening $\eta = \pi Tt \approx 0.41t$ introduced by the first fermionic frequency approximation.
We also find that the magnitude of the slope of the imaginary part of the self-energy increases upon increasing the density. This is consistent with the decreasing quasi-particle weight obtained from analytical continuation.

\begin{figure}%
\includegraphics[width=\linewidth]{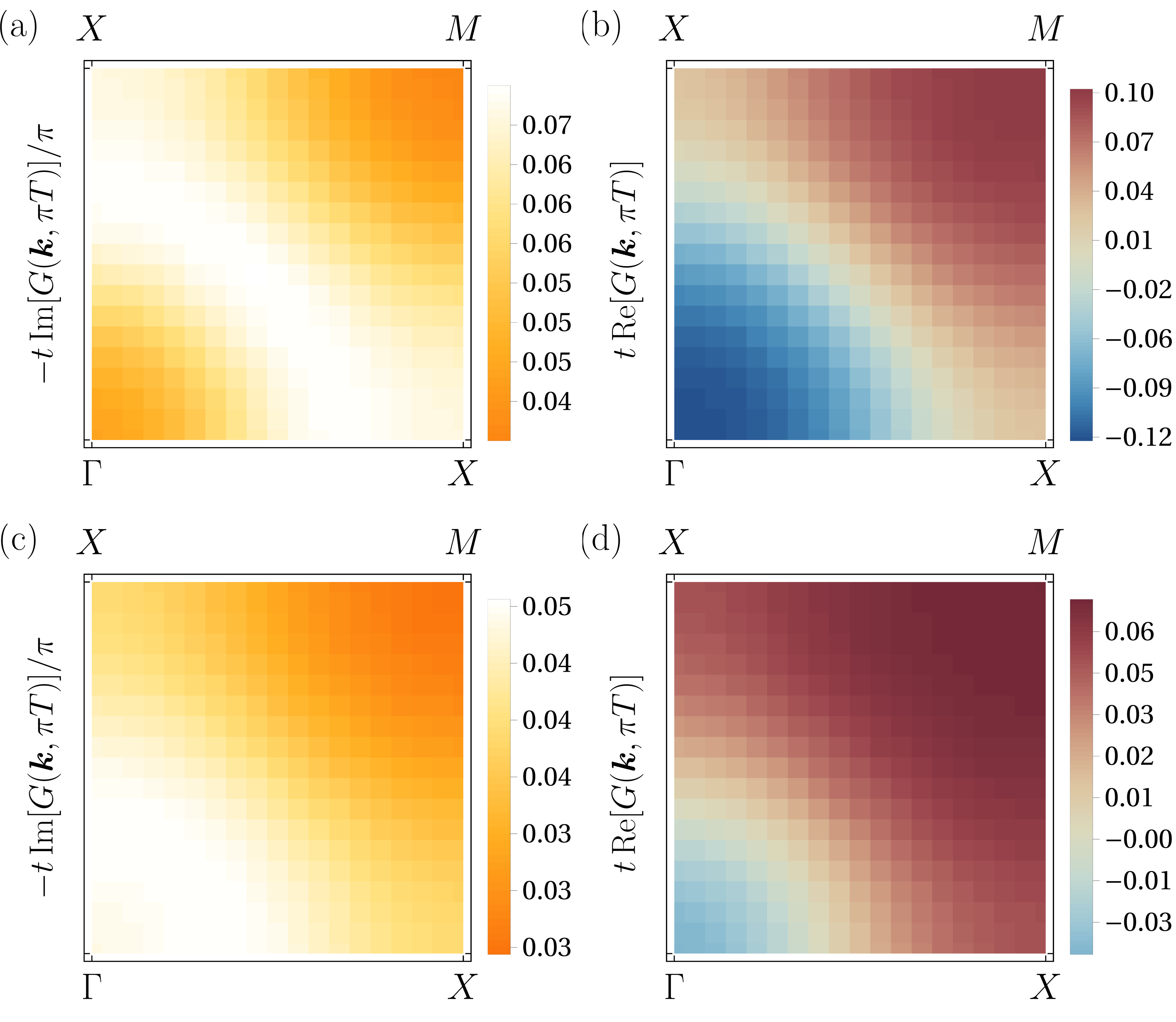}
\caption{(a) Imaginary and (b) real parts of the holon propagator evaluated at the first Matsubara frequency $ \omega_0 = \pi T$, 
in the first ferromagnetic regime at density $n = 0.75$ and $T = 0.53t$.
(c) and (d) show the same in the second ferromagnetic regime at density $n = 0.85$ and $T = 0.65t$.}
\label{fig:firstfrequency075}
\end{figure}%
%
%
%
%
%
%
%
%
%
%
%
%
%
At higher densities, 
the broadening $ \eta = \pi T $ becomes too disruptive because of the higher temperatures,
and we cannot obtain any meaningful results without analytical continuation. 
The Fermi surfaces in Figs.~\ref{fig:firstfrequency075} 
consequently deviate drastically from those in Fig.~\ref{fig:tripleplotn085T065}. 
In fact, 
the topology of the Fermi surface at $n = 0.75$ is not even reproduced. 
Fig.~\ref{fig:firstfrequency075} instead corresponds to a result obtained by the Pad\'{e} approximation at some higher temperature.

\renewcommand{\theequation}{E\arabic{equation}}

\section{Numerical errors of Matsubara summations} 
\label{app:Merror}
Our flow equations describe the evolution of the $t$ model, 
starting at infinite temperature and eventually ending up at some finite low temperature.  
Formally, 
the hierarchy of FRG flow equations in Fig.~\ref{fig:su2flow} is exact 
and is able to describe the entire finite-temperature disordered phase. 
However, 
in practice truncation and numerical implementation will limit the lowest temperature reachable. This is reflected in the phase diagram shown in Fig.~\ref{fig:PhaseDiag}.
At densities $n \geq n_{c1}$ we observe exponential growth in the magnetic susceptibility.
This naturally leads to a breakdown of the numerics and gives a lower boundary for the temperature we can reach. 
However, 
a similar behavior is also found at very small densities. 
We cannot attribute this to any divergence of a physical quantity.
Instead we find that the issue likely stems from the ill-convergence of Matsubara sums.
\begin{figure}%
\includegraphics[width=\linewidth]{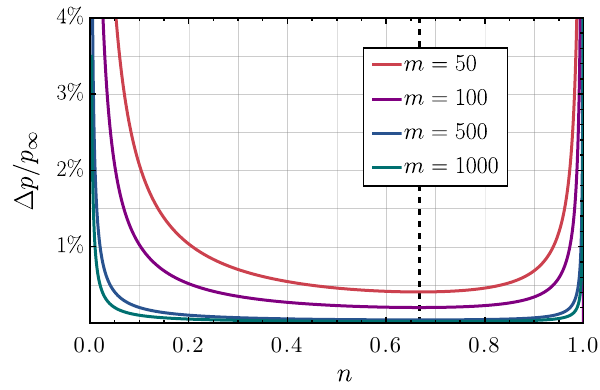}
\caption{Relative error $\Delta p / p_{\infty} = ( p_m - p_{\infty} ) / p_{\infty} $ of the Matsubara summation~\eqref{eq:erroreq} as function of density $ n $. 
The sum has been evaluated using $m = 50$, $100$, $500$ and $1000$ positive frequencies for $ p_m $,
and analytically for $ p_{\infty} $. 
The dashed black line marks $n = 2/3$ where $\mu_0 = 0$ and coincides with the smallest numerical error.}
\label{fig:errortotal}
\end{figure}%
To this end we show in Fig.~\ref{fig:errortotal} the error $\Delta p $ of numerical evaluation of
\begin{equation} \label{eq:erroreq}
p_m = \frac{1}{\beta}\sum_{\omega = -\omega_m}^{\omega_m} G_0^2(\omega) ,
\end{equation}
where $ G_0(\omega) = Z / ( i \omega + \mu_0 ) $ is the propagator in the atomic limit.
Fig.~\ref{fig:errortotal} reveals that the error depends heavily on the value of the bare chemical potential $\mu_0(n)$ given in Eq.~\eqref{eq:mu0}, 
which diverges logarithmically for both $n \rightarrow 0$ and $n \rightarrow 1$. 
Hence,
prohibitively many Matsubara frequencies are necessary to evaluate the sum with high accuracy both for very small and very large densities.
Nonetheless Eq.~(\ref{eq:erroreq}) is a convergent Matsubara sum. 
A finite bandwidth $ W $ inevitably exacerbates this problem at low temperatures.

Fortunately,
these considerations present only a worst-case scenario, 
which itself is \textit{not} present in the actual evaluation of the flow equations. 
Integrals involving only atomic propagators appear only in the initial conditions, 
which are evaluated analytically. 
Matsubara summations calculated during the flow are stabilized by hopping and self-energy corrections and the accompanying momentum integration.
Additionally, 
the flowing chemical potential decreases in magnitude for both low and high density upon lowering the temperature; see Fig.~\ref{fig:muofn} (b).
Still,
we expect remnants of this behavior to affect our Matsubara summations to some degree and the limit of very low and very high densities should be considered with this mind.

Another potential source of numerical error is the regularized Matsubara summation
\begin{equation}\label{eq:densityappendix}
	n ( \beta ) = \frac{2}{3}\left[2\int_K \cos(\omega 0^+) G_{ \Lambda = 1 } (K)+1\right].
\end{equation}
The counter term $\delta\mu_\Lambda$ introduced in Eq.~\eqref{eq:counter-term} should guarantee 
$ n ( 0 ) = n ( \beta ) $ 
via its flow equation \eqref{eq:counter-term}. 
\begin{figure}%
\includegraphics[width=\linewidth]{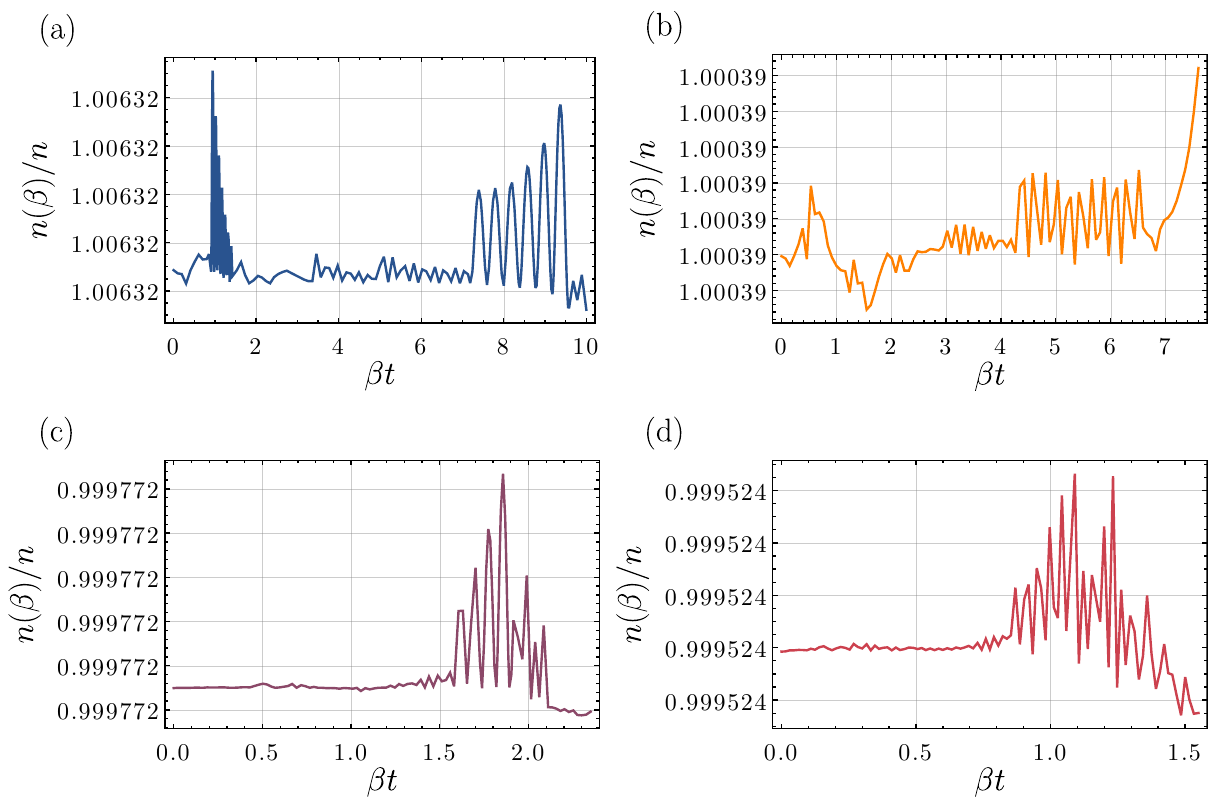}
\caption{Numerical evaluation of the integral~\eqref{eq:densityappendix} for the densities 
$(a)$ $n = 0.2$; 
$(b)$ $n = 0.56$; 
$(c)$ $n = 0.75$; 
$(d)$ $n = 0.85$. 
A constant offset $\mathcal{O}(10^{-3})$ is present at all densities. The breakdown of the flow is visible in form of a slope appearing at the final $\beta t$ values.}
\label{fig:countertermerror}
\end{figure}%
However, 
within our numerical evaluation an error of order $10^{-3}$ is present, as shown in Fig.~\ref{fig:countertermerror}. This is a not a statistical fluctuation, but a constant offset, which could accumulate during the flow.

Finally,
we consider the systematic error $ \delta $ of the normalization of the spectral function.
Since it is to good approximation $ \bm{k} $-independent,
we define it as 
\begin{equation}
\delta = Z -
\int_{ - \infty }^\infty \textrm{d} \omega A ( \bm{k} = \bm{0} , \omega ) . 
\end{equation} 
\begin{figure}%
\includegraphics[width=\linewidth]{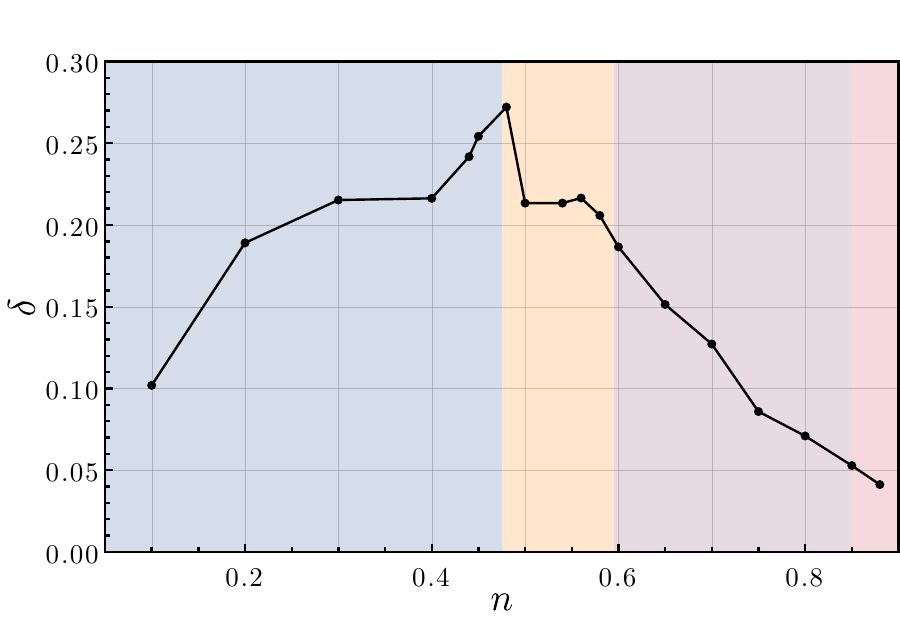}
\caption{Systematic error $ \delta $ of the normalization of the spectral function at the temperatures
where also the Luttinger volume $ n_L $ in Fig.~\ref{fig:LVviolation} was extracted.
Connecting lines are guides for the eye.
The background colors indicate the magnetic ground state in the color scheme of the phase diagram in Fig.~\ref{fig:PhaseDiag}.}
\label{fig:ZDeltaError}
\end{figure}%
Unfortunately, relative to density
this error can become rather large at low temperatures,
as seen in Fig.~\ref{fig:ZDeltaError}.
Therefore,
it is crucial to correct for it in the discussion of Luttinger's theorem in Sec.~\ref{sec:LTviolation}.

%
%
%
%
%
%
%
%
%
%
%
%
%
%
%
%
%
%


\begin{thebibliography}{99}
%
%
\bibitem{Kopietz2010}
P. Kopietz, L. Bartosch, and F. Sch\"{u}tz, \textit{Introduction to the Functional Renormalization Group} 
\href{https://doi.org/10.1007/978-3-642-05094-7}{(Springer, Berlin, 2010)}.
%
%
\bibitem{Metzner2012}
W. Metzner, M. Salmhofer, C. Honerkamp, V. Meden, and K. Sch\"{o}nhammer,
\textit{Functional renormalization group approach to correlated fermion systems},
\href{https://doi.org/10.1103/RevModPhys.84.299}
{Rev. Mod. Phys. \textbf{84}, 299 (2012)}.
%
%
\bibitem{Dupuis2021}
N. Dupuis, L. Canet, A. Eichhorn, W. Metzner, J. M. Pawlowski, M. Tissier, and N. Wschebor, 
\textit{The nonperturbative functional renormalization group and its applications}, 
\href{https://doi.org/10.1016/j.physrep.2021.01.001}
{Phys. Rep. \textbf{910}, 1 (2021)}.
%
%
%
%
\bibitem{Salmhofer2001}
M. Salmhofer and C. Honerkamp, 
\textit{Fermionic renormalization group flows},
\href{https://doi.org/10.1143/PTP.105.1}{Prog. Theor. Phys. \textbf{105}, 1 (2001)}.
%
%
\bibitem{Honerkamp2001}
C. Honerkamp and M. Salmhofer, 
\textit{Temperature-flow renormalization group  and the competition between superconductivity and ferromagnetism}, 
\href{https://doi.org/10.1103/PhysRevB.64.184516}{Phys. Rev. B \textbf{64}, 184516 (2001)}.
%
%
\bibitem{Kopietz2001}
P. Kopietz and T. Busche, 
\textit{Exact renormalization group flow equations for nonrelativistic fermions: Scaling toward the Fermi surface},
\href{https://doi.org/10.1103/PhysRevB.64.155101}{Phys. Rev. B \textbf{64}, 155101 (2001)}.
%
%
\bibitem{Husemann2009}
C. Husemann, and M. Salmhofer, 
{\it{Efficient parametrization of the vertex function, $\Omega$ scheme, and
the $t, t^\prime$ Hubbard model at van Hove filling}},
\href{https://doi.org/10.1103/PhysRevB.79.195125}{Phys. Rev. B {\bf{79}}, 195125 (2009)}.
%
%
\bibitem{Vilardi2017}
D. Vilardi, C. Taranto, and W. Metzner,
\textit{Nonseparable frequency dependence of the two-particle vertex in interacting fermion systems},
\href{https://doi.org/10.1103/PhysRevB.96.235110}{Phys. Rev. B \textbf{96}, 235110 (2017)}.
%
%
\bibitem{Vilardi2019}
D. Vilardi, C. Taranto, and W. Metzner, \textit{Antiferromagnetic and $d$-wave paring correlations in the strongly interacting two-dimensional Hubbard model from the functional renormalization group}, 
\href{https://doi.org/10.1103/PhysRevB.99.104501}{Phys. Rev. B \textbf{99}, 104501 (2019)}.
%
%
\bibitem{Halboth2000}
C. J. Halboth and W. Metzner, 
\textit{d-Wave Superconductivity and Pomeranchuk Instability in the Two-Dimensional Hubbard Model},
\href{https://doi.org/10.1103/PhysRevLett.85.5162}{Phys. Rev. Lett. \textbf{85}, 5162 (2000)}.
%
%
\bibitem{Halboth2000b}
C. J. Halboth and W. Metzner, \textit{Renormalization group analysis of the two-dimensional Hubbard model}, 
\href{https://doi.org/10.1103/PhysRevB.61.7364}{Phys. Rev. B \textbf{61}, 7364 (2000)}.
%
%
\bibitem{Salmhofer2004}
M. Salmhofer, C. Honerkamp, W. Metzner, and O. Lauscher,
\textit{Renormalization Group Flows into Phases with Broken Symmetry},
\href{https://doi.org/10.1143/PTP.112.943}{Prog. Theor. Phys. \textbf{112}, 943 (2004)}.
%
%
\bibitem{Ossadnik2008}
M. Ossadnik, C. Honerkamp, T. M. Rice, and M. Sigrist,
\textit{Breakdown of Landau Theory in Overdoped Cuprates near the Onset of Superconductivity},
\href{https://doi.org/10.1103/PhysRevLett.101.256405}{Phys. Rev. Lett. \textbf{101}, 256405 (2008)}.
%
%
\bibitem{Husemann2012}
C. Husemann, K.-U. Giering, and M. Salmhofer, 
\textit{Frequency-dependent vertex functions of the $(t, t^\prime )$ Hubbard model at weak coupling},
\href{https://doi.org/10.1103/PhysRevB.85.075121}{Phys. Rev. B \textbf{85}, 075121 (2012)}.
%
%
\bibitem{Taranto2014}
C. Taranto, S, Andergassen, J. Bauer, K. Held, A. Katanin, W. Metzner, G. Rohringer, and A. Toschi,
\textit{From Infinite to Two Dimensions through the Functional Renormalization Group}, 
\href{https://doi.org/10.1103/PhysRevLett.112.196402}{Phys. Rev. Lett. \textbf{112}, 196402 (2014)}.
%
%
\bibitem{Lichtenstein2017}
J. Lichtenstein, D. S. de la Pe\~na, D. Rohe, E. D. Napoli, C. Honerkamp, and S. A. Maier,
\textit{High-performance functional Renormalizaton group calculations for interacting fermions},
\href{https://doi.org/10.1016/j.cpc.2016.12.013}{Computer Physics Communications \textbf{213}, 100 (2017)}.
%
%
\bibitem{Honerkamp2018}
C. Honerkamp, 
\textit{Efficient vertex parametrization for the constrained functional renormalization group for effective low-energy interactions in multiband systems},
\href{https://doi.org/10.1103/PhysRevB.98.155132}{Phys. Rev. B \textbf{98}, 155132 (2018)}.
%
%
\bibitem{Ehrlich2020}
J. Ehrlich and C. Honerkamp,
\textit{Functional renormalization group for fermion lattice models in three dimensions: Application to the Hubbard model on the cubic lattice}, 
\href{https://doi.org/10.1103/PhysRevB.102.195108}{Phys. Rev. B \textbf{102}, 195108 (2020)}.
%
%
\bibitem{Hille2020}
C. Hille, F. B. Kugler, C. J. Eckhardt, Y.-Y. He, A. Kauch, C. Honerkamp, A. Toschi, and S. Andergassen, 
\textit{Quantitative functional renormalization grouo description of the two-dimensional Hubbard model}, 
\href{https://doi.org/10.1103/PhysRevResearch.2.033372}{Phys. Rev. Res. \textbf{2}, 033372 (2020)}.
%
%
\bibitem{Honerkamp2022}
C. Honerkamp, D. M. Kennes, V. Meden, M. M. Scherer, and R. Thomale,
\textit{Recent developments in the functional renormalization group approach to correlated electron systems},
\href{https://doi.org/10.1140/epjb/s10051-022-00463-1}{Eur. Phys. J. B \textbf{95}, 205 (2022)}.
%
%
\bibitem{Rueckriegel2023}
A. R\"{u}ckriegel, J. Arnold, R. Krämer, and P. Kopietz,
\textit{Functional renormalization group without functional integrals: Implementing Hilbert space projections for strongly correlated electrons via Hubbard X-operators},
\href{https://doi.org/10.1103/PhysRevB.108.115104}{Phys. Rev. B \textbf{108}, 115104 (2023)}.
%
%
\bibitem{Auerbach1994}
A. Auerbach,
\textit{Interacting Electrons and Quantum Magnetism},
\href{https://doi.org/10.1007/978-1-4612-0869-3}{(Springer, Berlin, 1994)}.
%
%
\bibitem{Ovchinnikov2004}
S. G. Ovchinnikov and V. V. Val'kov, 
\textit{Hubbard Operators in the Theory of Strongly Correlated Electrons}, 
\href{https://doi.org/10.1142/p314}{(Imperial College Press, London, 2004)}.
%
%
%
%
\bibitem{Izyumov1988}
Y. A. Izyumov and Y. N. Skryabin, 
\textit{Statistical Mechanics of Magnetically Ordered Systems}, 
\href{https://doi.org/10.1063/1.2810417}{(Springer, Berlin, 1988)}.
%
%
%
%
\bibitem{Fulde1995}
P. Fulde, 
\textit{Electron Correlations in Molecules and Solids},
\href{https://doi.org/10.1007/978-3-642-57809-0}{(Springer, Berlin,  Third Enlarged Edition, 1995)}.
%
%
\bibitem{Hubbard1963}
J. Hubbard,
\textit{Electron correlations in narrow energy bands},
\href{https://doi.org/10.1098/rspa.1963.0204}{Proc. Roy. Soc. (London) A \textbf{276}, 238 (1963)}.
%
%
\bibitem{Fazekas1999}
P. Fazekas, {\it{Lecture Notes on Electron Correlation and Magnetism}},
\href{https://doi.org/10.1142/2945}{(World Scientific,  Singapore, 1999)}.
%
%
%
%
\bibitem{Virosztek1990}
A. Virosztek and J. Ruvalds,
\textit{Nested-Fermi-liquid theory},
\href{https://doi.org/10.1103/PhysRevB.42.4064}
{Phys. Rev. B \textbf{42}, 4064 (1990)}.
%
%
\bibitem{Schaefer2015}
T. Sch\"{a}fer, F. Geles, D. Rost, G. Rohringer, E. Arrigoni, K. Held, N. Bl\"{u}mer, M. Aichhorn, and A. Toschi,
\textit{Fate of the false Mott-Hubbard transition in two dimensions},
\href{https://doi.org/10.1103/PhysRevB.91.125109}
{Phys. Rev. B \textbf{91}, 125109 (2015)}.
%
%
\bibitem{Rohringer2016}
G. Rohringer and A. Toschi,
\textit{Impact of nonlocal correlations over different energy scales: A dynamical vertex approximation study},
\href{https://doi.org/10.1103/PhysRevB.94.125144}
{Phys. Rev. B \textbf{94}, 125144 (2016)}.
%
%
\bibitem{Simkovic2020}
F. \v{S}imkovic  IV, J. P. F. LeBlanc, Aaram J. Kim, Youjin Deng, N. V. Prokof'ev, B. V. Svistunov, and E. Kozik,
\textit{Extended Crossover from a Fermi Liquid to a Quasiantiferromagnet in the Half-Filled 2D Hubbard Model},
\href{https://doi.org/10.1103/PhysRevLett.124.017003}
{Phys. Rev. Lett. \textbf{124}, 017003 (2020)}.
%
%
\bibitem{Kim2020}
A. J. Kim, F. \v{S}imkovic  IV, and E. Kozik,
\textit{Spin and Charge Correlations across the Metal-to-Insulator Crossover in the Half-Filled 2D Hubbard Model},
\href{https://doi.org/10.1103/PhysRevLett.124.117602}
{Phys. Rev. Lett. \textbf{124}, 117602 (2020)}.
%
%
%
%
\bibitem{Shankar1994}
R. Shankar,
\textit{Renormalization-group approach to interacting fermions},
\href{https://doi.org/10.1103/RevModPhys.66.129}{Rev. Mod. Phys. \textbf{66}, 129 (1994)}.
%
%
\bibitem{Svistunov2010}
K. Van Houcke, E. Kozik, N. Prokof'ev, and B. Svistunov,
\textit{Diagrammatic Monte Carlo},
\href{https://doi.org/10.1016/j.phpro.2010.09.034}{Phys. Procedia \textbf{6}, 95 (2010)}.
%
%
\bibitem{Wetterich1993}
C. Wetterich, 
\textit{Exact evolution equation for the effective potential}, 
\href{https://doi.org/10.1016/0370-2693(93)90726-X}
{Phys. Lett. B \textbf{301}, 90 (1993)}.
%
%
\bibitem{Berges2002}
J. Berges, N. Tetradis, and C. Wetterich,
\textit{Non-perturbative renormalization flow in quantum field theory and statistical physics},
\href{https://doi.org/10.1016/S0370-1573(01)00098-9}{Phys. Rep. \textbf{363}, 223 (2002)}.
%
%
\bibitem{Pawlowski2007}
J. M. Pawlowski, 
\textit{Aspects of the functional renormalisation group},
\href{https://doi.org/10.1016/j.aop.2007.01.007}{Ann. Phys. \textbf{322}, 2831 (2007)}.
%
%
\bibitem{Shastry2010}
B. S. Shastry,
\textit{Extremely correlated quantum liquids},
\href{https://doi.org/10.1103/PhysRevB.81.045121}{Phys. Rev. B \textbf{81}, 045121 (2010)}.
%
%
\bibitem{Chiappe1993}
G. Chiappe, E. Louis, J. Gal\'{a}n, F. Guinea, and J. A. Verg\'{e}s,
\textit{Ground-state properties of the U=\ensuremath{\infty} Hubbard model on a $4\times4$ cluster},
\href{https://doi.org/10.1103/PhysRevB.48.16539}{Phys. Rev. B \textbf{48}, 16539 (1993)}.
%
%
\bibitem{footnote_holon_and_Xoperators}
Both in the literature (see, for example, Ref.~[\onlinecite{Ovchinnikov2004}]) 
and in our previous work \cite{Rueckriegel2023}, 
the language of Hubbard X-operators
$ X_i^{ a b } = \ket{ i , a } \bra{ i , b } $
is often used to formulate the strong-coupling problem.
Here, 
$ a, b \in \{ 0 , \uparrow , \downarrow \} $ label the three states of the atomic Hilbert space. 
As we only require a small portion of the set of
X-operators we introduce the more concise notation of holon operators, 
$h_{i\sigma} = X_i^{0\sigma}$, 
$h_{i\sigma}^{\dagger} = X_i^{\sigma0}$ and 
$ \tilde{n}_{i\sigma} = X_i^{\sigma \sigma}$.
%
%
\bibitem{Rice1988}
F. C. Zhang and T. M. Rice,
\textit{Effective Hamiltonian for the superconducting Cu oxides},
\href{https://doi.org/10.1103/PhysRevB.37.3759}{Phys. Rev. B \textbf{37}, 3759(R) (1988)}.
%
%
%
%
\bibitem{Dessau1993}
D. S. Dessau, Z.-X. Shen, D. M. King, D. S. Marshall, L. W. Lombardo, P. H. Dickinson,
A. G. Loeser, J. DiCarlo, C.-H. Park, A. Kapitulnik, and W. E. Spicer,
\textit{Key features in the measured band structure of ${\mathrm{Bi}}_{2}{\mathrm{Sr}}_{2}{\mathrm{CaCu}}_{2}{\mathrm{O}}_{8+\mathrm{\ensuremath{\delta}}}$: Flat bands at ${\mathit{E}}_{\mathit{F}}$ and Fermi surface nesting},
\href{https://doi.org/10.1103/PhysRevLett.71.2781}{Phys. Rev. Lett. \textbf{71}, 2781 (1993)}.
%
%
\bibitem{Norman1998}
M. R. Norman, H. Ding, M. Randeria, J. C. Campuzano, T. Yokoya, T. Takeuchi, T. Takahashi, T. Mochiku, K. Kadowaki, P. Guptasarma, and D. G. Hinks, 
\textit{Destruction of the Fermi surface in underdoped high-$T_c$ superconductors},
\href{https://doi.org/10.1038/32366}{Nature \textbf{392}, 157 (1998)}.
%
%
\bibitem{Shen2005}
K. M. Shen, F. Ronning, D. H. Lu, F. Baumberger, N. J. C. Ingle, W. S. Lee, W. Meevasana, Y. Kohsaka, M. Azuma, M. Takano, H. Takagi, and Z.-X. Shen, 
\textit{Nodal Quasiparticles and Antinodal Charge Ordering in ${\mathrm{Ca}}_{2-x}\mathrm{Na}_{x}\mathrm{CuO}_{2}\mathrm{Cl}_2$},
\href{https://www.science.org/doi/10.1126/science.1103627}{Science \textbf{307}, 901 (2005)}.
%
%
\bibitem{Kanigel2007}
A. Kanigel, U. Chatterjee, M. Randeria, M. R. Norman, S. Souma, M. Shi1, Z. Z. Li, H. Raffy, and J. C. Campuzano,
\textit{Protected Nodes and the Collapse of Fermi Arcs in High-${T}_{c}$ Cuprate Superconductors},
\href{https://doi.org/10.1103/PhysRevLett.99.157001}{Phys. Rev. Lett. \textbf{99}, 157001 (2007)}.
%
%
%
%
%
\bibitem{Anderson2008}
P. W. Anderson,
\textit{Hidden Fermi liquid: The secret of high-$ T_c $ cuprates},
\href{https://doi.org/10.1103/PhysRevB.78.174505}{Phys. Rev. B \textbf{78}, 174505 (2008)}.
%
%
\bibitem{Anderson2009}
P. W. Anderson and P. A. Casey,
\textit{Transport anomalies of the strange metal: Resolution by hidden Fermi liquid theory},
\href{https://doi.org/10.1103/PhysRevB.80.094508}{Phys. Rev. B \textbf{80}, 094508 (2009)}.
%
%
\bibitem{Casey2011}
P. A. Casey and P. W. Anderson,
\textit{Hidden Fermi Liquid: Self-Consistent Theory for the Normal State of High-$ T_c $ Superconductors},
\href{https://doi.org/10.1103/PhysRevLett.106.097002}{Phys. Rev. Lett. \textbf{106}, 097002 (2011)}.
%
%
\bibitem{Yang2011}
H.-B. Yang, J. D. Rameau, Z.-H. Pan, G. D. Gu, P. D. Johnson, H. Claus, D. G. Hinks, and T. E. Kidd,
\textit{Reconstructed Fermi Surface of Underdoped ${\mathrm{Bi}}_{2}{\mathrm{Sr}}_{2}{\mathrm{CaCu}}_{2}{\mathrm{O}}_{8+\ensuremath{\delta}}$ Cuprate Superconductors},
\href{https://doi.org/10.1103/PhysRevLett.107.047003}{Phys. Rev. Lett. \textbf{107}, 047003 (2011)}.
%
%
\bibitem{Keimer2015}
B. Keimer, S. A. Kivelson, M. R. Norman, S. Uchida, and J. Zaanen,
\textit{From quantum matter to high-temperature superconductivity in copper oxides},
\href{https://doi.org/10.1038/nature14165}{Nature \textbf{518}, 179 (2015)}.
%
%
\bibitem{Sachdev2025}
S. Sachdev,
\textit{The foot, the fan, and the cuprate phase diagram: Fermi-volume-changing quantum phase transitions},
\href{https://doi.org/10.1016/j.physc.2025.1354707}{Physica C \textbf{633}, 1354707 (2025)}.
%
%
\bibitem{Sherman2003}
A. Sherman and M. Schreiber,
\textit{Two-dimensional t-J model at moderate doping},
\href{https://doi.org/10.1140/epjb/e2003-00090-x}{Eur. Phys. J. B \textbf{32}, 203 (2003)}.
%
\bibitem{Simkovic2024}
F. \v{S}imkovic IV, R. Rossi, A. Georges, and M. Ferrero,
\textit{Origin and fate of the pseudogap in the doped Hubbard model},
\href{https://doi.org/10.1126/science.ade9194}{Science \textbf{385}, eade9194 (2024)}.
%
%
\bibitem{Nagaoka1965}
Y. Nagaoka, {\it{Ground state of correlated electrons in a narrow almost half-filled s-band}},
\href{https://doi.org/10.1016/0038-1098(65)90266-8}{Solid State Commun. {\bf{3}}, 409 (1965)}.
%
%
\bibitem{Nagaoka1966}
Y. Nagaoka, 
{\it{Ferromagnetism in a Narrow, Almost Half-Filld s Band}},
\href{https://doi.org/10.1103/PhysRev.147.392}{Phys. Rev. {\bf{147}}, 392 (1966)}.
%
%
\bibitem{Kollar1996}
M. Kollar, R. Strack, and D. Vollhardt,
{\it{Ferromagnetism in correlated electron systems: Generalization of Nagaoka’s theorem}}, 
\href{https://doi.org/10.1103/PhysRevB.53.9225}{Phys. Rev. B {\bf{53}}, 9225 (1996)}.
%
%
\bibitem{Tasaki1998}
H. Tasaki, {\it{From Nagaoka's Ferromagnetism to Flat-Band Ferromagnetism and Beyond}}, 
\href{https://doi.org/10.1143/PTP.99.489}{Prog. Theor. Phys. {\bf{99}}, 489 (1998)}.
%
%
%
%
\bibitem{Roth1967}
L. M. Roth,
\textit{Spin wave stability of the ferromagnetic state for a narrow s-band},
\href{https://doi.org/10.1016/0022-3697(67)90285-5}{Journal of Physics and Chemistry of Solids, \textbf{28}, 1549 (1967)}.
%
%
\bibitem{Brinkman1970}
W. F. Brinkman and T. M. Rice,
\textit{Single-Particle Excitations in Magnetic Insulators},
\href{https://doi.org/10.1103/PhysRevB.2.1324}{Phys. Rev. B \textbf{2}, 1324 (1970)}.
%
%
\bibitem{Shastry1990}
B. S. Shastry, H. R. Krishnamurthy, and P. W. Anderson,
\textit{Instability of the Nagaoka ferromagnetic state of the $ U = \infty $ Hubbard model},
\href{https://doi.org/10.1103/PhysRevB.41.2375}{Phys. Rev. B \textbf{41}, 2375 (1990)}. 
%
%
\bibitem{Izyumov1990}
Yu. A. Izyumov and B. M. Letfulov,
\textit{A diagram technique for Hubbard operators: the magnetic phase diagram in the (t-J) model},
\href{https://doi.org/10.1088/0953-8984/2/45/005}{J. Phys.: Condens. Matter \textbf{2} 8905 (1990)}.
%
%
\bibitem{Elser1990}
A. G. Basile and V. Elser,
\textit{Stability of the ferromagnetic state with respect to a single spin flip: Variational calculations for the U=\ensuremath{\infty} Hubbard model on the square lattice},
\href{https://doi.org/10.1103/PhysRevB.41.4842}{Phys. Rev. B \textbf{41}, 4842(R) (1990)}.
%
%
\bibitem{Kotrla1990}
M. Kotrla and V. Drchal, 
\textit{Mean-Field Solution of Strongly Correlated Systems Using Hubbard Atomic Operators},
\href{https://onlinelibrary.wiley.com/doi/pdf/10.1002/pssb.2221570215}{phys. stat. sol. (b) \textbf{167}, 635 (1990)}.
%
%
\bibitem{Aruiac1990}
J. C. Anglès d'Auriac, B. Doucot, and R. Rammal,
\textit{Infinite-U Hubbard model in the large-spin regime: exact diagonalization study},
\href{https://doi.org/10.1088/0953-8984/3/22/009}{J. Phys.: Condens. Matter \textbf{3}, 3973 (1991)}.
%
%
\bibitem{vonderLinden1991}
W. von der Linden and D. M. Edwards,
\textit{Ferromagnetism in the Hubbard model},
\href{https://doi.org/10.1088/0953-8984/3/26/014}{J. Phys.: Condens. Matter \textbf{3}, 4917 (1991)}.
%
%
\bibitem{Putikka1992}
W. O. Putikka, M. U. Luchini, and M. Ogata,
\textit{Ferromagnetism in the two-dimensional $ t $-$ J $ model},
\href{https://doi.org/10.1103/PhysRevLett.69.2288}{Phys. Rev. Lett. \textbf{69}, 2288 (1992)}.
%
%
\bibitem{Hanisch1993}
Th. Hanisch and E. M\"{u}ller-Hartmann,
\textit{Ferromagnetism in the Hubbard model: instability of the Nagaoka state on the square lattice},
\href{https://doi.org/10.1002/andp.19935050407}{Ann. Physik \textbf{2}, 381 (1993)}.
%
%
\bibitem{Zotos1993}
M. W. Long and X. Zotos,
\textit{Hole-hole correlations in the $ U = \infty $ limit of the Hubbard model and the stability of the Nagaoka state},
\href{https://doi.org/10.1103/PhysRevB.48.317}{Phys. Rev. B \textbf{48}, 317 (1993)}.
%
%
\bibitem{Wurth1995}
P. Wurth and E. Müller-Hartmann,
\textit{Ferromagnetism in the Hubbard model: Spin waves and instability of the Nagaoka state},
\href{https://doi.org/10.1002/andp.19955070206}{ Ann. Phys. \textbf{507}, 144 (1995)}.
%
%
\bibitem{Obermeier1997}
T. Obermeier, T. Pruschke, and J. Keller,
\textit{Ferromagnetism in the large-$U$ Hubbard model},
\href{https://doi.org/10.1103/PhysRevB.56.R8479}{Phys. Rev. B \textbf{56}, R8479(R) (1997)}.
%
%
\bibitem{Kuzmin1997}
E. V. Kuz’min,
\textit{The ground state problem in the infinite-U Hubbard model},
\href{https://doi.org/10.1134/1.1130126}{Phys. Solid State \textbf{39}, 169–178 (1997)}.
%
%
\bibitem{Becca2001}
F. Becca and S. Sorella,
\textit{Nagaoka Ferromagnetism in the Two-Dimensional Infinite-$ U $ Hubbard Model},
\href{https://doi.org/10.1103/PhysRevLett.86.3396}{Phys. Rev. Lett. \textbf{86}, 3396 (2001)}.
%
%
\bibitem{Zitzler2002}
R. Zitzler, Th. Pruschke, and R. Bulla,
\textit{Magnetism and phase separation in the ground state of the Hubbard model},
\href{https://doi.org/10.1140/epjb/e2002-00180-3}{Eur. Phys. J. B \textbf{27}, 473-481 (2002)}.
%
%
\bibitem{Coleman2002}
P. Coleman and C. Pépin,
\textit{Supersymmetric approach to the infinite U Hubbard model},
\href{https://doi.org/10.1016/S0921-4526%2801%2901501-0}{Physica B: Cond. Mat. \textbf{312}, 539 (2002)}.	
%
%
\bibitem{Park2008}
H. Park, K. Haule, C. A. Marianetti, and G. Kotliar,
\textit{Dynamical mean-field theory study of Nagaoka ferromagnetism},
\href{https://doi.org/10.1103/PhysRevB.77.035107}{Phys. Rev. B \textbf{77}, 035107 (2008)}.
%
%
%
\bibitem{Kumar2008}
B. Kumar,
\textit{Canonical representation for electrons and its application to the Hubbard model},
\href{https://doi.org/10.1103/PhysRevB.77.205115}
{Phys. Rev. B \textbf{77}, 205115 (2008)}.
%
%
%
%
\bibitem{Baroni2011}
G. Carleo, S. Moroni, F. Becca, and S. Baroni,
\textit{Itinerant ferromagnetic phase of the Hubbard model},
\href{https://doi.org/10.1103/PhysRevB.83.060411}{Phys. Rev. B \textbf{83}, 060411(R) (2011)}.
%
%
\bibitem{Liu2012}
L. Liu, H. Yao, E. Berg, S. R. White, and S. A. Kivelson,
\textit{Phases of the Infinite $U$ Hubbard Model on Square Lattices},
\href{https://doi.org/10.1103/PhysRevLett.108.126406}{Phys. Rev. Lett. \textbf{108}, 126406 (2012)}.
%
%
\bibitem{Maska2012}
M. M. Ma\'{s}ka, 
M. Mierzejewski,
E. A. Kochetov,
L. Vidmar,
J. Bon\v{c}a, and
O. P. Sushkov,
\textit{Effective approach to the Nagaoka regime of the two-dimensional $ t $-$ J $ model},
\href{https://doi.org/10.1103/PhysRevB.85.245113}{Phys. Rev. B \textbf{85}, 245113 (2012)}.
%
%
\bibitem{Kochetov2017}
I. Ivantsov, A. Ferraz, and E. Kochetov,
\textit{Breakdown of the Nagaoka phase at finite doping},
\href{https://doi.org/10.1103/PhysRevB.95.155115}{Phys. Rev. B \textbf{95}, 155115 (2017)}.
%
%
\bibitem{Blesio2019}
G. G. Blesio, M. G. Gonzalez, and F. T. Lisandrini,
\textit{Magnetic phase diagram of the infinite-$U$ Hubbard model with nearest- and next-nearest-neighbor hoppings},
\href{https://doi.org/10.1103/PhysRevB.99.174411}{Phys. Rev. B \textbf{99}, 174411 (2019)}.
%
%
\bibitem{Morera2023}
I. Morera, M. Kanász-Nagy, T. Smolenski, L. Ciorciaro, A. Imamoğlu, and E. Demler,
\textit{High-temperature kinetic magnetism in triangular lattices},
\href{https://doi.org/10.1103/PhysRevResearch.5.L022048}{Phys. Rev. Research \textbf{5}, L022048 (2023)}.
%
%
\bibitem{Samajdar2024a}
R. Samajdar and R. N. Bhatt,
\textit{Polaronic mechanism of Nagaoka ferromagnetism in Hubbard models},
\href{https://doi.org/10.1103/PhysRevB.109.235128}{Phys. Rev. B \textbf{109}, 235128 (2024)}.
%
%
\bibitem{Newby2025}
R. C. Newby and E. Khatami,
\textit{Finite-temperature kinetic ferromagnetism in the square-lattice Hubbard model},
\href{https://doi.org/10.1103/PhysRevB.111.245120}{Phys. Rev. B \textbf{111}, 245120 (2025)}.
%
%
\bibitem{Sharma2025}
P. Sharma, Y. Peng, D. N. Sheng, H. J. Changlani, and Y. Wang,
\textit{Instability of Nagaoka State and Quantum Phase Transition via Kinetic Frustration Control},
\href{https://doi.org/10.48550/arXiv.2508.08410}{arXiv:2508.08410 [cond-mat.str-el]}.
%
%
\bibitem{Lieb1989}
E. H. Lieb,
\textit{Two theorems on the Hubbard model},
\href{https://doi.org/10.1103/PhysRevLett.62.1201}{Phys. Rev. Lett. \textbf{62}, 1201 (1989)};
Erratum 
\href{https://doi.org/10.1103/PhysRevLett.62.1927.5}{Phys. Rev. Lett. \textbf{62}, 1927 (1989)}.
%
%
%
%
\bibitem{Cheuk2016}
L. W. Cheuk, M. A. Nichols, K. R. Lawrence, M. Okan, H. Zhang, E. Khatami, N. Trivedi, T. Paiva, M. Rigol, and M. W. Zwierlein,
\textit{Observation of Spatial Charge and Spin Correlations in the 2D Fermi-Hubbard Model},
\href{https://doi.org/10.1126/science.aag3349}{Science \textbf{353}, 1260 (2016)}.
%
%
\bibitem{Dehollain2019}
J. P. Dehollain, U. Mukhopadhyay, V. P. Michal, Y. Wang, B. Wunsch, C. Reichl, W. Wegscheider, M. S. Rudner, E. Demler, and L. M. K. Vandersypen,
\textit{Nagaoka ferromagnetism observed in a quantum dot plaquette},
\href{https://doi.org/10.1038/s41586-020-2051-0}{Nature \textbf{579}, 528 (2020)}.
%
%
\bibitem{Bohrdt2021}
A. Bohrdt, L. Homeier, C. Reinmoser, E. Demler, and F. Grusdt,
\textit{Exploration of doped quantum magnets with ultracold atoms},
\href{https://doi.org/10.1016/j.aop.2021.168651}{Annals of Physics, \textbf{435}, 168651 (2021)}.
%
%
\bibitem{Spar2021}
B. M. Spar, E. Guardado-Sanchez, S. Chi, Z. Z. Yan, and W. S. Bakr,
\textit{Realization of a Fermi-Hubbard Optical Tweezer Array},
\href{https://doi.org/10.1103/PhysRevLett.128.223202}{Phys. Rev. Lett. \textbf{128}, 223202 (2022)}.
%
%
\bibitem{Lebrat2024}
M. Lebrat, M. Xu, L. H. Kendrick, A. Kale, Y. Gang, P. Seetharaman, I. Morera, E. Khatami, E. Demler, and M. Greiner,
\textit{Observation of Nagaoka Polarons in a Fermi-Hubbard Quantum Simulator},
\href{https://doi.org/10.1038/s41586-024-07272-9}{Nature \textbf{629}, 317 (2024)}.
%
%
\bibitem{Prichard2024}
M. L. Prichard, B. M. Spar, I. Morera, E. Demler, Z. Z. Yan, and W. S. Bakr,
\textit{Directly imaging spin polarons in a kinetically frustrated Hubbard system}, 
\href{https://doi.org/10.1038/s41586-024-07356-6}{Nature \textbf{629}, 323 (2024)}.
%
%
\bibitem{Kendrick2025}
L. H. Kendrick, A. Kale, Y. Gang, A. D. Deters, M. Lebrat, A. W. Young, and M. Greiner,
\textit{Pseudogap in a Fermi-Hubbard quantum simulator},
\href{https://doi.org/10.48550/arXiv.2509.18075}{arXiv:2509.18075}.
%
%
%
%
%
%
\bibitem{Metzner1991}
W. Metzner,
\textit{Linked-cluster expansion around the atomic limit of the Hubbard model},
\href{https://doi.org/10.1103/PhysRevB.43.8549}{Phys. Rev. B \textbf{43}, 8549 (1991)}.
%
%
\bibitem{Zaitsev1976}
R. O. Zaitsev,
\textit{Diagram technique and gas approximation in the Hubbard model},
Zh. Eksp. Teor. Fiz. \textbf{70}, 1100 (1976),
\href{http://www.jetp.ras.ru/cgi-bin/e/index/e/43/3/p574?a=list}{[Sov. Phys. JETP \textbf{43}, 574 (1976)]}.
%
%
\bibitem{Izyumov1992}
Yu. A. Izyumov, B. M. Letfulov, E. V. Shipitsyn, M. Bartkowiak, and K. A. Chao,
\textit {Theory of strongly correlated electron systems on the basis of a diagrammatic technique for Hubbard operators},
\href{https://doi.org/10.1103/PhysRevB.46.15697}{Phys. Rev. B \textbf{46}, 15697 (1992)}.
%

%
%
\bibitem{Pairault1998}
S. Pairault, D. Sénéchal, and A.-M. S. Tremblay,
\textit{Strong-Coupling Expansion for the Hubbard Model},
\href{https://doi.org/10.1103/PhysRevLett.80.5389}{Phys. Rev. Lett. \textbf{80}, 5389 (1998)}.
%
%
\bibitem{Pairault2000}
S. Pairault, D. Sénéchal, and A.-M. S. Tremblay,
\textit{Strong-coupling perturbation theory of the Hubbard model},
\href{https://doi.org/10.1007/s100510070253}{Eur. Phys. J. B \textbf{18}, 85 (2000)}.
%
%
%
%
\bibitem{Izyumov2005}
Yu. A. Izyumov, N. I. Chaschin, D. S. Alexeev, and F. Mancini,
\textit{A generating functional approach to the Hubbard model},
\href{https://doi.org/10.1140/epjb/e2005-00166-7}{Eur. Phys. J. B \textbf{45}, 69 (2005)}.
%
%
\bibitem{Perepelitsky2015}
E. Perepelitsky and B. S. Shastry,
\textit{Diagrammatic $\lambda$ series for extremely correlated Fermi liquids},
\href{https://doi.org/10.1016/j.aop.2015.03.010}{Ann. Phys. \textbf{357}, 1 (2015)}.
%
%
\bibitem{Carlstroem2021}
J. Carlstr\"{o}m,
\textit{Strong-coupling diagrammatic Monte Carlo technique for correlated fermions and frustrated spins},
\href{https://doi.org/10.1103/PhysRevB.103.195147}{Phys. Rev. B \textbf{103}, 195147 (2021)}.
%
%
\bibitem{Izyumov1994}
Yu. A. Izyumov, B.M. Letfulov, and E.V. Shipitsyn,
\textit{Electronic states in the t-J model near magnetic phase transitions},
Zh. Eksp. Teor. Fiz., \textbf{105} 1357, (1994),
[\href{http://jetp.ras.ru/cgi-bin/e/index/e/78/5/p731?a=list}{Sov. Phys. JETP \textbf{78}, 731 (1994)}].
%
%
\bibitem{Khatami2013}
E. Khatami, D. Hansen, E. Perepelitsky, M. Rigol, and B. S. Shastry,
\textit{Electronic spectral properties of the two-dimensional infinite-$ U $ Hubbard model},
\href{https://doi.org/10.1103/PhysRevB.87.161120}{Phys. Rev. B \textbf{87}, 161120(R) (2013)}.
%
%
\bibitem{Khatami2014}
E. Khatami, E. Perepelitsky, M. Rigol, and B. S. Shastry,
\textit{Linked-cluster expansion for the Green's function of the infinite-$ U $ Hubbard model},
\href{https://doi.org/10.1103/PhysRevE.89.063301}{Phys. Rev. E \textbf{89}, 063301 (2014)}.
%
%
\bibitem{Mai2018}
P. Mai and B. S. Shastry,
\textit{Extremely correlated Fermi liquid of the $ t - J $ model in two dimensions},
\href{https://doi.org/10.1103/PhysRevB.98.205106}{Phys. Rev. B \textbf{98}, 205106 (2018)}.
%
%
\bibitem{Carlstroem2025}
J. Carlstr\"{o}m,
\textit{High-Temperature Phase Separation and Charge-Magnon Liquid in Kinetic Antiferromagnets},
\href{https://doi.org/10.1103/d9c2-yr7j}{Phys. Rev. Lett. \textbf{135}, 106702 (2025)}.
%
%
\bibitem{Rohringer2018}
G. Rohringer, H. Hafermann, A. Toschi, A. A. Katanin, A. E. Antipov, M. I. Katsnelson, A. I. Lichtenstein, A. N. Rubtsov and K. Held,
\textit{Diagrammatic routes to nonlocal correlations
	beyond dynamical mean field theory},
\href{https://doi.org/10.1103/RevModPhys.90.025003}{Rev. Mod. Phys. \textbf{90}, 025003 (2018)}.
%
%
%
%
\bibitem{Arnold2025}
J. Arnold, P. Kopietz, and A. R\"{u}ckriegel,
\textit{Bad metal behavior and Lifshitz transition of a Nagaoka ferromagnet},
\href{https://doi.org/10.48550/arXiv.2510.01909}{arXiv:2510.01909v3 [cond-mat.str-el]}.
%
%
%
%
%
%
\bibitem{Haule2003}
K. Haule, A. Rosch, J. Kroha, and P. W\"{o}lfle,
\textit{Pseudogaps in the $ t $-$ J $ model: An extended dynamical mean-field theory study},
\href{https://doi.org/10.1103/PhysRevB.68.155119}{Phys. Rev. B \textbf{68}, 155119 (2003)}.
%
%
\bibitem{Wang2018}
Y. Wang, B. Moritz, C.-C. Chen, T. P. Devereaux, and K. Wohlfeld,
\textit{Influence of magnetism and correlation on the spectral properties of doped Mott insulators},
\href{https://doi.org/10.1103/PhysRevB.97.115120}{Phys. Rev. B \textbf{97}, 115120 (2018)}.
%
%
\bibitem{White2001}
S. R. White and I. Affleck,
\textit{Density matrix renormalization group analysis of the Nagaoka polaron in the two-dimensional $ t $-$ J $ model},
\href{https://doi.org/10.1103/PhysRevB.64.024411}{Phys. Rev. B \textbf{64}, 024411 (2001)}.
%
%
%
%
\bibitem{Luttinger1960}
J. M. Luttinger,
\textit{Fermi Surface and Some Simple Equilibrium Properties of a System of Interacting Fermions},
\href{https://doi.org/10.1103/PhysRev.119.1153}{Phys. Rev. \textbf{119}, 1153 (1960)}.
%
%
\bibitem{Oshikawa2000}
M. Oshikawa,
\textit{Topological Approach to Luttinger's Theorem and the Fermi Surface of a Kondo Lattice},
\href{https://doi.org/10.1103/PhysRevLett.84.3370}{Phys. Rev. Lett. \textbf{84}, 3370 (2000)}.
%
%
\bibitem{Seki2017}
K. Seki and S. Yunoki,
\textit{Topological interpretation of the Luttinger theorem},
\href{https://doi.org/10.1103/PhysRevB.96.085124}{Phys. Rev. B \textbf{96}, 085124 (2017)}.
%
%
\bibitem{Stephan1991}
W. Stephan and P. Horsch,
\textit{Fermi surface and dynamics of the t-J model at moderate doping},
\href{https://doi.org/10.1103/PhysRevLett.66.2258}{Phys. Rev. Lett. \textbf{66}, 2258 (1991)}.
%
%
%
%
%
\bibitem{Putikka1998}
W. O. Putikka, M. U. Luchini, and R. R. P. Singh,
\textit{Violation of Luttinger's Theorem in the Two-Dimensional $ t $-$ J $ Model},
\href{https://doi.org/10.1103/PhysRevLett.81.2966}{Phys. Rev. Lett. {\bf{81}}, 2966 (1998)}.
%
%
\bibitem{Kokalj2007}
J. Kokalj and P. Prelov\v{s}ek,
\textit{Luttinger sum rule for finite systems of correlated electrons},
\href{https://doi.org/10.1103/PhysRevB.75.045111}{Phys. Rev. B {\bf{75}}, 045111 (2007)}.
%
%
\bibitem{Singh1992}
R. R. P. Singh and R. L. Glenisiter,
\textit{Momentum distribution function for the two-dimensional t-J model},
\href{https://doi.org/10.1103/PhysRevB.46.14313}{Phys. Rev. B \textbf{46}, 14313(R) (1992)}.
%
%
\bibitem{Kozik2015}
E. Kozik, M. Ferrero, and A. Georges,
\textit{Non-existence of the Luttinger-Ward functional and misleading convergence of skeleton diagrammatic series for Hubbard-like models},
\href{https://doi.org/10.1103/PhysRevLett.114.156402}{Phys. Rev. Lett. \textbf{114}, 156402 (2015)}.
%
%
\bibitem{Quinn2018}
E. Quinn,
\textit{Splitting of electrons and violation of the Luttinger sum rule},
\href{https://doi.org/10.1103/PhysRevB.97.115134}{Phys. Rev. B \textbf{97}, 115134 (2018)}.
%
%
\bibitem{Shastry2019}
B. S. Shastry,
\textit{Fermi Surface Volume of Interacting Systems},
\href{https://doi.org/10.1016/j.aop.2019.03.016}{Annals of Physics \textbf{405}, 155 (2019)}.
%
%
\bibitem{Osborne2021}
I. Osborne, T. Paiva, and N. Trivedi,
\textit{Broken Luttinger theorem in the two-dimensional Fermi-Hubbard model},
\href{https://doi.org/10.1103/PhysRevB.104.235122}{Phys. Rev. B \textbf{104}, 235122 (2021)}.
%
%
\bibitem{footnote_reg}
The $ 2 / 3 $ prefactor of $ \delta \mu_\Lambda $
in the symmetric regularization 
is a direct consequence of the holon algebra \eqref{eq:holon_algebra},
which demands that
$ \sum_\sigma n_{ i \sigma }
= 1 - \sum_\sigma h_{ i \sigma } h_{ i \sigma }^\dagger / 2 $.
Hence,
$ \delta \mu_\Lambda
\sum_\sigma n_{ i \sigma }
%
%
= ( \delta \mu_\Lambda / 3 )
\left(
\sum_\sigma
h_{ i \sigma }^\dagger h_{ i \sigma }
+ 2
\sum_\sigma
n_{ i \sigma }
\right)
%
%
= ( \delta \mu_\Lambda / 3 )
\left(
\sum_\sigma h_{ i \sigma }^\dagger h_{ i \sigma }
- 
\sum_\sigma h_{ i \sigma } h_{ i \sigma }^\dagger
+ 2
\right) $.
%
%
\bibitem{Goll2020}
R. Goll, A. R\"{u}ckriegel, and P. Kopietz,
\textit{Zero-magnon sound in quantum Heisenberg ferromagnets},
\href{https://doi.org/10.1103/PhysRevB.102.224437}{Phys.~Rev.~B \textbf{102}, 224437 (2020)}.
%
%
\bibitem{Rueckriegel2025}
A. R\"{u}ckriegel,
\textit{Functional renormalization group for quantum spins},
(Habilitation thesis, Goethe-Universit\"{a}t Frankfurt am Main, 2025).
%
%
\bibitem{footnote_beta_previous}
In our previous work \cite{Rueckriegel2023},
this problem was circumvented at $ T = 0 $ by a judicious choice of a chemical potential counter term.
This is no longer possible at $ T \neq 0 $.
%
%
\bibitem{Profe2022}
J. B. Profe and D. M. Kennes,
{\it{TU$^2$ FRG; a scalable approach for truncated unity functional renormalization group in generic fermionic models}},
\href{https://doi.org/10.1140/epjb/s10051-022-00316-x}{Eur. Phys. J. B {\bf{95}}, 60 (2022)}.
%
%
\bibitem{footnote_d-wave}
The form factor that is expected to be most important is the $d$-wave one, 
$f_d(\bm{k}) = \cos k_x - \cos k_y$, 
for the superconducting channel \cite{Husemann2009,Vilardi2017}. 
Including it, the flow equations remain solvable, 
though no additional contributions to the superconducting channel, 
besides the $s$-wave portion, are generated as long as channel-mixing contributions are neglected.
%
%
%
\bibitem{Kohn1965}
W. Kohn and J. M. Luttinger,
\textit{New Mechanism for Superconductivity},
\href{https://doi.org/10.1103/PhysRevLett.15.524}
{Phys. Rev. Lett. \textbf{15}, 524 (1965)}.
%
%
%
%
%
%
\bibitem{Katanin04}
A. A. Katanin, 
\textit{Fulfillment of Ward identities in the functional renormalization group approach}, 
\href{https://doi.org/10.1103/PhysRevB.70.115109}{Phys. Rev. B \textbf{70}, 115109 (2004)}.
%
%
\bibitem{footnote_tG_sum}
While the Matsubara sum in
$ \int_K t_{ \bm{k} } G_\Lambda ( K ) $
is  formally not convergent,
it does not depend on the regularization because
its regularization-dependent part is proportional to $ t_{ i i } = 0 $.
%
%
%
%
\bibitem{Beach2000}
K. S. D. Beach, R. J. Gooding, and F. Marsiglio,
\textit{Reliable Pad\'{e} analytical continuation method based on a high-accuracy symbolic computation algorithm},
\href{https://doi.org/10.1103/PhysRevB.61.5147}{Phys. Rev. B \textbf{61}, 5147 (2000)}.
%
%
\bibitem{Schoett2016}
J. Sch\"{o}tt, I. L. M. Locht, E. Lundin, O. Gr\r{a}n\"{a}s, O. Eriksson, and I. Di Marco,
\textit{Analytic continuation by averaging Pad\'{e} approximants},
	\href{https://doi.org/10.1103/PhysRevB.93.075104}{Phys. Rev. B \textbf{93}, 075104 (2016)}.
%
\bibitem{Chakravarty1989}
S. Chakravarty, B. I. Halperin, and D. Nelson, 
{\it{Two-dimensional quantum Heisenberg antiferromagnet at low temperatures}},
\href{https://doi.org/10.1103/PhysRevB.39.2344}{Phys. Rev. B {\bf{39}}, 2344 (1989)}.
%
%
%
%
\bibitem{Vilk1997}
Y.M. Vilk, and A.-M.S. Tremblay,
\textit{Non-perturbative many-body approach to the Hubbard model and single-particle pseudogap},
\href{https://doi.org/10.1051/jp1:1997135 }
{J. Phys. I France \textbf{7}, 1309 (1997)}.
%
%
%
%
\bibitem{Sachdev2011}
S. Sachdev, 
\textit{Quantum Phase Transitions}, 
\href{https://doi.org/10.1017/CBO9780511973765}{(Cambridge University Press, Cambridge, UK, 2011)}.
%
%
\bibitem{Schafer2020}
T. Sch\"{a}fer, N. Wentzell, F. Šimkovic IV, Y.-Y. He, C. Hille, M. Klett, C. J. Eckhardt, B. Arzhang, V. Harkov, F.-M. L. Régent, A. Kirsch, Y. Wang, A. J. Kim, E. Kozik, E. A. Stepanov, A. Kauch, S. Andergassen, P. Hansmann, D. Rohe, Y. M. Vilk, J. P. F. LeBlanc, S. Zhang, A.-M. S. Tremblay, M. Ferrero, O. Parcollet, and A. Georges,
\textit{Tracking the Footprints of Spin Fluctuations: A MultiMethod, MultiMessenger Study of the Two-Dimensional Hubbard Model},
\href{https://doi.org/10.1103/PhysRevX.11.011058}{Phys. Rev. X \textbf{11}, 011058 (2021)}.
%
%
%
%
%
\bibitem{Mermin1966}
N. D. Mermin and H. Wagner,
\textit{Absence of Ferromagnetism or Antiferromagnetism in One- or Two-Dimensional Isotropic Heisenberg Models},
\href{https://doi.org/10.1103/PhysRevLett.17.1133}{Phys. Rev. Lett. \textbf{17}, 1133 (1966)}.
%
%
\bibitem{Hohenberg1967}
P. C. Hohenberg,
\textit{Existence of Long-Range Order in One and Two Dimensions},
\href{https://doi.org/10.1103/PhysRev.158.383}{Phys. Rev. \textbf{158}, 383 (1967)}.
%
%
%
%
%
%
%
%
\bibitem{Sandvik2010}
A. W. Sandvik,
{\it Computational Studies of Quantum Spin Systems},
\href{https://doi.org/10.1063/1.3518900}{AIP Conf. Proc. {\bf 1297}, 135 (2010)}.
%
%
\bibitem{Shenker1980}
S. H. Shenker and J. Tobochnik,
\textit{Monte Carlo renormalization-group analysis of the classical Heisenberg model in two dimensions},
\href{https://doi.org/10.1103/PhysRevB.22.4462}{Phys. Rev. B \textbf{22}, 4462 (1980)}.
%
%
\bibitem{Kopietz1989}
P. Kopietz and S. Chakravarty,
\textit{Low-temperature behavior of the correlation length and the susceptibility of a quantum Heisenberg ferromagnet in two dimensions},
\href{https://doi.org/10.1103/PhysRevB.40.4858}{Phys. Rev. B \textbf{40}, 4858 (1989)}.
%
%
%
%
\bibitem{Chubukov2012}
A. V. Chubukov and D. L. Maslov,
\textit{First-Matsubara-frequency rule in a Fermi liquid. I. Fermionic self-energy},
\href{https://doi.org/10.1103/PhysRevB.86.155136}{Phys. Rev. B \textbf{86}, 155136 (2012)}.
%
%
\bibitem{Zitko2013}
R. \v{Z}itko, D. Hansen, E. Perepelitsky, J. Mravlje, A. Georges, and B. S. Shastry,
\textit{Extremely correlated Fermi liquid theory meets dynamical mean-field theory: Analytical insights into the doping-driven Mott transition},
\href{https://doi.org/10.1103/PhysRevB.88.235132}{Phys. Rev. B \textbf{88}, 235132 (2013)}.
%
\bibitem{Emery1995}
V. J. Emery and S. A. Kivelson, {\it{Superconductivity in Bad Metals}},
\href{https://doi.org/10.1103/PhysRevLett.74.3253}{Phys. Rev. Lett. {\bf{74}}, 3253 (1995)}.
%
\bibitem{Deng2013}
X. Deng, J. Mravlje, R. Žitko, M. Ferrero, G. Kotliar, and A. Georges,
\textit{How Bad Metals Turn Good: Spectroscopic Signatures of Resilient Quasiparticles},
\href{https://doi.org/10.1103/PhysRevLett.110.086401}{Phys. Rev. Lett. \textbf{110}, 086401 (2013)}.
%
%
%
%
\bibitem{Kanamori1963}
J. Kanamori,
\textit{Electron Correlation and Ferromagnetism of Transition Metals},
\href{https://doi.org/10.1143/PTP.30.275}{Prog. Theor. Phys. \textbf{30}, 275 (1963)}.
%
%
%
%
\bibitem{Mielke1991a}
A. Mielke,
\textit{Ferromagnetic ground states for the Hubbard model on line graphs},
\href{https://doi.org/10.1088/0305-4470/24/2/005}{J. Phys. A: Math. Gen. \textbf{24}, L73 (1991)}. 
%
%
\bibitem{Mielke1991b}
A. Mielke,
\textit{Ferromagnetism in the Hubbard model on line graphs and further considerations},
\href{https://doi.org/10.1088/0305-4470/24/14/018}{J. Phys. A: Math. Gen. \textbf{24}, 3311 (1991)}. 
%
%
\bibitem{Tasaki2003}
H. Tasaki, 
\textit{Ferromagnetism in the Hubbard Model: A Constructive Approach}, \href{https://doi.org/10.1007/s00220-003-0952-z}{Commun. Math. Phys. \textbf{242}, 445 (2003)}. 
%
%
\bibitem{Hu2025}
H. Hu, O. Vafek, K. Haule, and B. A. Bernevig,
\textit{Ferromagnetism vs. Antiferromagnetism in Narrow-Band Systems: Competition Between Quantum Geometry and Band Dispersion},
\href{https://doi.org/10.1103/zdyq-3m9x}{Phys. Rev. Lett. \textbf{136}, 256505 (2026)}.
%
%
\bibitem{Volovik2016}
G. E. Volovik,
\textit{Topological Lifshitz transitions},
\href{https://doi.org/10.1063/1.4974185}{Low Temp. Phys. \textbf{43}, 47–55 (2017)}.
%
%
%
%
\bibitem{Galan1992}
J. Gal\'{a}n, F. Guinea, and J. A. Verg\'{e}s,
G. Chiappe, and E. Louis,
\textit{Nonconventional behavior of the one-band Hubbard Hamiltonian in two dimensions},
\href{https://doi.org/10.1103/PhysRevB.46.3163}{Phys. Rev. B \textbf{46}, 3163(R) (1992)}.
%
%
\bibitem{Louis1993}
E. Louis and G. Chiappe,
J. Gal\'{a}n, F. Guinea, and J. A. Verg\'{e}s,
\textit{Wave-function renormalization constant for the one-band Hubbard Hamiltonian in two dimensions},
\href{https://doi.org/10.1103/PhysRevB.48.426}{Phys. Rev. B \textbf{48}, 426 (1993)}.
%
%
\bibitem{Dagotto1994}
E. Dagotto,
\textit{Correlated electrons in high-temperature superconductors},
\href{https://doi.org/10.1103/RevModPhys.66.763}{Rev. Mod. Phys. \textbf{66}, 763 (1994)}.
%
%
%
\bibitem{Shastry2013}
B. S. Shastry,
\textit{Extremely correlated Fermi liquids: The formalism},
\href{https://doi.org/10.1103/PhysRevB.87.125124}{Phys. Rev. B \textbf{87}, 125124 (2013)}.
%
%
%
\bibitem{footnote_occupation}
We use the spectral function representation \eqref{eq:n_via_A} to compute the occupation numbers shown in Fig.~\ref{fig:nwalk} 
to avoid having to subtract the systematic error $ \delta $ of Eq.~\eqref{eq:n_error}.
Since the cancellation is already built in exactly in Eq.~\eqref{eq:n_via_A},
we prefer it here.
For the gradient 
$ \nabla_{ \bm{k} } n ( \bm{k} ) $,
this constant offset is however irrelevant. 
%
%
\bibitem{gude}
J. Arnold,
P. Kopietz, and
A. R\"{u}ckriegel,
\textit{Wolfram Mathematica} notebook, 
Goethe University Data Repository (GUDe),
2026,
\href{https://gude.uni-frankfurt.de/handle/gude/722}
{https://gude.uni-frankfurt.de/handle/gude/722}.
%
%
	%
%
	%
	%
	\bibitem{Kitatani2025}
	M. Kitatani, Y. Nomura, S. Sakai, and R. Arita,
	\textit{Luttinger surface and exchange splitting induced by ferromagnetic fluctuations},
	\href{https://doi.org/10.48550/arXiv.2509.21034}{arXiv:2509.21034 [cond-mat.str-el]}.
	%
	\bibitem{Wu2017}
	W. Wu, M. Ferrero, A. Georges, and E. Kozik,
	\textit{Controlling Feynman diagrammatic expansions: physical nature of the pseudo gap in the two-dimensional Hubbard model},
	\href{https://doi.org/10.1103/PhysRevB.96.041105}{Phys. Rev. B \textbf{96}, 041105 (2017)}.
	%
	%
	\bibitem{Lihm2025}
	J.-M. Lihm, D. Kiese, S.-S. B. Lee, and F. B. Kugler,
	\textit{The finite-difference parquet method: Enhanced electron-paramagnon scattering opens a pseudogap},
	\href{https://doi.org/10.1073/pnas.2525308123}{PNAS, \textbf{123}, e2525308123 (2026)}.
%
%
%
%
%
%
%
%
	%
	%
	%
	%
%
%
	%
	%
	%
	%
	%
	%
	%
	%
	%
	%
	%
	%
	%
	%
	%
	%
	%
	%
	%
	%
%
%
%
%
	%
	%
	%
	%
	%
	%
	%
	%
	%
	%
	%
	%
	%
	%
	%
	%
	%
	%
	%
	%
	%
	%
	%
	%
	%
	%
	%
	%
	%
	%
	%
	%
	%
%
%
%

%
%
%
%
\end{thebibliography}
\end{document}